\documentclass{lmcs}


\usepackage{hyperref}
\usepackage{xspace}
\usepackage{etoolbox}
\usepackage{mathtools}
\usepackage[textsize=scriptsize]{todonotes}
\usepackage{mathbbol}
\usepackage{amsmath,amssymb,amsthm}
\usepackage{stmaryrd}
\usepackage{enumitem}

\usepackage{tikz}

\usetikzlibrary{automata}
\usetikzlibrary{arrows}
\usetikzlibrary{backgrounds}
\usetikzlibrary{positioning}
\usetikzlibrary{fit}
\usetikzlibrary{calc}
\usetikzlibrary{decorations}
\usetikzlibrary{decorations.pathmorphing}
\usetikzlibrary{shapes}
\usetikzlibrary{patterns,decorations.pathreplacing}

\newenvironment{romanenumerate}
  {\begin{enumerate}[label=(\roman*)]}
  {\end{enumerate}}

\newenvironment{alphaenumerate}
  {\begin{enumerate}[label=(\alph*)]}
  {\end{enumerate}}

\newcommand{\proofsubparagraph}[1]{\par\smallskip\noindent\textbf{#1.}\ }
\newtoggle{versionlong}

\toggletrue{versionlong}

\newcommand{\cut}[1]{}

\newcommand{\set}[1]{\{ #1 \}}

\newcommand{\pair}[2]{(#1,#2)}
\newcommand{\triple}[3]{(#1,#2,#3)}

\newcommand{\Nat}{\ensuremath{\mathbb{N}}}
\newcommand{\Rat}{\ensuremath{\mathbb{Q}}}
\newcommand{\Rea}{\ensuremath{\mathbb{R}}}
\newcommand{\Zed}{\ensuremath{\mathbb{Z}}}

\newcommand{\powerset}[1]{{\mathcal P}(#1)}

\newcommand{\card}[1]{|#1|}

\newcommand{\dlwedge}{\sqcap}
\newcommand{\dlvee}{\sqcup}
\newcommand{\dlneg}{\neg}
\newcommand{\aconcept}{C}
\newcommand{\aconceptbis}{D}
\newcommand{\aconceptter}{E}
\newcommand{\ainter}{\mathcal{I}}

\newcommand{\aidomain}{\Delta^{\ainter}}

\newcommand{\inter}[1]{#1^{\ainter}}

\newcommand{\aconceptname}{A}
\newcommand{\aconceptnamebis}{B}
\newcommand{\conceptnames}{{\sf N}_{\mathbf{C}}}
\newcommand{\acfeature}{f}
\newcommand{\cfeatures}{{\sf N}_{\mathbf{CF}}}
\newcommand{\arolename}{r}

\newcommand{\arolenamebis}{s}

\newcommand{\rolenames}{{\sf N}_{\mathbf{R}}}
\newcommand{\frolenames}{{\sf N}_{\mathbf{FR}}}
\newcommand{\individualnames}{{\sf N}_{\mathbf{I}}}
\newcommand{\arolepath}{p}
\newcommand{\arp}{\arolepath}

\newcommand{\aontology}{\mathcal{O}}
\newcommand{\atbox}{\mathcal{T}}
\newcommand{\aabox}{\mathcal{A}}

\newcommand{\ALC}{\mathcal{A}\mathcal{L}\mathcal{C}}
\newcommand{\ALCO}{\mathcal{A}\mathcal{L}\mathcal{C}\mathcal{O}}
\newcommand{\ALCI}{\mathcal{A}\mathcal{L}\mathcal{C}\mathcal{I}}
\newcommand{\ALCOI}{\mathcal{A}\mathcal{L}\mathcal{C}\mathcal{O}\mathcal{I}}

\newcommand{\ALCOF}{\mathcal{A}\mathcal{L}\mathcal{C}\mathcal{O}\mathcal{F}}

\newcommand{\ALCOCD}{\mathcal{A}\mathcal{L}\mathcal{C}\mathcal{O}^{{\rm CD}+}}
\newcommand{\acdomain}{\mathcal{D}}
\newcommand{\adomain}{\ensuremath{\mathbb{D}}}

\newcommand{\exrest}[2]{\exists #1 . #2} 
\newcommand{\varest}[2]{\forall #1 . #2}

\newcommand{\aind}{d}
\newcommand{\aindbis}{e}

\newcommand{\exptime}{\ensuremath{\text{\textup{\textsc{ExpTime}}}}\xspace}

\newcommand {\ptime} {\textsc{PTime}\xspace}

\newcommand{\aset}{X}

\newcommand{\asetter}{Z}

\newcommand{\aautomaton}{{\mathbb A}}
\newcommand{\aautomatonbis}{{\mathbb B}}

\newcommand{\egdef}{\stackrel{\mbox{\begin{tiny}def\end{tiny}}}{=}} 
\newcommand{\equivdef}{\stackrel{\mbox{\begin{tiny}def\end{tiny}}}{\equivaut}} 
\newcommand{\equivaut}{\;\Leftrightarrow\;}

\newcommand{\var}{{\rm VAR}}

\newcommand{\amap}{\mathfrak{f}}
\newcommand{\amapbis}{\mathfrak{g}}

\newcommand{\aalphabet}{\mathbf{\Sigma}}     

\newcommand{\locations}{Q}
\newcommand{\alocation}{q}

\newcommand{\aletter}{\ensuremath{\mathtt{a}}}

\newcommand{\variables}{\var}

\newcommand{\defstyle}[1]{{\emph{#1}}}

\newcommand{\interval}[2]{[#1,#2]}

\newcommand{\apath}{\pi}
\newcommand{\arun}{\rho}

\newcommand{\atree}{t}
\newcommand{\adatatree}{\mathbb{t}}
\newcommand{\anode}{n}
\newcommand{\anodebis}{m}

\newcommand{\alang}{{\rm L}}

\newcommand{\avaluation}{\mathfrak{v}}
\newcommand{\aassignment}{\avaluation}

\makeatletter
\def\mylab#1#2{%
  {\textbf{#1}}%
  \begingroup%
    \def\@currentlabel{\textbf{#1}}%
    \phantomsection\label{#2}%
  \endgroup%
}
\makeatother

\newcommand{\transitions}{\delta}

\newcommand{\acons}{\Theta}

\newcommand{\atransition}{\mathtt{t}}

\newcommand{\size}[1]{size(#1)}

\newcommand{\neproblem}[1]{{\rm NE(}#1{\rm)}}

\newcommand{\degree}{\mathtt{d}}

 \newcommand{\treeconstraints}[1]{{\rm Cons}(#1)}

\newcommand{\maxconstraintsize}[1]{\mathtt{MCS}(#1)}

\newcommand{\csp}[1]{\textsf{CSP}(#1)}
\newcommand{\rootlocation}{\ensuremath{\text{\textcircled{$\varepsilon$}}}}
\newcommand{\rootnode}{\rootlocation}

\newcommand{\globalinfo}{\gabs}
\newcommand{\parentinfo}{\mathsf{PI}}

\newcommand{\act}{\mathsf{ACT}}
\newcommand{\slinks}{\mathsf{SLinks}}

\newcommand{\dir}[1]{{dir}(#1)}

\newcommand{\gabs}{\mathsf{GA}}                    
\newcommand{\labs}{\mathsf{LA}}                    
\newcommand{\pabs}{\mathsf{PA}}                    
\newcommand{\gabsset}{\mathcal{GA}}                
\newcommand{\labsset}{\mathcal{LA}}                
\newcommand{\pabsset}{\mathcal{PA}}                

\newcommand{\cabs}{\mathsf{CA}}      
\newcommand{\cabsset}{\mathcal{CA}}    

\newenvironment{tightcenter}{%
  \setlength\topsep{1pt}
  \setlength\parskip{2pt}
  \begin{center}
}{%
  \end{center}
}

\newcommand{\bigstt}[1]{\todo[inline,color=red!25]{{\bf S. to T.:} #1}}

\newcommand{\acsystem}{S} 

\newcommand{\cdrestriction}[3]{#1 \ #2 . \ #3}

\newcommand{\cassertion}[2]{#1:#2} 
\newcommand{\rassertion}[3]{(#1,#2):#3} 

\newcommand{\aindname}{a}
\newcommand{\aindnamebis}{b}
\newcommand{\aindnameter}{c}

\newcommand{\adatum}{\mathbb{d}}
\newcommand{\avalue}{\adatum}

\newcommand{\atuple}{\vec{z}}

\newcommand{\registers}[2]{{\rm REG}(#1,#2)}
\newcommand{\aregister}{x}
\newcommand{\aregisterbis}{y}

\newcommand{\atype}{\tau}

\newcommand{\atoms}[1]{\mathsf{Atoms}(#1)}
\newcommand{\literals}[1]{\mathsf{Literals}(#1)}
\newcommand{\symbtypes}[1]{\mathsf{STypes}(#1)}
\newcommand{\satsymbtypes}[1]{\mathsf{SatSTypes}(#1)}

\newcommand{\apolynomial}{\mathtt{p}}
\newcommand{\aquantifier}{\mathcal{Q}}
\newcommand{\lex}{<_{lex}}
\newcommand{\prefix}{<_{prefix}}

\newcommand{\actype}{T}
\newcommand{\ctypes}[1]{\mathsf{CTypes}(#1)}
\newcommand{\subconcepts}[1]{\mathsf{Sub}(#1)}

\newcommand{\symblinks}{sl}
\newcommand{\aactvector}{act} 

\newcommand{\aecdconcept}{\aconcept^{\star}}

\newcommand{\localabstraction}[2]{locabs(#1,#2)}
\newcommand{\arole}{R}

\newcommand{\inverse}[1]{#1^{-}}
\newcommand{\roles}[1]{{\sf Roles}(#1)} 
\newcommand{\anominal}{\{ \hspace{-0.02in} \aindname \hspace{-0.02in}\}}
\newcommand{\anominalbis}{\{ \hspace{-0.02in} \aindnamebis \hspace{-0.02in}\}}

\tikzset{every picture/.style={scale=0.9,semithick,transform shape,initial text={}}}

\tikzset{->,>=stealth',shorten >=1pt}

\tikzset{every state/.style={draw=black!20,thin,shape=ellipse,inner sep=3pt,minimum size=4mm,shade,ball color=white}}
\tikzset{whitedot/.style={circle,inner sep=0.8pt,minimum size=4mm,shade,ball color=white}}
\tikzset{blackdot/.style={circle,inner sep=0.8pt,minimum size=4mm,shade,ball color=black}}
\tikzset{graydot/.style={circle,inner sep=0.8pt,minimum size=4mm,shade,ball color=gray}}
\tikzset{dot/.style={whitedot}}
\tikzset{markeddot/.style={blackdot}}
\tikzset{autstate/.style={draw,shape=circle,inner sep=3pt,minimum size=4mm}}

\tikzset{allnode/.style={draw,shape=rectangle,minimum size=5mm,shade,top color=black!20,bottom color=black!5}}
\tikzset{exnode/.style={draw,shape=rounded rectangle,minimum size=5mm,shade,top color=black!5,bottom color=black!20}}

\begin{document}

\title[Robustness of Constraint Automata for DLs with Concrete Domains]{Robustness of Constraint Automata for Description Logics with Concrete Domains}



\thanks{The authors gratefully acknowledge financial support from Laboratoire M\'ethodes Formelles (LMF) and
 Graduate School of Computer Science, Universit\'e Paris-Saclay (GS-ISN)}


\author[S.~Demri]{St\'ephane Demri\lmcsorcid{0000-0002-3493-2610}}
\author[T.~Gu]{Tianwen Gu\lmcsorcid{0009-0009-7976-5973}}

\address{Universit\'e Paris-Saclay, CNRS, ENS Paris-Saclay, Laboratoire M\'ethodes Formelles, 91190 Gif-sur-Yvette, France}	






\begin{abstract}
Decidability or complexity issues about the consistency problem
for description logics with concrete domains have already been analysed
with tableaux-based or type elimination methods. Concrete domains
in ontologies are  essential to consider concrete objects
and predefined relations.
In this work, we expose an automata-based approach leading
to the optimal upper bound \exptime,
that is designed by enriching the transitions with symbolic constraints.
We show that the nonemptiness problem for such automata
belongs to \exptime if the concrete domains satisfy
a few simple properties. Then, we provide a reduction from
the consistency problem for ontologies,
yielding \exptime-membership. 
Thanks to the expressivity of constraint automata,
the results are extended  to 
additional ingredients such as  inverse  roles, functional role names
and constraint assertions,
while  maintaining \exptime-membership, which  illustrates the robustness of the approach.
\end{abstract}

\maketitle

\section{Introduction}
\label{section-introduction}

\paragraph*{Landmark results for description logics with  concrete domains.}
In description logics with concrete domains, the elements of 
the interpretation domain are enriched with tuples of
values from the concrete domain, 
see e.g.~\cite{Baader&Hanschke91,Lutz02,Lutz03,Lutz&Milicic07,Baaderetal17},
enabling reasoning about concrete information. 
The first decidability result for a large class of concrete domains $\acdomain$ can be traced back to~\cite{Lutz&Milicic07} in which
the decidability of the consistency problem with $\omega$-admissible concrete domains
is established using tableaux-based decision procedures (apart from the very first decidability result
in~\cite[Theorem~4.3]{Baader&Hanschke91}).
Though this was a genuine breakthrough in understanding
the description logic $\ALC$ with concrete domains, a few restrictions were required:
the set of binary relations is finite; $\acdomain$ satisfies the patchwork property~\cite[Definition~3]{Lutz&Milicic07}
which corresponds to the amalgamation property; and it satisfies compactness, see e.g.~\cite[Section~6.1]{Baader&Lutz06}
and~\cite[Definition~4]{Lutz&Milicic07}
  which corresponds to homomorphism $\omega$-compactness.
  Furthermore,
  constraint domain restrictions (a.k.a. CD-restrictions)
  are of the
  form $\cdrestriction{\aquantifier}{v_1: \arp_1, v_2: \arp_2}{P_1(v_1,v_2) \vee \cdots
    \vee P_k(v_1,v_2)}$~\cite[Definition~6]{Lutz&Milicic07},
  where $\aquantifier \in \set{\exists,\forall}$ and
  the $P_i$'s are binary predicates.
  Moreover, no ABoxes in the consistency problem are handled in~\cite{Lutz&Milicic07}.
  Among the famous
  concrete domains captured by~\cite{Lutz&Milicic07},
  it is worth mentioning  Allen's interval algebra and
  the region connection calculus
RCC8, see more details in~\cite[Section~2]{Lutz&Milicic07}.

Though the works~\cite{Lutz&Milicic07,CarapelleTurhan16} mainly focus on decidability, 
a recent remarkable breakthrough was achieved in~\cite{Borgwardt&DeBortoli&Koopmann24},
which shows that the knowledge base consistency problem for $\ALC$ with $\omega$-admissible
concrete domains (with more conditions)
belongs
to \exptime, refining the decidability results; see also~\cite{Baaderetal25}.
This complexity result nicely extends
the type elimination algorithm for $\ALC$ that leads already to \exptime,
based on the technique using elimination of Hintikka sets established
in~\cite{Pratt79}, see also~\cite[Section~6.8]{Blackburn&deRijke&Venema01},
\cite[Section~2.3]{Marx06}
and~\cite[Section~4.2]{PrattHartmann23}.

\paragraph*{Our motivations.}
The well-known automata-based approach, see e.g.~\cite{Buchi62,Vardi&Wolper86,Vardi06},
consists of reducing logical problems to
automata-based problems to take advantage of results and decision procedures from
automata theory, see also~\cite[Chapter~14]{Demri&Goranko&Lange16}.
Typically,  a reduction is designed and one
relies on automata-based decision procedures. 
Among the expected properties,
the reduction should be conceptually simple even if
showing its correctness requires some substantial work.
Ideally, the
computational
complexity of the automata-based target problems should be
well-understood, so that the reduction yields a tight upper bound
on the complexity of the logical problem. 
Preferably, the reduction should lead to the optimal complexity for the logical
problem.
 
In this paper,
we would like to follow an automata-based approach for reasoning in
the description logics $\ALCO$ with  concrete
    domains
    and regain the \exptime-membership from~\cite{Borgwardt&DeBortoli&Koopmann24}
    (see also the recent work~\cite{Baaderetal25}) based
    on the approach
    from~\cite{Demri&Quaas23bis,Demri&Quaas25}.
    Type elimination and tableaux-based techniques are definitely worth being developed
    but we would like to expose an alternative approach with
    comprehensively developed formal methods that are equally accessible to a wide audience. 
Our motivations definitely rest on the use of
    concrete domains as
    the automata-based approach for description logics (without concrete domains)
  with automata over finite alphabets is already advocated in~\cite[Section~3.2]{Baader09}, see
  also~\cite{Baaderetal03bis} and~\cite[Section~3.6.1]{Baader&Horrocks&Sattler08}. This is the 
  research agenda we wish to pursue with {\em constraint} automata,
  see also~\cite{Kartzow&Weidner15,Peterler&Quaas22,Demri&Quaas23bis}.

\paragraph*{Our contributions.}
  Our first task  is to
  introduce a class of constraint automata parameterised by concrete domains
  that accept {\em infinite data trees} and the
  scope of the constraints slightly extends the scope involved
  in the constraint automata in~\cite{Demri&Quaas23bis,Demri&Quaas25} (Section~\ref{section-tgca}).
  The constraints involved in the transitions have a simple and natural form
  (constraints between siblings are allowed)
  that
  shall lead to a forthcoming smooth reduction from the consistency problem, though
  technical difficulties still arise.
  In order to  get $k$-\exptime-membership of the nonemptiness problem,
  the concrete domains satisfy (C1.1) the completion property~\cite{Demri&DSouza07,Bhaskar&Praveen24} (resp.
  (C1.1.1) the amalgamation property and (C1.1.2)
  homomorphism $\omega$-compactness~\cite{Borgwardt&DeBortoli&Koopmann24}),
  (C2) the arity of predicates is bounded,
  (C3.$k$) the constraint
  satisfaction problem for the concrete domain $\acdomain$
  is in $k$-\exptime
  and (C4) equality is part of the relations.
  We provide a parameterised complexity analysis
  that is helpful to characterise then the complexity of the consistency problem for the
  description logic $\ALCO(\acdomain)$.
  More precisely, we design a reduction to the nonemptiness
  problem for B\"uchi tree automata,
  known to be solved in
  quadratic time, see e.g.~\cite{Chatterjee&Henzinger12}.
  Along the paper, we are particularly interested in determining
  which conditions are really needed for each part of our investigations.

  Then, we design a reduction from the consistency problem for
  the description logics
  $\ALCO(\acdomain)$ with concrete domains $\acdomain$ satisfying the above conditions
  into the nonemptiness problem
  for constraint automata parameterised by $\acdomain$, leading to \exptime-membership if (C3.1) holds,
  which is known to be optimal,
  see e.g.~\cite{Borgwardt&DeBortoli&Koopmann24,Baaderetal25}.
  We mix standard ingredients to translate modal/temporal/description
  logics into tree automata with a quite natural handling of CD-restrictions.
  A few complications arise to handle
  nominals (see e.g.~\cite{Tobies00} and the notion of guess
  in~\cite[Definition~6]{Sattler&Vardi01} about the $\mu$-calculus
  with converse, universality modality and nominals), and the fact that
  concrete features admit partial interpretations. 
  It is notable that we allow slightly more general classes of concrete domains than those
  in~\cite{Borgwardt&DeBortoli&Koopmann24} (see Section~\ref{section-concrete-domains} for a brief discussion about the
  differences). By way of example, we allow more general concrete domain restrictions
  and finiteness of the set
  of relations is not always required.
  Furthermore, we are a bit more liberal with the set of CD-restrictions, which allows us
  to get rid of the condition JEPD (jointly exhaustive pairwise disjoint),
  see e.g.~\cite{Borgwardt&DeBortoli&Koopmann24}.
  Often, we introduce notions that have counterparts
  in~\cite{Borgwardt&DeBortoli&Koopmann24,Baaderetal25} but tailored to our approach.

Constraint automata allow us to add ingredients
  and to adapt smoothly the translation, for instance by
  adding inverse roles while giving up the nominals
  (as hinted possible in~\cite[Section~5]{Borgwardt&DeBortoli&Koopmann24}),
  functional role names and constraint assertions
  (see Section~\ref{section-beyond-alc} and Section~\ref{section-extensions}).
  The extension with inverse roles, though technically challenging, perfectly
  illustrates the versatility of our approach.
  Section~\ref{section-integers} also handles $\ALCO$ over the integers.
  Though this concrete domain does not satisfy the forthcoming condition (C1), we can still
  take advantage of developments in Section~\ref{section-encoding-consistency} and
  in~\cite{Demri&Quaas25}.
  In short,
  we revisit the complexity results about description logics with concrete
  domains from~\cite{Borgwardt&DeBortoli&Koopmann24,Baaderetal25} by advocating an automata-based
  approach, while slightly
  refining some results (use of the completion property)
  and establishing a few new ones (inverse roles).\\

{\em This paper is the extended version of the conference paper~\cite{Demri&Gu26}.}

\section{Preliminaries: Towards the Consistency Problem for Ontologies}
\label{section-preliminaries}
In this section, we introduce the concrete domains $\acdomain$ and the few assumptions
involved in this document. Then, we define the description logics
$\ALCO(\acdomain)$ parameterised by
the concrete domain $\acdomain$.
After recalling the definition of the consistency problem for ontologies
(a.k.a. knowledge
bases), we conclude this section by presenting several additional features that
are handled in Section~\ref{section-extensions}, by taking advantage of our
modular automata-based approach. 

\subsection{Concrete Domains Under Scrutiny}
\label{section-concrete-domains}
A \defstyle{concrete domain} is a relational structure
\(
\acdomain = (\adomain, P_1^\adomain, P_2^\adomain, \dots),
\)
where \(\adomain\) is a nonempty set and each 
\(P_i^\adomain \subseteq \adomain^{k_i}\) 
interprets a predicate symbol \(P_i\) of arity \(k_i\).
Elements of $\adomain$ are written $\adatum_0, \adatum_1, \ldots$. 
Let \(\var = \{v_1, v_2, \dots\}\) be a countably infinite set of variables.
A \defstyle{(Boolean) constraint} is an expression following the grammar below:
\[
\acons ::= P(v_1,\dots,v_k) \mid \lnot \acons \mid \acons
\land \acons \mid \acons \lor \acons,
\]
with \(P\) a predicate symbol of arity $k$ and \(v_1, \dots, v_k \in \var\) (repetitions allowed).
A \defstyle{valuation} is a total function 
\(\avaluation: \var \to \adomain\).
Satisfaction of atomic formulas is defined by
\[
\avaluation \models P(v_1,\dots,v_k) \ \  \equivdef \ \ 
(\avaluation(v_1),\dots,\avaluation(v_k)) \in P^\adomain,
\]
and extended to Boolean combinations in the usual way.
A \defstyle{constraint system}  $\acsystem$
is a set
of 
literals of the form either \(P(v_1,\dots,v_k)\) or  \( \neg P(v_1,\dots,v_k)\).
We write $\var(\acsystem)$ to denote the set of variables occurring
in
$\acsystem$. 
Given $\aset \subseteq \var$, we write $\acsystem|_{\aset}$ to denote the subset
of $\acsystem$ made of literals using only variables from the set $\aset$. 
A valuation \(\avaluation: \var \to \adomain\) satisfies the system \(\acsystem\), 
written \(\avaluation \models \acsystem\), if it satisfies every literal in \(\acsystem\).
The \defstyle{constraint satisfaction problem} for \(\acdomain\), denoted 
\(\csp{\acdomain}\), takes as an input
a finite constraint system \(\acsystem\)
and asks whether there is a valuation $\avaluation$
such that  \(\avaluation \models \acsystem\). 
\cut{
\begin{quote}
  \textbf{Input:} A finite constraint system \(\acsystem\).\\[1mm]
  \textbf{Question:} Does there exist a valuation
  \(\avaluation: \var \to \adomain\) 
such that \(\avaluation \models \acsystem\)?
\end{quote}}

In this document, we assume that the concrete domains satisfy a few properties
to guarantee at least decidability of the logical decision problems, and
$k$-\exptime-membership if possible.
The concrete domain $\acdomain$ satisfies the \defstyle{completion
  property} $\equivdef$
for every finite constraint system \(\acsystem\)
and for every valuation
    \(\avaluation: \aset \to \adomain\) 
    with \(\aset \subseteq \var(\acsystem)\), if \(\avaluation\)
    satisfies the restriction
     \(\acsystem|_{\aset}\) and \(\acsystem\) is satisfiable, then there
     is an extension \(\avaluation': 
    \var(\acsystem) \to \adomain\) such that \(\avaluation'|_{\aset} = \aassignment\)
    and \(\avaluation' \models \acsystem\).

    A constraint system $\acsystem$ is \defstyle{complete} with respect to
    some set $\mathcal{P}$ of predicate symbols
    and some set $\aset$ of variables $\equivdef$
    no other predicate symbol or variable occurs in
    $\acsystem$ and for all $P \in \mathcal{P}$ of arity $k$ and
    $v_1, \ldots, v_k \in \aset$, we have
    either $P(v_1, \ldots, v_k) \in \acsystem$ or $\neg P(v_1, \ldots, v_k) \in \acsystem$
    (but not both).
    
    A concrete domain 
    $\acdomain$ satisfies the
    \defstyle{amalgamation property} $\equivdef$
    \begin{itemize}
    \item either $\acdomain$ has a finite set $\mathcal{P}_{\acdomain}$ of predicate symbols 
      and for all finite constraint systems $\acsystem$, $\acsystem'$
      such that $\acsystem|_{\var(\acsystem) \cap \var(\acsystem')} = \acsystem'|_{\var(\acsystem) \cap \var(\acsystem')}$
      and $\acsystem|_{\var(\acsystem) \cap \var(\acsystem')}$ is complete w.r.t.
      $\mathcal{P}_{\acdomain}$ and $\var(\acsystem) \cap \var(\acsystem')$, 
      we have $(\bigtriangleup)$
      $\acsystem$ and $\acsystem'$ are
      satisfiable (separately) iff
      $\acsystem \cup \acsystem'$ is satisfiable,
    \item or $\acdomain$ has an  infinite set $\mathcal{P}_{\acdomain}$ of predicate symbols and
      for all finite sets $\mathcal{P} \subseteq \mathcal{P}_{\acdomain}$,
      for all finite constraint systems $\acsystem$, $\acsystem'$ built over
      $\mathcal{P}$ such that
      $\acsystem|_{\var(\acsystem) \cap \var(\acsystem')} = \acsystem'|_{\var(\acsystem) \cap \var(\acsystem')}$
      and $\acsystem|_{\var(\acsystem) \cap \var(\acsystem')}$ is complete w.r.t.
      $\mathcal{P}$ and $\var(\acsystem) \cap \var(\acsystem')$, 
    we have ($\bigtriangleup$). 
    \end{itemize}
    Similar definitions about amalgamation can be found
    in~\cite{Lutz&Milicic07,Rydval22,Baader&Rydval22,Borgwardt&DeBortoli&Koopmann24,Baaderetal25}.
    Note that the involved constraint systems are finite, and unlike~\cite{Lutz&Milicic07,Rydval22},
    we do not assume completeness of the constraint systems,
    see also the related notion of symbolic type in
  Section~\ref{section-reduction-to-bta}.

  Finally,  
      a concrete domain 
      $\acdomain$
      is
      \defstyle{homomorphism \(\omega\)-compact}
      (see e.g.~\cite[Section~2]{Borgwardt&DeBortoli&Koopmann24}) 
      $\equivdef$ 
      if every finite $\acsystem' \subseteq \acsystem$ of
      a  countable constraint system $\acsystem$ is satisfiable,
      then $\acsystem$ itself is satisfiable.

      We are ready to define the main properties satisfied by  concrete domains
      involved herein.
      \begin{description}
      \item[(C1)] The concrete domain $\acdomain$ satisfies
        \begin{enumerate}
        \item the completion property {\em or}
        \item it satisfies the amalgamation property {\em and} is
          homomorphism $\omega$-compact.\\
          \hfill \ (from local to global)
        \end{enumerate}
      \item[(C2)] There is $k_0$ such that all the relations in $\acdomain$ have arity
        bounded by $k_0$.\\ 
        \hfill \ (bounded arity)
      \item[(C3.$k$)] $\csp{\acdomain}$ is in $k$-\exptime ($k \geq 1$). \hfill
        ($k$-\exptime CSP problem)
      \item[(C4)] $\acdomain$ contains the equality relation.
        \hfill (equality in the concrete domain)
      \end{description}

The condition~(C4) is a bit less general than the condition JD (jointly diagonal)
from~\cite{Borgwardt&DeBortoli&Koopmann24,Baader&Rydval22} but simpler to check.
The completion property has  been already considered
in the literature many times, see e.g.~\cite{Balbiani&Condotta02,Demri&DSouza07,Bhaskar&Praveen24,Dauvier&Filiot&Reynier26},
sometimes under the term ``global consistency'',
see e.g.~\cite[Definition~2.3]{Dechter92}
and~\cite{Lutz&Milicic07}.
By way of example,  concrete domains satisfying the completion property
include $\triple{\Rea}{<}{=}$,
$\triple{\Rat}{<}{=}$ and $\pair{\adomain}{=}$ for
some infinite set $\adomain$.
The concrete domain RCC8 with space regions in $\Rea^2$ equipped with
topological relations between spatial regions~\cite{Wolter&Zakharyaschev00}
also satisfies the completion property,
see e.g.~\cite[Section~2.1]{Lutz&Milicic07} and~\cite{Demri&DSouza07}. 
This also applies to the
Allen's temporal concrete domain 
$\acdomain_{A} = \pair{I_{\Rat}}{(R_i)_{i \in \interval{1}{13}}}$,
where $I_{\Rat}$ is the set of closed intervals $\interval{r}{r'} \subseteq \Rat$
and $(R_i)_{i \in \interval{1}{13}}$ is the family of 13 Allen's relations~\cite{Allen83},
see also~\cite[Section~2]{Lutz&Milicic07}. Observe that
RCC8 and Allen's interval algebra also satisfy (C2), (C3.1) and (C4),
see e.g.~\cite[Section~2]{Lutz&Milicic07}. 
The completion property being one alternative
in the condition (C1) is a natural condition that happens to
be sufficient for our needs.
\iftoggle{versionlong}{
However, the satisfaction of (C1), (C2), (C3.1) and (C4) excludes 
the concrete domains $\triple{\Nat}{<}{=}$
and $\triple{\Zed}{<}{=}$ that are nevertheless handled
in~\cite{Labai&Ortiz&Simkus20,Labai21,Demri&Quaas25}, see also Section~\ref{section-integers}.
}{
However, this excludes 
the concrete domain $\triple{\Nat}{<}{=}$
that is nevertheless handled
in~\cite{CarapelleTurhan16,Labai&Ortiz&Simkus20,Labai21,Demri&Quaas25}.
}
Indeed, $\Nat$ (resp. $\Zed$) does not satisfy the completion property
and is not homomorphism $\omega$-compact.
A simple property can  be already established.

\begin{lem}
\label{lemma-cp-implies-ap}
If 
a concrete domain \(\acdomain\) satisfies the completion property, then 
it  satisfies
the amalgamation property.
\end{lem}

\begin{proof}
  Let $\mathcal{P}_{\acdomain}$ be the set of predicate symbols of the constraint domain
  $\acdomain$. We treat uniformly the case $\mathcal{P}_{\acdomain}$ is finite and
  the case $\mathcal{P}_{\acdomain}$  is infinite. So, let $\mathcal{P}  = \mathcal{P}_{\acdomain}$
  if $\mathcal{P}_{\acdomain}$ is finite, and some finite non-empty
  $\mathcal{P}  \subseteq \mathcal{P}_{\acdomain}$ otherwise.
  
  Let \(\acsystem_1\) and \(\acsystem_2\) be finite constraint systems
  built over $\mathcal{P}$, 
  and assume that
  $\acsystem_1|_{V_0} = \acsystem_2|_{V_0}$
  and $\acsystem_1|_{V_0}$ is complete w.r.t.  $\mathcal{P}$ and $V_0$
  with  \(V_0 \egdef \var(\acsystem_1) \cap \var(\acsystem_2)\). 
We show both directions.

\proofsubparagraph{(\(\Rightarrow\)) If \(\acsystem_1\) and \(\acsystem_2\) are satisfiable, 
then \(\acsystem_1 \cup \acsystem_2\) is satisfiable}

Let \(\acsystem_0 \egdef \acsystem_1|_{V_0} = \acsystem_2|_{V_0}\) be the constraint system
restricted to the variables in $V_0$. 
Since both \(\acsystem_1\) and \(\acsystem_2\) are satisfiable and agree on shared variables, 
\(\acsystem_0\) is satisfiable.

Let \(\avaluation_0: V_0 \to \adomain\) be a valuation such that 
\(\avaluation_0 \models \acsystem_0\).
By the completion property, 
\begin{itemize}
  \item there exists \(\avaluation_1: \var(\acsystem_1) \to \adomain\) 
    such that \(\avaluation_1|_{V_0} = \avaluation_0\) and \(\avaluation_1 \models \acsystem_1\),
  \item there exists \(\avaluation_2: \var(\acsystem_2) \to \adomain\) 
    such that \(\avaluation_2|_{V_0} = \avaluation_0\) and \(\avaluation_2 \models \acsystem_2\).
\end{itemize}
Define \(\avaluation: \var(\acsystem_1) \cup \var(\acsystem_2) \to \adomain\) by:
\[
\avaluation(v) \egdef
\begin{cases}
\avaluation_1(v) & \text{if } v \in \var(\acsystem_1), \\
\avaluation_2(v) & \text{if } v \in \var(\acsystem_2).
\end{cases}
\]
Note that this is well-defined since \(\avaluation_1\) and \(\avaluation_2\) agree on \(V_0\).
Hence, \(\avaluation \models \acsystem_1 \cup \acsystem_2\), so the union is satisfiable.

\proofsubparagraph{(\(\Leftarrow\)) If \(\acsystem_1 \cup \acsystem_2\) is satisfiable, 
then \(\acsystem_1\) and \(\acsystem_2\) are satisfiable}

Let \(\avaluation: \var(\acsystem_1) \cup \var(\acsystem_2) \to \adomain\) 
be a valuation satisfying \(\acsystem_1 \cup \acsystem_2\). Then
$\avaluation \models \acsystem_1$ and
$\avaluation \models \acsystem_2$, whence \(\acsystem_1\) and \(\acsystem_2\) are satisfiable.
\end{proof}

The converse implication is not known to hold in general; that is, 
whether the amalgamation property (or some variant) implies the completion property remains open. 
More broadly, the exact relationships between the completion property, the amalgamation property, 
and \(\omega\)-compactness are not fully understood.
By contrast with~\cite{Borgwardt&DeBortoli&Koopmann24,Baaderetal25},
observe that
the satisfaction of (C1), (C2), (C3.1) and (C4) allows the concrete domains to have an infinite
set of relations as in $\triple{\Rat}{<}{(=_{q})_{q \in \Rat}}$.
The condition JEPD from~\cite{Borgwardt&DeBortoli&Koopmann24} is
not required in our framework and (C4) implies JD (jointly diagonal)
from~\cite{Borgwardt&DeBortoli&Koopmann24}.
Unlike the completion property in (C1), the second disjunct of (C1) is often assumed in the
literature,
see e.g.~\cite{Lutz&Milicic07,Rydval22} and
the \exptime-$\omega$-admissible concrete domains
in~\cite{Borgwardt&DeBortoli&Koopmann24,Baaderetal25}.
\subsection{Reasoning about Ontologies with Concrete Objects}
\label{section-consistency-problem}

In this section, we define
the description logic $\ALCO(\acdomain)$ (over the concrete domain
$\acdomain$) defined as the description logic $\ALC(\acdomain)$
from~\cite[Section 2]{Borgwardt&DeBortoli&Koopmann24} (see also~\cite{Baaderetal25})
except that
the assumptions about $\acdomain$ may differ and the concrete domain restrictions (a.k.a.
CD-restrictions) are defined in a slightly more general way.
Let
\(\conceptnames = \set{\aconceptname, \aconceptnamebis, \ldots}\),
\(\rolenames = \set{\arolename, \arolenamebis, \ldots}\),
$\individualnames = \set{\aindname, \aindnamebis, \aindnameter, \ldots}$, 
and \(\cfeatures = \set{\acfeature_0, \acfeature_1, \ldots}\)
be countable sets of
\defstyle{concept names},
\defstyle{role names},
\defstyle{individual names} 
and
\defstyle{concrete features}, respectively.
Following the assumptions from~\cite{Borgwardt&DeBortoli&Koopmann24},
a \defstyle{role path} $\arolepath$ is either a concrete feature $\acfeature$
or the concatenation $\arolename \cdot \acfeature$ of a role name $\arolename$
and a concrete feature $\acfeature$. 
The set of \defstyle{complex concepts} in \(\ALCO(\acdomain)\) is defined as follows.
\[
  \aconcept \coloneqq {} \quad 
  \aconceptname \mid \anominal \mid \top \mid \bot \mid
  \dlneg \aconcept \mid 
  \aconcept \dlwedge \aconcept  \mid
  \aconcept  \dlvee \aconcept \mid
  \exrest{\arolename}{\aconcept}  \mid
  \varest{\arolename}{\aconcept} \mid
  \]
\[
  \cdrestriction{\exists}{v_1: \arp_1, \dots, v_k: \arp_k}{\acons(v_1,\dots,v_k)}
  \mid
  \cdrestriction{\forall}{v_1: \arp_1, \dots, v_k: \arp_k}{\acons(v_1,\dots,v_k)},
\]
\cut{
\begin{align*}
  \aconcept \coloneqq {} \quad &
  \aconceptname \mid \anominal \mid \top \mid \bot \mid
  \dlneg \aconceptname \mid 
  \aconcept \dlwedge \aconcept  \mid
  \aconcept  \dlvee \aconcept \mid
  \exrest{\arolename}{\aconcept}  \mid
  \varest{\arolename}{\aconcept}  \\
  & \mid
  \cdrestriction{\exists}{v_1: \arp_1, \dots, v_k: \arp_k}{\acons(v_1,\dots,v_k)}
  \mid
  \cdrestriction{\forall}{v_1: \arp_1, \dots, v_k: \arp_k}{\acons(v_1,\dots,v_k)},
\end{align*}
}
where \(\aconceptname \in \conceptnames\),
$\aindname \in \individualnames$,
\(\arolename \in \rolenames\),
and $\acons(v_1,\dots,v_k)$ is a $\acdomain$-constraint built over the {\em distinct}
variables $v_1, \ldots, v_k$ (not necessarily an atomic formula). 
The variables \(v_i\)  bind values retrieved via the paths \(\arp_i\).
The expression $\anominal$ is known as a nominal and is interpreted by a singleton,
namely the one containing the interpretation of the individual name $\aindname$. 
For instance,
let $\acdomain = (\Rea, <, =)$, and suppose $\mathsf{Patient} \in \conceptnames$, 
$\mathsf{age} \in \cfeatures$, and $\mathsf{hasBrother} \in \rolenames$. 
The concept
\(
\mathsf{Patient} \dlwedge \cdrestriction{\forall}{v_1: \mathsf{age}}{v_1 < 18}
\)
describes patients younger than 18, while
\[
\cdrestriction{\exists}{v_1: \mathsf{age},\; v_2: \mathsf{hasBrother} \cdot \mathsf{age}}{v_2 < v_1}
\]
captures individuals older than at least one of their brothers.

A \defstyle{general concept inclusion} (GCI) is an expression
of the form
\(\aconcept \sqsubseteq \aconceptbis\), where $\aconcept$, $\aconceptbis$
are $\ALCO(\acdomain)$
concepts. A \defstyle{role assertion} (resp. \defstyle{concept assertion})
is an expression
of the form
$\rassertion{\aindname}{\aindnamebis}{\arolename}$
(resp. $\cassertion{\aindname}{\aconcept}$).
An \defstyle{ontology} is a pair  \(\aontology = (\atbox, \aabox)\),
where $\atbox$ is a finite set of GCIs and $\aabox$ is a finite set of assertions
(the letter `$\mathcal{O}$' is usually used to denote the presence of nominals
in description logics, but no confusion will be possible).
As we shall provide complexity analyses, let us briefly describe how the size
of ontologies is defined.
The size of
an ontology $\aontology$, written $\size{\aontology}$, is defined
as the sum of the size of $\atbox$, written $\size{\atbox}$, and the size of $\aabox$,
written $\size{\aabox}$. The size of a TBox is
defined as the sum of the sizes of
  its GCIs, whereas the size of an ABox is defined as the sum of the sizes of its
  assertions. Assuming that the size of a concept $\aconcept$, written $\size{\aconcept}$,
  is defined as the sum between
  the number of its subconcepts and the number of its subconstraints occurring in CD-restrictions,
  the size of a GCI is the sum of the size of its two concepts.
  Moreover, the size of a role assertion is defined as one,
  and the size of a concept assertion is defined as the size of its concept. 

An \defstyle{interpretation} \(\ainter = (\aidomain, \inter{\cdot})\) consists of
a nonempty (interpretation) domain \(\aidomain\) (its elements
are written $\aind, \aindbis, \ldots$)
and an interpretation function
$\inter{\cdot}$ such that 
\begin{itemize}
\itemsep 0 cm 
      \item \(\inter{\aconceptname} \subseteq \aidomain\) for all \(\aconceptname \in \conceptnames\),
      \item \(\inter{\arolename} \subseteq \aidomain \times \aidomain\) for all 
            \(\arolename \in \rolenames\),
      \item $\inter{\aindname} \in \aidomain$ for all $\aindname \in \individualnames$.
          \item \(\inter{\acfeature}: \aidomain \to \adomain\) is a partial function, for
           all \(\acfeature 
      \in \cfeatures\).
    \item Given $\aind \in \aidomain$, 
      \(\inter{\arp}(\aind)\) denotes either \(\set{\inter{\acfeature}(\aind)}\) for 
        \(\arp = \acfeature\) or \(\set{\inter{\acfeature}(\aindbis) \mid
          \pair{\aind}{\aindbis} \in \inter{\arolename}}\)
        for \(\arp = \arolename \cdot \acfeature\).
        Such sets might be empty in case $\inter{\acfeature}$ is undefined
        for the involved elements of $\aidomain$. 
    \end{itemize}
The semantics
below is  standard, except CD-restrictions
    admit arbitrary constraints, see also~\cite[Definition~1]{Rydval22}.
\cut{
    {\small
      \begin{tightcenter}
      $\inter{\top} \egdef \aidomain, \  \inter{\bot} \egdef \varnothing, \ 
  \inter{\anominal} \egdef \set{\inter{\aindname}}, \
  \inter{(\dlneg \aconceptname)} \egdef \aidomain \setminus \inter{\aconceptname},
  \inter{(\aconcept \dlwedge \aconceptbis)} \egdef \inter{\aconcept} \cap \inter{\aconceptbis}, 
  \inter{(\aconcept \dlvee \aconceptbis)} = \inter{\aconcept} \cup \inter{\aconceptbis},$ \\
  $\inter{\exrest{\arolename}{\aconcept}} \egdef \set{\, \aind \in \aidomain \mid \exists \aindbis:\, 
  \pair{\aind}{\aindbis} \in \inter{\arolename},\ \aindbis \in \inter{\aconcept} \,},
\inter{\varest{\arolename}{\aconcept}}\egdef \set{\, \aind \in \aidomain \mid \forall \aindbis:\, 
  \pair{\aind}{\aindbis} \in \inter{\arolename} \Rightarrow \aindbis \in \inter{\aconcept}\,},$ \\
  $\inter{
  (\cdrestriction{\exists}{v_1: \arp_1, \dots, v_k: \arp_k}{\acons})
} 
\egdef \left\{\, \aind \in \aidomain \;\middle|\; \exists \ (\avalue_1,\dots,\avalue_k) \in \prod \inter{\arp_j}(\aind):\ 
[v_1 \mapsto \avalue_1, \dots, v_k \mapsto \avalue_k] \models \acons \,\right\},$\\
      $\inter{
  (\cdrestriction{\forall}{v_1: \arp_1, \dots, v_k: \arp_k}{\acons})} 
\egdef \left\{\, \aind \in \aidomain \;\middle|\; \forall \ (\avalue_1,\dots,\avalue_k) \in \prod \inter{\arp_j}(\aind):\ 
       [v_1 \mapsto \avalue_1, \dots, v_k \mapsto \avalue_k] \models \acons \,\right\}$.
      \end{tightcenter}
    }
}
{\small
  \begin{align*}
    &
    \inter{\top} \egdef \aidomain, \  \inter{\bot} \egdef \varnothing, \ 
  \inter{\anominal} \egdef \set{\inter{\aindname}}, \
  \inter{(\dlneg \aconcept)} \egdef \aidomain \setminus \inter{\aconcept},
  \inter{(\aconcept \dlwedge \aconceptbis)} \egdef \inter{\aconcept} \cap \inter{\aconceptbis}, 
\inter{(\aconcept \dlvee \aconceptbis)} = \inter{\aconcept} \cup \inter{\aconceptbis}, \\
&
\inter{(\exrest{\arolename}{\aconcept})} \egdef \set{\, \aind \in \aidomain \mid \exists \aindbis:\, 
  \pair{\aind}{\aindbis} \in \inter{\arolename},\ \aindbis \in \inter{\aconcept} \,},
\inter{(\varest{\arolename}{\aconcept})}\egdef \set{\, \aind \in \aidomain \mid \forall \aindbis:\, 
  \pair{\aind}{\aindbis} \in \inter{\arolename} \ \mbox{implies} \ \aindbis \in \inter{\aconcept}\,}, \\
&
\inter{
  (\cdrestriction{\exists}{v_1: \arp_1, \dots, v_k: \arp_k}{\acons})
} 
\egdef \\
& \left\{\, \aind \in \aidomain \;\middle|\; \mbox{there is} \ (\avalue_1,\dots,\avalue_k) \in \prod \inter{\arp_j}(\aind):\ 
[v_1 \mapsto \avalue_1, \dots, v_k \mapsto \avalue_k] \models \acons \,\right\},\\
 &
\inter{
  (\cdrestriction{\forall}{v_1: \arp_1, \dots, v_k: \arp_k}{\acons})
} 
\egdef \\
& \left\{\, \aind \in \aidomain \;\middle|\; \mbox{for all} \ (\avalue_1,\dots,\avalue_k) \in \prod \inter{\arp_j}(\aind):\ 
[v_1 \mapsto \avalue_1, \dots, v_k \mapsto \avalue_k] \models \acons \,\right\}.
  \end{align*}
  }
By way of example, given the graphical representation below of some interpretation $\ainter$,
we have $\inter{\bigl(\cdrestriction{\exists}{v_1: r_1\acfeature_1,\ v_2: r_2\acfeature_2}{v_1 < v_2}\bigr)}
= \set{\aind_1, \aind_4}$.
\begin{center}
\scalebox{0.8}{
   \begin{tikzpicture}[->,>=stealth',shorten >=1pt,auto,node distance=4cm,thick,node/.style={circle,draw,scale=0.7}, roundnode/.style={circle, black, draw=black},
   rectnode/.style={rectangle, black, draw=black}, 
   roundrectnode/.style={rounded corners=.99mm, black, draw=black}]
\begin{scope}[yshift=-1.5cm, scale=1.2, transform shape]
\node[state] (a_1) at (0,0) {$\aind_1$}; 
\node[state] (a_2) at (2,0) {$\aind_2$};
\node[state] (a_4) at (2,-2) {$\aind_4$}; 
\node[state] (a_3) at (0,-2) {$\aind_3$};
\node[left=0.5mm of a_1] {$\acfeature_1=1\atop \acfeature_2=3$};
\node[right=0.5mm of a_2] {$\acfeature_1=1\atop \acfeature_2=2$};
\node[left=0.5mm of a_3] {$\acfeature_1\text{ undef.}\atop \acfeature_2\text{ undef.}$};
\node[right=0.5mm of a_4] {$\hspace{-0.5cm} \acfeature_1=0 \atop \acfeature_2\text{ undef.}$};

\path [->] (a_1)  edge[bend left]  node [above,sloped] {$\arolename_1$} (a_4);
\path [->] (a_4)  edge[bend left]  node [above] {$\arolename_2$} (a_1);
\path [->] (a_1)  edge[bend left]  node [above] {$\arolename_2$} (a_2);
\path [->] (a_4)  edge[bend right]  node [right] {$\arolename_1$} (a_2); 
\path [->] (a_1)  edge[bend right]  node [left] {$\arolename_1$} (a_3);
\end{scope}
\end{tikzpicture}}
\end{center}
An interpretation \(\ainter\) \defstyle{satisfies} the ontology \(\aontology = (\atbox, \aabox)\)
(written $\ainter \models \aontology$)
$\equivdef$  \(\inter{\aconcept} \subseteq \inter{\aconceptbis}\) for all 
    \(\aconcept \sqsubseteq \aconceptbis \in \atbox\), 
    \(\inter{\aindname} \in \inter{\aconcept}\) for all \(\cassertion{\aindname}{\aconcept} \in \aabox\)
    and 
   \(\pair{\inter{\aindname}}{\inter{\aindnamebis}} \in \inter{\arolename}\) for all 
   \(\rassertion{\aindname}{\aindnamebis}{\arolename} \in \aabox\).

   The \defstyle{consistency problem for $\ALCO(\acdomain)$}
   consists in determining for some  ontology
   \(\aontology = (\atbox, \aabox)\)
   whether there exists some interpretation that satisfies it.
   Recall that  $\cassertion{\aindname}{\aconcept}$
   (resp. $\rassertion{\aindname}{\aindnamebis}{\arolename}$)
   can be encoded by $\anominal \sqsubseteq \aconcept$
   (resp. $\anominal \sqsubseteq \exrest{\arolename} \anominalbis$).
   Herein, we design an {\em automata-based decision procedure} 
for solving the fundamental consistency problem of $\ALCO$ with a concrete domain $\acdomain$
satisfying the conditions (C1), (C2), (C3.1), (C4), and {\em get \exptime-membership}.
The consistency problem for $\ALC$ (without concrete domains) can take
advantage of the {\em finite model property}, see e.g.~\cite[Section 4.2]{Baaderetal17},
but this property does not hold for $\ALCO(\acdomain)$.
For instance, within $\ALCO(\Rat, <)$, the GCI $(\top \sqsubseteq \cdrestriction{\exists}{
  v_1: \acfeature_1, v_2: \arolename \cdot \acfeature_1}{v_1 < v_2})$ can be satisfied
only by interpretations with infinite domains $\aidomain$. 
As usual, we write $\ALC(\acdomain)$ to denote
the fragment of
$\ALCO(\acdomain)$ without nominals.
We find it natural to handle nominals in the language of
complex concepts as ontologies admit ABoxes (see also~\cite{Baaderetal25}). However,
there is a price to pay. 
It is  challenging to handle
nominals  with an automata-based
approach as only one node of the accepted trees should correspond to the interpretation
of each nominal. Furthermore, infinite interpretations
(see the example above for $\ALCO(\Rat, <)$)
and satisfying $\top \sqsubseteq \exrest{\arolename}{\anominal}$
lead to interpretations such that $\inter{\aindname}$ has an
{\em infinite amount} of incoming $\arolename$-edges.

The first general result about the complexity
of $\ALC$ with concrete domains is recalled below, refining the decidability results from~\cite{Lutz&Milicic07}
(in~\cite{Borgwardt&DeBortoli&Koopmann24} without nominals,
in~\cite{Baaderetal25} with nominals).
This is obtained  by using the type elimination 
method initially designed for PDL, see
e.g.~\cite{Pratt79} and~\cite[Section 5.1.2]{Baaderetal17}.

\begin{prop} \label{proposition-dl-2024} \cite[Section~3]{Borgwardt&DeBortoli&Koopmann24}
\cite[Theorem~2]{Baaderetal25}
The consistency problem for $\ALCO(\acdomain)$ is in \exptime if
$\acdomain$ has a finite family of relations,  satisfies the second disjunct of (C1),
(C3.1), the conditions JEPD, JD and the CD-restrictions are only of the form 
$\cdrestriction{\aquantifier}{v_1: \arp_1, \dots, v_k: \arp_k}{P(v_1,\dots,v_k)}$.
\end{prop}

Note that~\cite[Theorem~8]{Borgwardt&DeBortoli&Koopmann24} does not involve nominals
unlike~\cite[Theorem~2]{Baaderetal25} and, role and concept assertions  can be encoded
by GCIs thanks to the nominals. 

Given a set $\aset$ of individual names, 
an interpretation $\ainter$ satisfies the unique name assumption (UNA) w.r.t.
$\aset$ iff for all distinct individual names
$\aindname \neq \aindname' \in \aset$, 
we have 
$\inter{\aindname} \neq \inter{\aindname'}$. Below, we state
a standard result that allows us to deal with interpretations
satisfying (UNA) only, as we are interested in complexity classes
equal or above \exptime.
Indeed, suppose
that $\ainter \models \aontology$ and $\equiv$ be the
equivalence relation on $\set{\aindname_0, \ldots, \aindname_{\eta-1}}$
(individual names in $\aontology$)
such that $\aindname \equiv \aindnamebis$
iff $\inter{\aindname} = \inter{\aindnamebis}$. We write $[\aindname]$ to denote
the equivalence class of $\aindname$. We can build a new
ontology $\aontology_{\equiv} = \pair{\atbox_{\equiv}}{\aabox_{\equiv}}$
as the quotient of $\aontology$ by the equivalence
relation $\equiv$: in each concept, the nominal $\anominal$ is replaced
by $\{ \hspace{-0.02in} [\aindname] \hspace{-0.02in}\}$
and each individual name $\aindname$ in the assertions
is replaced by $[\aindname]$.
One can show that
$\aontology$ is consistent iff
there is some equivalence relation
$\equiv$ on $\set{\aindname_0, \ldots, \aindname_{\eta-1}}$
such that 
$\aontology_{\equiv}$ can be satisfied by an interpretation
satisfying (UNA) w.r.t. $\set{[\aindname_j]
  \mid j \in \interval{0}{\eta-1}}$. The equivalence classes
$[\aindname_j]$ are viewed as new individual names.

\subsection{Beyond \texorpdfstring{$\ALCO$}{} with Concrete Domains}
\label{section-beyond-alc}
We briefly present several ingredients 
that can be found in description logics, see e.g.~\cite{Baaderetal17}.
Section~\ref{section-extensions} is dedicated to showing how the
automata-based approach for $\ALCO(\acdomain)$
developed in Section~\ref{section-encoding-consistency} can be
adapted to these additional features.
We write $\ALCI(\acdomain)$ to denote the
extension of $\ALC(\acdomain)$ (without nominals)
with inverse roles. The roles are either  $\arolename$ or $\arolename^{-}$ where $\arolename
\in \rolenames$ and $\inter{(\arolename^{-})}$ is equal to the converse of $\inter{\arolename}$.
Herein, inverse roles can be found in restrictions as in $\exrest{\arolename^{-}}{\aconcept}$
and in CD-restrictions as in
$\cdrestriction{\forall}{v_1: \acfeature_1, v_2: \arolename^{-} \acfeature_2}{\acons(v_1,v_2)}$.
We write $\ALCOF(\acdomain)$  to denote the extension of $\ALCO(\acdomain)$
with functional role names, see e.g.~\cite{Lutz01b}.
We introduce the set $\frolenames$ of functional role
names (a.k.a. abstract features) and the roles in $\ALCOF(\acdomain)$ are the role names
from $\rolenames \cup \frolenames$. If $\arolename \in \frolenames$ and $\ainter$ is
an interpretation of $\ALCOF(\acdomain)$, then $\inter{\arolename}$
is weakly functional. Functional role names can be used anywhere.
We may also enrich the set of assertions by allowing in  ontologies, expressions
 $\acons(\acfeature_1(\aindname_1), \ldots,\acfeature_k(\aindname_k))$ stating
constraints between the concrete feature values of the individual names
$\aindname_1, \ldots, \aindname_k$ (predicate assertions are usually restricted
to atomic constraints).
\cut{
\subparagraph{More CD-restrictions.} In~$\ALC(\acdomain)$, the way CD-restrictions
$\cdrestriction{\aquantifier}{v_1: \arp_1, \dots, v_k: \arp_k}{\acons}$ are interpreted
cannot force that two values associated to the variables $v_i$ and $v_j$ with $i \neq j$
are taken from the same element of the domain. Such a possibility can be
found in the description logics in~\cite{Labai&Ortiz&Simkus20,Demri&Quaas23bis}
but not in~\cite{Borgwardt&DeBortoli&Koopmann24}.
We write $\ALCOCD(\acdomain)$  to denote the extension of $\ALCO(\acdomain)$
by adding CD-restrictions of the form
\[
\cdrestriction{\aquantifier}{
  \{v_1^1: \arp_1^1, \dots, v_{k_1}^1: \arp_{k_1}^1\},
  \ldots,
  \{v_1^N: \arp_1^N, \dots, v_{k_N}^N: \arp_{k_N}^N\}
}{\acons(v_1^1,\dots,v_{k_N}^N)},\]
where for all $i$, if $k_i > 1$ then all $\arp_1^i, \dots,\arp_{k_i}^i$ are
feature paths of length two  related to the same role name
and all the values $v_1^i, \ldots, v_{k_i}^i$ are taken from
the {\em same}  successor element.
When $k_i = 1$, the curly parentheses can be omitted;  
the fragment $\ALC(\acdomain)$ corresponds to the case $k_1 = \cdots = k_N =1$. 
}
\section{Tree Global Constraint Automata}
\label{section-tgca}

In this section, we introduce a class of tree constraint automata parameterised
by concrete domains $\acdomain$ that accept infinite data trees in which
every node is labelled by a finite tuple of data values from $\adomain$.
This is slightly
more general than the automata models from~\cite{Demri&Quaas23bis},
not only because the definition does not assume a fixed concrete domain
but more importantly because  constraints between siblings are allowed
unlike what is done in~\cite{Demri&Quaas23bis,Demri&Quaas25}. 

\subsection{Definitions}

Unless otherwise stated, in the rest of this section, we assume that
the concrete domain $\acdomain$ satisfies the conditions (C1) and (C3.$k$)
for some $k \geq 1$
(meaning that (C2) and (C4) are not needed).
We introduce the class of tree global constraint automata
that accept sets of labelled trees of the form 
\(\adatatree : \interval{0}{\degree-1}^* \to \aalphabet \times \adomain^\beta\)
for some finite alphabet $\aalphabet$, for some $\degree \geq 1$ and  $\beta \geq 0$.
The transition relation of such automata expresses constraints between the $\beta$ values at 
a node and the values at its children. The automaton is equipped with a Büchi 
acceptance condition.
Formally,
a \defstyle{tree global constraint automaton} (in short, TGCA) over
some concrete domain
$\acdomain$ 
is a tuple
\(
\aautomaton = (\locations, \aalphabet, \degree, \beta, \locations_{in}, \transitions, F),
\)
where:
\begin{itemize}
\item \(\locations\) is a finite set of locations,
 \item  \(\locations_{in} \subseteq \locations\) 
   are initial locations, \(F \subseteq \locations\) are accepting locations,
\item
  \(\aalphabet\) is a finite alphabet,
  \item \(\degree \ge 1\) is the branching degree, and \(\beta \in \Nat\) the number of 
    registers (a.k.a. variables),
  \item \(\transitions \subseteq \locations \times \aalphabet \times \treeconstraints{\beta, \degree}
    \times \locations^\degree\) is the transition relation
    where \(\treeconstraints{\beta, \degree}\) 
    denotes the set of $\acdomain$-constraints built over
    the distinguished registers in 
    \(
    \registers{\beta}{\degree} \egdef 
    \set{\aregister_1, \dots, \aregister_\beta} \cup
    \set{\aregister_j^i \mid 1 \le j \le \beta,\ 0 \le i < \degree}\).
    The expression 
    \(\aregister_j^i\) refers to the \(j\)th register of the \(i\)th child.
    Arbitrary registers in $\registers{\beta}{\degree}$ are also referred to by $\aregisterbis_0,
    \aregisterbis_1, \aregisterbis_2, \ldots$.
    By way of example, a constraint in \(\treeconstraints{\beta, \degree}\)
    can be of the form $(\aregister_1 = \frac{1}{3} \wedge (\aregister_2^{0} = \aregister_3^{0})) \vee \aregister_1^{2} <
    \aregister_1^{3}$
    for some concrete domain based on the set of rational numbers and $\degree$ at least equal to four. 
\end{itemize}

\paragraph*{Runs.}
Let \(\adatatree : \interval{0}{\degree-1}^* \to \aalphabet \times \adomain^\beta\)
be a full infinite  \(\degree\)-ary 
data tree, with \(\adatatree(\anode) = (\aletter_{\anode}, \atuple_{\anode})\)
at each node $\anode$.
A \defstyle{run} of $\aautomaton$ 
on $\adatatree$ is a mapping \(\arun : \interval{0}{\degree-1}^* \to \transitions\) 
that maps each node \(\anode\) to a transition 
\((\alocation_{\anode}, \aletter_{\anode}, \acons_{\anode}, \alocation_{\anode \cdot 0},
\dots, \alocation_{\anode \cdot (\degree-1)})\) 
verifying the following conditions. 
\begin{romanenumerate}
  \item The source location of $\arun(\anode \cdot i)$ is $\alocation_{\anode \cdot i}$ 
  for all \(0 \le i < \degree\).
\item Let \(\avaluation_{\anode} : \registers{\beta}{\degree} \to
  \adomain\) be the valuation defined by
  \(
  \avaluation_{\anode}(\aregister_j) = \atuple_n[j] \text{ and }
  \avaluation_{\anode}(\aregister_j^i) = \atuple_{\anode \cdot i}[j],
  \)
  for all \(1 \le j \le \beta,\ 0 \le i < \degree\). Then \(\avaluation_{\anode}
  \models \acons_{\anode}\).
  \item The source location of \(\arun(\varepsilon) \text{ is in } \locations_{in}\).
  \end{romanenumerate}
Given a path \(\apath = j_1\cdot j_2\dots\) in $\arun$ starting from $\varepsilon$, we define
$\inf(\arun, \apath)$ to be the set of locations that appear infinitely often as the source
locations of the transitions in $\arun(\varepsilon)\arun(j_1)\arun(j_2)\dots$
A run $\arun$ is \defstyle{accepting} if, for all paths $\apath$ in
$\arun$ starting from $\varepsilon$, 
we have \(\inf(\arun, \apath) \cap F \neq \emptyset.\)
We write \(\alang(\aautomaton)\) to denote the set of data trees $\adatatree$ for which there exists
some accepting run of $\aautomaton$ on $\adatatree$.
The \defstyle{nonemptiness problem for TGCA}, written \neproblem{TGCA}, asks 
whether \(\alang(\aautomaton) \neq \emptyset\) for some TGCA $\aautomaton$.
Herein, we use B\"uchi acceptance conditions but
it is easy to consider
other acceptance conditions, apart from the fact that
such conditions
play no essential role in our investigations because we shall always assume that
$F = \locations$ (as in looping tree automata,
see e.g.~\cite[Section~3.6.1]{Baader&Horrocks&Sattler08}).
\subsection{Reduction to Nonemptiness for Büchi Tree Automata}
\label{section-reduction-to-bta}
To analyse the complexity of
the decision problem \neproblem{TGCA},
we construct a reduction to the nonemptiness problem 
for Büchi tree automata (BTA), known to be in \ptime~\cite{Vardi&Wolper86}.
B\"uchi tree automata are
TGCA with $\beta = 0$
and no $\acdomain$-constraints. 
Given
a fixed TGCA
$\aautomaton
= (\locations, \aalphabet, \degree, \beta, 
\locations_{in}, \transitions, F)\),
we write \(\mathcal{P}_\aautomaton = \{P_1, \dots, P_m\}\)
to denote either the full set of predicate symbols of $\acdomain$
if finite, or the {\em finite} set of predicate symbols from \(\acdomain\)
occurring in $\aautomaton$.

\paragraph*{Symbolic types.} 
To perform the reduction, we abstract
over concrete values in a TGCA run by using
\defstyle{symbolic types}, following the standard notion of types in first-order languages,
see e.g.~\cite[Section~3.1]{PrattHartmann23} and~\cite{Segoufin&Torunczyk11}. 
These are sets of literals that capture all atomic constraints relevant to the current node and its children. 
A forthcoming notion of 
projection compatibility ensures the coherence between parent and child types across the tree.
Let
$\atoms{\aautomaton}$ 
be the set of atomic 
formulae of the form
\(P(\aregisterbis_1, \ldots, \aregisterbis_k)\)
for some $\aregisterbis_1, \ldots, \aregisterbis_k \in \registers{\beta}{\degree}$
and \(P \in \mathcal{P}_\aautomaton\).
Similarly, let $\literals{\aautomaton}$
be the set of all literals over \(\atoms{\aautomaton}\), 
i.e., atomic formulae or their negations.
\iftoggle{versionlong}{
\begin{defi}[Symbolic type]
  A \defstyle{symbolic type} $\atype$ is a
  set
  \(\atype \subseteq \literals{\aautomaton}\)
such that
for every \(\acons \in \atoms{\aautomaton}\), either \(\acons \in \atype\) or 
\(\neg \acons \in \atype\), but not both. Let \(\symbtypes{\aautomaton}\) denote the set 
of all symbolic types.
\end{defi}

Symbolic types can be obviously understood as constraints made of conjunctions of its elements
built over the registers in~$\registers{\beta}{\degree}$.

\begin{defi}[Satisfiable symbolic type]
A symbolic type \(\atype\) is \defstyle{satisfiable} if there is a valuation 
\(\avaluation : \registers{\beta}{\degree} \to \adomain\) such that \(\avaluation \models \atype\). 
Let \(\satsymbtypes{\aautomaton}\) denote the set of all satisfiable symbolic types. 
\end{defi}
}{
 A \defstyle{symbolic type} $\atype$ is a
  set
  \(\atype \subseteq \literals{\aautomaton}\)
  such that
for every \(\acons \in \atoms{\aautomaton}\), either \(\acons \in \atype\) or 
\(\neg \acons \in \atype\), but not both. Let \(\symbtypes{\aautomaton}\) denote the set 
of all symbolic types.
Symbolic types can be
obviously understood as constraints made of conjunctions of its elements
built over the registers in~$\registers{\beta}{\degree}$. 
A symbolic type \(\atype\) is \defstyle{satisfiable} if there is
a valuation 
\(\avaluation : \registers{\beta}{\degree} \to \adomain\) such that \(\avaluation \models \atype\). 
Let \(\satsymbtypes{\aautomaton}\) denote the set of all satisfiable symbolic types. 
}

Computing \(\satsymbtypes{\aautomaton}\) can be done as soon as $\csp{\acdomain}$ is decidable
and in particular if the condition (C3.$k$) holds for some $k$. 

The properties of satisfiable types that we mainly use are stated in Lemma~\ref{lemma-types}
where $\acdomain \models \atype(\atuple, \atuple_0, \ldots, \atuple_{\degree-1})$
means that $\avaluation \models \bigwedge_{\acons \in \atype} \acons$
for the valuation $\avaluation$
 such that  $\avaluation(\aregister_j) = \atuple[j]$
and $\avaluation(\aregister_j^i) = \atuple_{i}[j]$,
  for all \(1 \le j \le \beta,\ 0 \le i < \degree\).

\iftoggle{versionlong}{
\begin{lem}[Three properties about types] \label{lemma-types} \ 
\begin{description}
\itemsep 0 cm   
   \item[(I)]
    Let $\atuple, \atuple_0, \ldots, \atuple_{\degree-1} \in \adomain^{\beta}$. There is a unique
     satisfiable type
    $\atype \in \satsymbtypes{\aautomaton}$
     such that $\acdomain \models \atype(\atuple, \atuple_0, \ldots, \atuple_{\degree-1})$.
\item[(II)] For every constraint $\acons$ built over
  the predicate symbols in $\mathcal{P}_\aautomaton$
  and the registers $\registers{\beta}{\degree}$, 
  there is a disjunction
  $\atype_1 \vee \cdots \vee \atype_{\gamma}$ logically equivalent to $\acons$
  and each $\atype_i$ belongs to $\satsymbtypes{\aautomaton}$ (empty disjunction stands for $\perp$).
\item[(III)] For all distinct types $\atype \neq \atype' \in \satsymbtypes{\aautomaton}$,
$\atype \wedge \atype'$ is not satisfiable. 
  \end{description}
\end{lem}
}{
\begin{lem}[Three properties about types] \label{lemma-types} 
(I) Let $\atuple, \atuple_0, \ldots, \atuple_{\degree-1} \in \adomain^{\beta}$. There is a unique
     satisfiable type
    $\atype \in \satsymbtypes{\aautomaton}$
     such that $\acdomain \models \atype(\atuple, \atuple_0, \ldots, \atuple_{\degree-1})$.
(II) For every constraint $\acons$ built over
  the predicate symbols in $\mathcal{P}_\aautomaton$
  and the registers $\registers{\beta}{\degree}$, 
  there is a disjunction
  $\atype_1 \vee \cdots \vee \atype_{\gamma}$ logically equivalent to $\acons$
  and each $\atype_i$ belongs to $\satsymbtypes{\aautomaton}$ (empty disjunction stands for $\perp$).
(III) For all distinct types $\atype \neq \atype' \in \satsymbtypes{\aautomaton}$,
$\atype \wedge \atype'$ is not satisfiable. 
\end{lem}
}

\cut{
In Lemma~\ref{lemma-types}, $\acdomain \models \atype(\atuple, \atuple_0, \ldots, \atuple_{\degree-1})$
means that $\avaluation \models \bigwedge_{\acons \in \atype} \acons$
for the valuation $\avaluation$
 such that  $\avaluation(\aregister_j) = \atuple[j]$
and $\avaluation(\aregister_j^i) = \atuple_{i}[j]$,
for all \(1 \le j \le \beta,\ 0 \le i < \degree\).
}
\iftoggle{versionlong}{
\begin{proof}
  (I) Let \(\atuple, \atuple_0, \ldots, \atuple_{\degree-1} \in \adomain^{\beta}\).  
  Define a valuation \(\avaluation\) on the variables \(\registers{\beta}{\degree}\) by
  \(
  \avaluation(\aregister_j) \egdef \atuple[j] \text{ and } 
  \avaluation(\aregister_j^i) \egdef \atuple_i[j] \text{ for all } 
  1 \le j \le \beta,\ 0 \le i < \degree.
  \)
  Let \(\atype\) be the set of all literals over the variables in \(\registers{\beta}{\degree}\) 
  defined as follows. For each predicate symbol \(P \in \mathcal{P}_\aautomaton\) of 
  arity \(k\), and each tuple \((\aregisterbis_1, \ldots, \aregisterbis_k) 
  \in \registers{\beta}{\degree}^k\), we include:
  \[
  \begin{cases}
  P(\aregisterbis_1, \ldots, \aregisterbis_k), & \text{if } \acdomain \models 
  P(\avaluation(\aregisterbis_1), \ldots, \avaluation(\aregisterbis_k)), \\
  \lnot P(\aregisterbis_1, \ldots, \aregisterbis_k), & \text{otherwise}.
  \end{cases}
  \]
  Then \(\atype\) is a satisfiable type: it consists only of literals that hold under \(\avaluation\). 
  Moreover, \(\atype\) is complete, i.e., for every atomic formulae 
  \(P(\aregisterbis_1, \ldots, \aregisterbis_k)\) 
  exactly one of \(P(\aregisterbis_1, \ldots, \aregisterbis_k)\) or 
  \(\lnot P(\aregisterbis_1, \ldots, \aregisterbis_k)\) belongs to \(\atype\). 
  Finally, the type \(\atype\) is uniquely determined by the valuation \(\avaluation\), 
  which is uniquely determined by \(\atuple, \atuple_0, \ldots, \atuple_{\degree-1}\). 
  Hence there is a unique satisfiable type \(\atype\) such that 
  \(\acdomain \models \atype(\atuple, \atuple_0, \ldots, \atuple_{\degree-1})\).\\
  (II) Every constraint $\acons$ built  over
  the predicate symbols in $\mathcal{P}_\aautomaton$
  and the variables $\registers{\beta}{\degree}$ can be put
  in {\em full} disjunctive normal form (using standard propositional reasoning), leading to a
  disjunction of types $\atype_{1} \vee \cdots \vee \atype_{\alpha}$.
  Removing the unsatisfiable types from $\atype_{1} \vee \cdots \vee \atype_{\alpha}$
  can be easily shown to lead to a logically equivalent disjunction, whence the
  property. \\ 
  (III) If $\atype \neq \atype'$, then there is some atomic constraint  that belongs to
  $\atype \cup \atype'$ as well as its negation. Hence,
  $\atype \wedge \atype'$ is not satisfiable. 
\end{proof}
 
}{
  The proof of Lemma~\ref{lemma-types} is by an easy verification
  and can be found in
 Appendix~\ref{appendix-proof-lemma-types}.
}
We write $\atype \models \acons$ to denote that
for all valuations $\avaluation$,
if $\avaluation \models \atype$, then 
$\avaluation \models \acons$.
If $\atype \in \satsymbtypes{\aautomaton}$, then $\atype \models \acons$
can be checked in polynomial time.
Let
 \(\atype, \atype_0, \dots, \atype_{\degree-1} \in \satsymbtypes{\aautomaton}\).
  We say that \(\atype\) 
  is \defstyle{projection-compatible} with \((\atype_0, \dots, \atype_{\degree-1})\)
  $\equivdef$ for each 
\(i < \degree\) and every predicate symbol \(P \in \mathcal{P}_\aautomaton\) of arity \(k\), 
and for all  tuples \((j_1, \dots, j_k) \in \interval{1}{\beta}^k\), we have:
\(
P(\aregister_{j_1}^i, \dots, \aregister_{j_k}^i) \in \atype\)
iff 
\(P(\aregister_{j_1}, \dots, \aregister_{j_k}) \in
\atype_i
\).
We recall that in a type $\atype$, registers associated to the
values of the children nodes are also present. Hence, projection-compatibility
simply reflects the fact that the constraints for such registers
in $\atype$ must be compatible with the constraints of the
registers in the children types.

\paragraph*{Construction of Büchi tree automata.}
We now define the Büchi tree automaton 
\[
\aautomatonbis = (\locations', \aalphabet, \transitions', \degree, \locations'_{\mathsf{in}}, F')
\]
that simulates the TGCA $\aautomaton$ using symbolic types.
The basic idea consists in abstracting data values in $\adomain$ by satisfiable
types from $\satsymbtypes{\aautomaton}$. Moreover, the type at a node needs to be
compatible with the children's types since a type expresses constraints between
the values at a node and those at its children. This is where
projection compatibility plays a crucial role. In this way, an infinite tree
accepted by the B\"uchi tree automaton can be viewed as an infinite constraint system
(each node has its own variables coming from $\beta$ local registers) 
for which satisfiability is required. Generally, additional conditions are needed
to guarantee satisfiability, see e.g.~\cite{Labai&Ortiz&Simkus20,Labai21,Demri&Quaas23bis}.
In the case of concrete domains satisfying the condition (C1), importantly, this is all we need
as shown below. First, let us  define
the B\"uchi tree automaton $\aautomatonbis$ with the remarkably simple  construction below. 
\begin{itemize}
   \item \(\locations' \egdef \locations \times \satsymbtypes{\aautomaton}\), 
   \item \(\locations'_{in} \egdef \locations_{in} \times \satsymbtypes{\aautomaton}\),
   \(F' \egdef F \times \satsymbtypes{\aautomaton}\),
   \item \(\transitions'\) contains the transition
   \(
   ((\alocation, \atype), \aletter, (\alocation_0, \atype_0), \dots, (\alocation_{\degree-1}, 
   \atype_{\degree-1}))
   \)
   \iftoggle{versionlong}{
   whenever:
   \begin{alphaenumerate}
     \item \((\alocation, \aletter, \acons, \alocation_0, \dots, \alocation_{\degree-1}) 
      \in \transitions\),
     \item \(\atype \models \acons\),
     \item \(\atype\) is projection-compatible with \((\atype_0, \dots, \atype_{\degree-1})\).
   \end{alphaenumerate}
   }{
     if (a)
     \((\alocation, \aletter, \acons, \alocation_0, \dots, \alocation_{\degree-1}) 
      \in \transitions\),
      (b) \(\atype \models \acons\)
      and (c) 
     \(\atype\) is projection-compatible with \((\atype_0, \dots, \atype_{\degree-1})\).
   }
\end{itemize}
To build effectively $\aautomatonbis$ from
$\aautomaton$, we only need the decidability of  $\csp{\acdomain}$ 
to determine the satisfiable symbolic types. 
We can establish that this construction preserves nonemptiness.
Again, apart from the decidability of $\csp{\acdomain}$, only the condition (C1) is required
to get the equivalence as the condition (C2) is mainly
relevant for computational purposes.

\begin{lem}[Soundness]
\label{lemma-tgca-soundness}
  \(\alang(\aautomaton) \neq \emptyset\) implies \(\alang(\aautomatonbis) \neq \emptyset\).
\end{lem}

The correctness proof in this direction does not require much about the
concrete domain $\acdomain$
since it mainly rests on the construction of
satisfiable types from tuples made of data values in $\adomain$.

\begin{proof}
Assume that \(\alang(\aautomaton) \neq \emptyset\).
There exist
a
tree \(\adatatree : \interval{0}{\degree-1}^* \to \aalphabet \times \adomain^\beta\)
and an accepting run \(\arun : \interval{0}{\degree-1}^* \to \transitions\)
on $\adatatree$ 
such that 
for every node \(\anode\),
\(\adatatree(\anode) = \pair{\aletter_{\anode}}{\atuple_{\anode}}\)
and \(\arun(\anode) = (\alocation_{\anode}, \aletter_{\anode},
                \acons_{\anode}, \alocation_{\anode \cdot 0}, \dots,
                \alocation_{\anode \cdot (\degree-1)}).
                \)
                We aim at defining a run \(\arun' : \interval{0}{\degree-1}^* \to \transitions'\)
                of  
                \(\aautomatonbis\) over the specific input tree $\atree$
                such that for every node $\anode$, \(\atree(\anode) \egdef \aletter_{\anode}\).

                For all nodes $\anode \in \interval{0}{\degree-1}^*$,
                we define a valuation $\avaluation_{\anode}:
                \registers{\beta}{\degree} \to \adomain$ as follows.
                We set 
    \(
    \avaluation_{\anode}(\aregister_j) \egdef
    \atuple_{\anode}[j] \text{, }
    \avaluation_{\anode}(\aregister_j^i) \egdef
    \atuple_{\anode \cdot i}[j]\text{ for } 1 \le j \le \beta,\ 0 \le i < \degree.
    \)
    We also define
    the symbolic type \(\atype_{\anode} \in \symbtypes{\aautomaton}\)
    with 
    \(
    \atype_{\anode} \egdef
    \set{\acons \in \literals{\aautomaton} \mid
      \avaluation_{\anode} \models \acons}
    \). 
    Thus, $\avaluation_{\anode} \models \atype_{\anode}$
    and $\atype_{\anode} \in \satsymbtypes{\aautomaton}$.
    Since \(\avaluation_{\anode} \models \acons_{\anode}\)
    ($\arun$ is a run, see condition (ii))
    and \(\atype_{\anode}\) includes all literals true under \(\avaluation_{\anode}\)
    (by definition), 
    it follows that \(\atype_{\anode} \models \acons_{\anode}\)
    by Lemma~\ref{lemma-types}.
    Indeed $\atype_{\anode} \models \atype_1 \vee \cdots \vee \atype_{\alpha}$
    iff $\atype_{\anode} = \atype_i$ for some $i$.
    
    In order to verify the satisfaction of the projection compatibility, observe that for all predicates
    \(P\) in $\mathcal{P}_\aautomaton$ of arity \(k\), and all index tuples 
    \(j_1, \dots, j_k \in \interval{1}{\beta}\), we have:
    \(
    \avaluation_{\anode}
    \models P(\aregister_{j_1}^i, \dots, \aregister_{j_k}^i)\)
    iff \(\avaluation_{\anode \cdot i} \models P(\aregister_{j_1}, \dots, \aregister_{j_k}).
    \)
    Thus,
    \(
    P(\aregister_{j_1}^i, \dots, \aregister_{j_k}^i) \in \atype_{\anode}\)
    iff \(P(\aregister_{j_1}, \dots, \aregister_{j_k}) \in \atype_{\anode \cdot i}
    \) by definition of $\atype_{\anode}$ and  $\atype_{\anode \cdot i}$.
    Therefore, the projection compatibility condition holds.

    Hence, the tuple
    \(
    ((\alocation_{\anode}, \atype_{\anode}), \aletter_{\anode},
    ((\alocation_{\anode \cdot 0}, \atype_{\anode \cdot 0}), \dots,
    (\alocation_{\anode \cdot (\degree-1)}, \atype_{\anode \cdot (\degree-1)})))\)
    belongs to the set $\transitions'$ of transitions
    and it is therefore a transition of 
     \(\aautomatonbis\). Let \(\arun'(\anode)\) be equal to such a transition.

     Finally, since \(\arun\) is accepting (i.e., every infinite path
     visits some state in \(F\) infinitely often), 
     and \(\arun'(\anode)\) carries \(\alocation_{\anode}\)
     as its first component, the Büchi acceptance condition 
    for \(\aautomatonbis\) is satisfied.
    Therefore, \(\arun'\) is an accepting run of \(\aautomatonbis\)
    on $\atree$, hence \(\alang(\aautomatonbis) \neq \emptyset\).
\end{proof}

\begin{lem}[Completeness]
\label{lemma-tgca-completeness}
  \(\alang(\aautomatonbis) \neq \emptyset\) implies \(\alang(\aautomaton) \neq \emptyset\).
\end{lem}

This direction is less obvious whose proof takes advantage
of the properties of the concrete domain $\acdomain$, namely the
satisfaction of (C1). 

\begin{proof}
Assume that \(\alang(\aautomatonbis) \neq \emptyset\).
  There exists a tree
  \(\atree : \interval{0}{\degree-1}^* \to \aalphabet\)
  and an accepting run
  \(\arun' : \interval{0}{\degree-1}^* \to \transitions'\)
  such that
  for each
  node \(\anode\),
  \(\atree(\anode) = \aletter_{\anode}\)
  and \(
    \arun'(\anode) = ((\alocation_{\anode}, \atype_{\anode}), \aletter_{\anode},
    (\alocation_{\anode \cdot 0}, \atype_{\anode \cdot 0}), 
    \dots, (\alocation_{\anode \cdot (\degree-1)}, \atype_{\anode \cdot (\degree-1)}))
    \).
    By definition of $\aautomatonbis$, this implies that for all nodes \(\anode\),
    there exists a transition
    \((\alocation_{\anode}, \aletter_{\anode}, \acons_{\anode}, \alocation_{\anode \cdot 0}, \dots,
    \alocation_{\anode \cdot (\degree-1)}) \in
    \transitions\)
    such that \(\atype_{\anode} \models \acons_{\anode}\),
    and for each \(i < \degree\), every predicate symbol \(P \in \mathcal{P}_{\aautomaton}\) of arity \(k\) and
    all indices \(j_1, \dots, j_k \in \{1, \dots, \beta\}\), we have
    \(
    P(\aregister_{j_1}^i, \dots, \aregister_{j_k}^i) \in \atype_{\anode}\)
    iff 
    \(P(\aregister_{j_1}, \dots, \aregister_{j_k}) \in \atype_{\anode \cdot i}.
    \)
    That is, projection compatibility holds.
    
    Now, we  construct a data tree \(\adatatree : \interval{0}{\degree-1}^* \to \aalphabet
    \times \adomain^\beta\) and a run 
    \(\arun : \interval{0}{\degree-1}^* \to \transitions\) of \(\aautomaton\), under either of the 
    assumptions on \(\acdomain\) from the condition (C1).
    First, always
    $\arun(\anode) \egdef
    (\alocation_{\anode}, \aletter_{\anode}, \acons_{\anode}, \alocation_{\anode \cdot 0}, \dots,
    \alocation_{\anode \cdot (\degree-1)})$ for some transition in $\transitions$
    as mentioned earlier.

    Let us start by providing the common ground to define $\adatatree$ and $\arun$. 
    Let $\prec$ be the well-founded total ordering on $\interval{0}{\degree-1}^*$ corresponding
    to the breadth-first search in the infinite tree $\interval{0}{\degree-1}^*$.
    More precisely, $\anode \prec \anode'$ iff either the depth of $\anode$
    is strictly smaller than the depth of $\anode'$ or $\anode \lex \anode'$,
    where $\lex$ is the lexicographical ordering on $\interval{0}{\degree-1}^*$
    as the nodes are also finite words over the finite alphabet $\interval{0}{\degree-1}$.
     We can assume that the nodes in $\interval{0}{\degree-1}^*$ are enumerated as follows:
    $
    \anode_0 \prec \anode_1 \prec \anode_2 \prec \cdots
    $
    with $\anode_0 = \varepsilon$ following a breadth-first search in
    $\interval{0}{\degree-1}^*$.
    
    Given the tree $\atree$ and the accepting run $\arun'$, we define
    the countable constraint system $\acsystem$ built over variables
    of the form $v_1^{\anode}, \ldots, v_{\beta}^{\anode}$ for some $\anode \in \interval{0}{\degree-1}^*$
    (which is fine because $\variables$ is  countably infinite).
    Intuitively, providing an interpretation for the variable
    $v_j^{\anode}$ shall amount to determine
    the value $\atuple_{\anode}[j]$ with $\adatatree(\anode) = \pair{\aletter_{\anode}}{\atuple_{\anode}}$
    (yet to be defined).
    Given a node $\anode \in \interval{0}{\degree-1}^*$ with $\arun'(\anode) =
    ((\alocation_{\anode}, \atype_{\anode}), \aletter_{\anode},
    (\alocation_{\anode \cdot 0}, \atype_{\anode \cdot 0}), \ldots,
    (\alocation_{\anode \cdot (\degree-1)}, \atype_{\anode \cdot (\degree-1)}))$, 
    we write $\acsystem^{\anode}$ to denote the constraint system
    obtained from $\atype_{\anode}$ by renaming the registers
    from $\atype_{\anode}$ by variables.
    More precisely,
    $\acsystem^{\anode}$ is obtained from
    $\atype_{\anode}$ by replacing any
    register $\aregister_j$ by the variable $v_j^{\anode}$
    and any register $\aregister_j^i$ by the variable $v_j^{\anode \cdot i}$.
    Observe that the constraint systems $\acsystem^{\anode}$ and $\acsystem^{\anode'}$
    share variables only if $\anode = \anode'$ or $\anode$ (resp. $\anode'$)
    is the parent of $\anode'$ (resp. $\anode$).
    We write $\acsystem^{\anode}_{*}$ to denote
    the finite constraint system
    $
    \bigcup_{\anodebis \preceq \anode} \acsystem^{\anodebis}
    $ obtained by accumulation,
    and $\acsystem =  \bigcup_{\anode \in \interval{0}{\degree-1}^*} \acsystem^{\anode}$
    (that can be understood as the limit constraint system).
    Since $\acsystem_{*}^{\anode_0} \subseteq \acsystem_{*}^{\anode_1} \subseteq
    \cdots$ is an infinite chain  such that
     $\acsystem = \bigcup_k \acsystem_{*}^{\anode_k}$,
    every finite subset of $\acsystem$ is included in some $\acsystem_{*}^{\anode_k}$.
    Below, we show that $\acsystem$ is satisfiable and this is witnessed by some
    valuation $\avaluation$.
    Hence, the data tree $\adatatree$ accepted by $\aautomaton$ must ensure that
    for all nodes $\anode$ and all $j \in \interval{1}{\beta}$,
    we have $\adatatree(\anode) = \pair{\aletter}{\atuple}$
    with $\atuple[j] = \avaluation(v_j^{\anode})$ by definition.
    Our main tasks consist in showing that $\acsystem$ is satisfiable
    and then construct the accepting run $\arun$ on the data tree $\adatatree$
    as defined above. The second part is easier to handle
    using the run $\arun'$, the constraint system $\acsystem$
    and the data tree $\adatatree$. Actually a definition is already provided above.
    We make a distinction according to the disjunct satisfied in the condition (C1).
    The structure of proof is illustrated in Figure~\ref{fig:two-constructions-lemma-3-5}.

    \begin{figure}[t]
    \centering
    \scalebox{1}{
    \begin{tikzpicture}[
      node distance=0.95cm,
      every node/.style={font=\scriptsize},
      box/.style={draw, rounded corners, align=center, inner sep=2.5pt},
      arr/.style={->, thick}
    ]

    \node[box] (run)
    {accepting run \(\arun'\) of \(\aautomatonbis\)\\
    with symbolic types \(\atype_{\anode}\)};

    \node[box, below=0.65cm of run] (systems)
    {local systems \(\acsystem^{\anode}\) and accumulated systems\\
    \(\acsystem^{\anode_i}_{*}
    =
    \bigcup_{\anode \preceq \anode_i}\acsystem^{\anode}\)};th

    \node[box, below left=0.85cm and 1.25cm of systems] (completion)
    {completion property};

    \node[box, below right=0.85cm and 1.25cm of systems] (ap)
    {amalgamation property\\
    and homomorphism \(\omega\)-compactness};

    \node[box, below=0.65cm of completion] (v0)
    {\(\avaluation_0 \models \acsystem^{\anode_0}_{*}\)};

    \node[box, below=0.55cm of v0] (v1)
    {\(\avaluation_1 \models \acsystem^{\anode_1}_{*}\)\\
    extends \(\avaluation_0\)};

    \node[box, below=0.55cm of v1] (vi)
    {\(\cdots\)};

    \node[box, below=0.55cm of vi] (vlim)
    {limit valuation \(\avaluation\)\\
    with \(\avaluation \models \acsystem\)};

    \node[box, below=0.65cm of ap] (s0)
    {\(\acsystem^{\anode_0}_{*}\) satisfiable};

    \node[box, below=0.55cm of s0] (s1)
    {\(\acsystem^{\anode_1}_{*}
    =
    \acsystem^{\anode_0}_{*}
    \cup
    \acsystem^{\anode_1}\)\\
    satisfiable by amalgamation};

    \node[box, below=0.55cm of s1] (si)
    {\(\cdots\)\\
    all \(\acsystem^{\anode_i}_{*}\) satisfiable};

    \node[box, below=0.55cm of si] (sglobal)
    {\(\acsystem=\bigcup_i \acsystem^{\anode_i}_{*}\)\\
    satisfiable by \(\omega\)-compactness};

    \node[box, below=0.8cm of $(vlim)!0.5!(sglobal)$] (tree)
    {define \(\adatatree(\anode)=
    \pair{\aletter_{\anode}}{\atuple_{\anode}}\), where
    \(\atuple_{\anode}[j]=\avaluation(v_j^{\anode})\)};

    \node[box, below=0.65cm of tree] (runA)
    {obtain an accepting run \(\arun\) of \(\aautomaton\)};

    \draw[arr] (run) -- (systems);
    \draw[arr] (systems) -- (completion);
    \draw[arr] (systems) -- (ap);

    \draw[arr] (completion) -- (v0);
    \draw[arr] (v0) -- (v1);
    \draw[arr] (v1) -- (vi);
    \draw[arr] (vi) -- (vlim);

    \draw[arr] (ap) -- (s0);
    \draw[arr] (s0) -- (s1);
    \draw[arr] (s1) -- (si);
    \draw[arr] (si) -- (sglobal);

    \draw[arr] (vlim) -- (tree);
    \draw[arr] (sglobal) -- (tree);
    \draw[arr] (tree) -- (runA);

    \end{tikzpicture}
    }
    \caption{The two constructions used in the proof of Lemma~\ref{lemma-tgca-completeness}.
    With the completion property, the valuation is constructed incrementally along
    the breadth-first enumeration
    \(\anode_0 \prec \anode_1 \prec \cdots\).
    With amalgamation and homomorphism \(\omega\)-compactness, one first proves
    that every accumulated finite system \(\acsystem^{\anode_i}_{*}\) is satisfiable,
    and then obtains a valuation of the whole countable system \(\acsystem\).}
    \label{fig:two-constructions-lemma-3-5}
    \end{figure}
        
    \proofsubparagraph{Case 1: \(\acdomain\) satisfies the completion property}
    Thanks to the satisfaction of the completion
    property, we shall guarantee that for all $i \geq 0$, we define
    the valuation $\avaluation_i$ such that
    $\avaluation_i \models \acsystem^{\anode_i}_{*}$ and
    $\avaluation_i$ is a restriction of $\avaluation_{i+1}$;
    only the new variables in $\acsystem^{\anode_{i+1}}_{*}$ but not in $\acsystem^{\anode_i}_{*}$
    need to get a value in $\adomain$.
    Hence, it makes sense to define $\avaluation$ as the limit valuation
    and it satisfies $\acsystem$ since it satisfies every
    constraint system $\acsystem_{*}^{\anode}$. 
    We construct the valuations inductively from the root.

    \textit{Base case.} 
    We have \(\atype_\varepsilon\) is satisfiable and  \(\atype_\varepsilon \models \acons_\varepsilon\).
    Therefore, there is a valuation satisfying $\atype_{\varepsilon}$ and thanks to the injective
    renaming we performed, 
    there is also a valuation $\avaluation_\varepsilon$ satisfying $\acsystem^{\varepsilon}$.
    As explained above, 
    $\adatatree(\varepsilon) \egdef \pair{\aletter_{\varepsilon}}{\atuple_{\varepsilon}}$
    and
    $\arun(\varepsilon) = (\alocation_\varepsilon, \aletter_\varepsilon,
    \acons_\varepsilon, \alocation_{0}, \dots, \alocation_{\degree-1})$
    with $\atuple_{\varepsilon}[j] \egdef \avaluation_{\varepsilon}(v_j^{\varepsilon})$ for all
    $j \in \interval{1}{\beta}$.
    Furthermore, for all $i \in \interval{0}{\degree-1}$,
    $\adatatree(i) \egdef \pair{\aletter_{i}}{\atuple_{i}}$
    with $\atuple_{i}[j] \egdef \avaluation_{\varepsilon}(v_j^{i})$ for all
    $j \in \interval{1}{\beta}$.
    Observe that $\arun(\varepsilon)$ is fine to satisfy forthcoming requirements about runs
    since \(\atype_\varepsilon \models \acons_\varepsilon\), $\avaluation_{\varepsilon}
    \models \acsystem^{\varepsilon}$ and therefore
    $\acdomain
    \models \acons_\varepsilon(\atuple_{\varepsilon}, \atuple_0, \ldots, \atuple_{\degree-1})$,
    so that 
    the condition (ii) for runs is satisfied at the (root) node $\anode = \varepsilon$. 

\textit{Inductive step.}
Suppose that for all $k \leq I$,
$\adatatree(\anode_k)$ and $\arun(\anode_k)$ are properly defined
and there is some valuation $\avaluation_{I}$ such that
$\avaluation_{I} \models \acsystem^{\anode_I}_{*}$ (therefore
$\acsystem^{\anode_I}_{*}$ is satisfiable). Recall that
the nodes in $\interval{0}{\degree-1}^*$ are enumerated along
the sequence $\anode_0, \anode_1, \anode_2, \ldots$. 
Let us define $\adatatree(\anode_{I+1})$, $\arun(\anode_{I+1})$
and $\avaluation_{I+1}$ such that
$\avaluation_{I+1} \models \acsystem^{\anode_{I+1}}_{*}$.
Since $I+1 > 0$, there is $J \leq I$ such that $\anode_{I+1} = \anode_{J} \cdot j_0$
for some $j_0 \in \interval{0}{\degree-1}$.
Observe that $\adatatree(\anode_{I+1})$ is therefore already defined. 

Since $\atype_{\anode_{J}}$ is projection-compatible
with $(\atype_{\anode_{J \cdot 0}}, \ldots,\atype_{\anode_{J \cdot (\degree-1)}})$
and $\avaluation_{I} \models \acsystem^{\anode_J}_{*}$
(since $\acsystem^{\anode_J}_{*} \subseteq \acsystem^{\anode_I}_{*}$),
$\avaluation_{I}$ satisfies all the
literals in $\acsystem^{\anode_{I+1}}$
involving only variables among $v_1^{\anode_{I+1}}, \ldots, v_{\beta}^{\anode_{I+1}}$. 
Since $\atype_{\anode_{I+1}}$ is satisfiable
and therefore $\acsystem^{\anode_{I+1}}$
(obtained by renaming from $\atype_{\anode_{I+1}}$) is
satisfiable, by the completion property, there is an extension
$\avaluation_{I+1}$ of $\avaluation_{I}$ such that
$\avaluation_{I+1} \models \acsystem^{\anode_{I+1}}$.
The variables that got a first value in $\avaluation_{I+1}$
are not present in $\acsystem^{\anode_{I}}_{*}$ and are of the form
$v_{j}^{\anode_{I+1} \cdot i}$. 
Consequently, we also got
$\avaluation_{I+1} \models \acsystem^{\anode_{I+1}}_{*}$.

As explained above, 
    $\arun(\anode_{I+1}) = (\alocation_{\anode_{I+1}}, \aletter_{\anode_{I+1}},
\acons_{\anode_{I+1}}, \alocation_{\anode_{I+1} \cdot 0}, \dots,
\alocation_{\anode_{I+1} \cdot (\degree-1)})$.
Furthermore, for all $i \in \interval{0}{\degree-1}$,
$\adatatree(\anode_{I+1} \cdot i) \egdef \pair{\aletter_{\anode_{I+1} \cdot i}}{\atuple_{
    \anode_{I+1} \cdot i}}$
with $\atuple_{\anode_{I+1} \cdot i}[j] \egdef \avaluation_{I+1}(v_j^{\anode_{I+1}
  \cdot i})$ for all
    $j \in \interval{1}{\beta}$.
Similarly to the base case,
$\arun(\anode_{I+1})$
is fine to satisfy forthcoming requirements about runs.

\proofsubparagraph{Case 2: \(\acdomain\) satisfies the amalgamation property and
  is homomorphism \(\omega\)-compact} We proceed a bit differently from the first case.

\textit{Base case.}  We have \(\atype_\varepsilon\) is satisfiable and
\(\atype_\varepsilon \models \acons_\varepsilon\).
    Thanks to the injective
    renaming we performed,  $\acsystem^{\varepsilon} = \acsystem^{\varepsilon}_{*}$ is satisfiable too.
    Again, we have 
    $\arun(\varepsilon) = (\alocation_\varepsilon, \aletter_\varepsilon,
    \acons_\varepsilon, \alocation_{0}, \dots, \alocation_{\degree-1})$.

    \textit{Inductive step.} Suppose that for all $k \leq I$,
    $\arun(\anode_k)$ is properly defined  and
    $\acsystem^{\anode_I}_{*}$ is satisfiable.
    Since $I+1 \geq 0$, there is $J \leq I$ such that $\anode_{I+1} = \anode_{J} \cdot j_0$
    for some $j_0 \in \interval{0}{\degree-1}$.
    Thanks to the injective
    renaming we performed,  $\acsystem^{\anode_{I+1}}$ is satisfiable.
    The only variables in $\var(\acsystem^{\anode_{I}}_{*}) \cap \var(\acsystem^{\anode_{I+1}})$
    are those of the form $v_1^{\anode_{I+1}}, \ldots, v_{\beta}^{\anode_{I+1}}$.
    We have the following properties.
    \begin{enumerate}
    \item The constraint system $\acsystem^{\anode_{I+1}}$ is satisfiable and built over $\mathcal{P}_{\aautomaton}$.
    \item The constraint system $\acsystem^{\anode_I}_{*}$ is satisfiable (by the induction hypothesis)
      and built over $\mathcal{P}_{\aautomaton}$.
    \item The only variables in $\var(\acsystem^{\anode_{I}}_{*}) \cap \var(\acsystem^{\anode_{I+1}})$
      are those of the form $v_1^{\anode_{I+1}}, \ldots, v_{\beta}^{\anode_{I+1}}$ and
      since $\atype_{\anode_{J}}$ is projection-compatible
      with $(\atype_{\anode_{J \cdot 0}}, \ldots,\atype_{\anode_{J \cdot (\degree-1)}})$,
      \[
      \acsystem^{\anode_{I}}_{*}|_{\set{v_1^{\anode_{I+1}}, \ldots, v_{\beta}^{\anode_{I+1}}}}  =
      \acsystem^{\anode_{I+1}}|_{\set{v_1^{\anode_{I+1}}, \ldots, v_{\beta}^{\anode_{I+1}}}},
      \]
      and  $\acsystem^{\anode_{I}}_{*}|_{\set{v_1^{\anode_{I+1}}, \ldots, v_{\beta}^{\anode_{I+1}}}}$
      is complete w.r.t. $\mathcal{P}_{\aautomaton}$ and $v_1^{\anode_{I+1}}, \ldots, v_{\beta}^{\anode_{I+1}}$.
      Indeed,  $\atype_{\anode_{J}}$ and  $\atype_{\anode_{J \cdot j_0}}$ are symbolic types, which leads to the completeness
      of $\acsystem^{\anode_{I}}_{*}|_{\set{v_1^{\anode_{I+1}}, \ldots, v_{\beta}^{\anode_{I+1}}}}$. 
    \end{enumerate}
    By the amalgamation property, we get
    $\acsystem^{\anode_{I}}_{*} \cup \acsystem^{\anode_{I+1}} =
    \acsystem^{\anode_{I+1}}_{*} $ is satisfiable too (whether $\acdomain$ has
    a finite set of relations or not, both cases are covered by our definition
    for amalgamation).
    Again, we require
    $\arun(\anode_{I+1}) = (\alocation_{\anode_{I+1}}, \aletter_{\anode_{I+1}},
\acons_{\anode_{I+1}}, \alocation_{\anode_{I+1} \cdot 0}, \dots,
\alocation_{\anode_{I+1} \cdot (\degree-1)})$.

The sequence $\acsystem_{*}^{\anode_0} \subseteq \acsystem_{*}^{\anode_1} \subseteq
    \cdots$ is an infinite chain  such that
    $\acsystem = \bigcup_k \acsystem_{*}^{\anode_k}$
    and therefore 
    every finite subset of $\acsystem$ is included in some $\acsystem_{*}^{\anode_k}$.
By using that $\acdomain$ is homomorphism $\omega$-compact,
we get that $\acsystem$ is satisfiable, say $\avaluation \models
\acsystem$ for some valuation $\avaluation$.
As hinted earlier, for all nodes $\anode$ and all $j \in \interval{1}{\beta}$,
    we have $\adatatree(\anode) \egdef \pair{\aletter_{\anode}}{\atuple}$
    with $\atuple[j] \egdef \avaluation(v_j^{\anode})$.
   
    \proofsubparagraph{Conclusion}
    In either case, we construct a  data tree
    \(\adatatree : \interval{0}{\degree-1}^* \to \aalphabet \times \adomain^\beta\) 
    and a corresponding run \(\arun : \interval{0}{\degree-1}^* \to \delta\) of the TGCA \(\aautomaton\),
    such that for every 
    node \(\anode\), the transition condition is satisfied:
    \(\arun(\anode) = (\alocation_{\anode}, \aletter_{\anode},
    \acons_{\anode}, \dots)\) and the associated valuation 
    satisfies \(\acons_{\anode}\).
    Moreover, since  the run
    \(\arun'\) of \(\aautomatonbis\) visits accepting locations from
    \(F \times \satsymbtypes{\aautomaton}\) 
    infinitely often along every branch, and \(\arun\) preserves the
    elements in $\locations$ (typically, the locations $\alocation_{\anode}$)
    it follows that \(\arun\) visits the locations
    in \(F\) infinitely often. Hence, the Büchi acceptance condition of \(\aautomaton\) is satisfied.
    Therefore, \(\alang(\aautomaton) \neq \emptyset\). \qedhere 
\end{proof}

In the proof of Lemma~\ref{lemma-tgca-completeness}, we have seen that
each disjunct of the condition (C1) leads to a way to
construct the valuation $\avaluation$ on which the data values
for the data trees accepted by $\aautomaton$ are defined.
If $\acdomain$ satisfies the completion property,
then 
the valuation $\avaluation$ is built incrementally while visiting
$\interval{0}{\degree-1}^*$ with a  breadth-first
search. By contrast, assuming the second disjunct of (C1)
allows us to define an infinite chain of constraint systems
(thanks to the amalgamation property) and then to obtain
the existence of $\avaluation$ as $\acdomain$ is
homomorphism $\omega$-compact. Today, it is unclear to us
what are the exact relationships between these two options;
it is not excluded that model-theoretical developments from~\cite[Chapter~4]{Rydval22}
could be helpful but we were unable to draw conclusive arguments.

\subsection{Parameterised Complexity Analysis}
It remains to perform  a parameterised analysis of the complexity
of the nonemptiness problem for TGCA.
We warn the reader that the analysis
is rather standard but this needs to be performed with care.
Given a TGCA $\aautomaton$, let
us consider the parameters below. 
\begin{itemize}
  \item \(m = \card{\mathcal{P}_\aautomaton}\).
        As a consequence of the definition of $\mathcal{P}_{\aautomaton}$,
        if $\acdomain$ has an infinite set of
        relations, $m$ is the number of predicate symbols occurring in
        $\aautomaton$.
        Otherwise, $m$ is the number of relations in $\acdomain$.
  \item \(\beta\) is the number of registers, or equivalently the dimension
    of the tuples of data values in data trees,
  \item \(\degree\) is the branching degree of the trees,
  \item \(k_0\) is the maximum arity of predicates in $\card{\mathcal{P}_\aautomaton}$ (fixed if (C2) holds true),
  \item \(\maxconstraintsize{\aautomaton}\) is defined as the maximal size of
    any 
    constraint \(\acons\) appearing in a transition of \(\aautomaton\).
\end{itemize}

\paragraph*{Construction of types}
Each type is defined over the set $\registers{\beta}{\degree}$,
containing \(\beta(\degree + 1)\) registers. The number of atomic
formulae 
over \(\registers{\beta}{\degree}\) is bounded by:
\(
|\atoms{\aautomaton}| \le m \times (\beta(\degree+1))^{k_0}.
\)
Each type corresponds to a complete assignment of truth values to these atoms,
hence:
\[
\card{\satsymbtypes{\aautomaton}} \le
\card{\symbtypes{\aautomaton}} =
2^{\card{\atoms{\aautomaton}}} 
\le 2^{m \cdot (\beta(\degree+1))^{k_0}}.
\]

The satisfiability status of each type can be checked by invoking
a decision procedure dedicated to the constraint satisfaction problem
\(\csp{\acdomain}\)
on a constraint system of size
\(\card{\atoms{\aautomaton}}\).
Let us suppose that \(\csp{\acdomain}\) can be solved in
time $\mathcal{O}(2^{\apolynomial_1(n) - n})$
for some polynomial $\apolynomial_1$ if (C3.1) is assumed.
The use of the awkward polynomial $\apolynomial_1(n) - n$ shall be convenient
for the final expression. 
Each check 
runs in time
\(
\mathcal{O}(2^{\apolynomial_1(\card{\atoms{\aautomaton}}) - \card{\atoms{\aautomaton}}})
\).
Thus, the total time for constructing the set $\satsymbtypes{\aautomaton}$ is:
\begin{equation}
\label{equation-sattypes}
2^{m \times (\beta(\degree+1))^{k_0}} \times
\mathcal{O}(2^{\apolynomial_1( m \times (\beta(\degree+1))^{k_0}) - m \times (\beta(\degree+1))^{k_0} })
= \mathcal{O}(2^{\apolynomial_1( m \times (\beta(\degree+1))^{k_0})}).
\end{equation}
\cut{
\(
2^{m \cdot (\beta(\degree+1))^{k_0}} \cdot 2^{\mathcal{O}(\apolynomial_1(m \times (\beta(\degree+1))^{k_0}))}
= 2^{\mathcal{O}(\apolynomial_1(m \times (\beta(\degree+1))^{k_0}))}.
\)
}

\paragraph*{Construction of the BTA}
The locations  in the B\"uchi tree automaton $\aautomatonbis$
are of the form \(\pair{\alocation}{\atype}\) 
with $\alocation \in \locations$ and \(\atype \in \satsymbtypes{\aautomaton}\).
So, 
\[
\card{\locations'} =
\card{\locations} \times \card{\satsymbtypes{\aautomaton}} \le \card{\locations} 
\cdot 2^{m \times (\beta(\degree+1))^{k_0}}.
\]
A transition in the B\"uchi tree automaton is derived from a TGCA transition
\((\alocation, \aletter, \acons, \alocation_0, \dots, \alocation_{\degree-1})\) 
and symbolic types \(\atype, \atype_0, \dots, \atype_{\degree-1}\) satisfying the following conditions.
\begin{itemize}
  \item \(\atype \models \acons\). This can be checked in time in 
    \(\mathcal{O}(\maxconstraintsize{\aautomaton})\),
  \item \(\atype\) is projection-compatible with
    \((\atype_0, \ldots, \atype_{\degree-1})\).
    This 
    can be 
    checked 
    in time linear in \(\card{\atoms{\aautomaton}}\) and $\degree$. Hence, checking projection-compatibility 
    takes time in \(\mathcal{O}(\degree \times \card{\atoms{\aautomaton}})\).
\end{itemize}
Thus, the total time to decide whether a given transition belongs to \(\transitions'\) belongs to 
\[
\mathcal{O}\left(\card{\transitions} \times \left(\maxconstraintsize{\aautomaton} + \degree \times
\card{\atoms{\aautomaton}} \right)\right),
\]
and the number of such transitions is:
\[
\card{\transitions'} \le \card{\locations}^{\degree+1} \times \card{\aalphabet} \times 
\card{\satsymbtypes{\aautomaton}}^{\degree+1}.
\]
Thus the total time for constructing $\transitions'$ is:
\begin{equation}
\label{equation-bta-construction}
\mathcal{O}\left(
\card{\locations}^{\degree+1} \times |\aalphabet| \times 2^{(\degree+1) \times m \times (\beta(\degree+1))^{k_0}} 
\times \card{\transitions} \times (\maxconstraintsize{\aautomaton} + \degree \times m \times (\beta(\degree+1))^{k_0})
\right).
\end{equation}

\paragraph*{Nonemptiness check}
The nonemptiness problem for Büchi tree automata is decidable in time
polynomial  in the size of automata (actually in quadratic time).
Let us assume that it can be solved in time $\mathcal{O}(\apolynomial_2(n)-n)$
for some other polynomial \(\apolynomial_2\).
Again, the use of the awkward polynomial $\apolynomial_2(n) - n$ shall be convenient
for the final expression. 
Since \(\card{\transitions'}\) dominates \(\card{\locations'}\),
the cost of this step is:

{\footnotesize
\begin{equation}
\label{equation-nonemptiness}
\mathcal{O}(\apolynomial_2(\card{\transitions'}) - \card{\transitions'})
= \mathcal{O}\left(\apolynomial_2\left(\card{\locations}^{\degree+1} \times \card{\aalphabet}
\times 2^{(\degree+1) 
\times m \times (\beta(\degree+1))^{k_0}}\right)
- \card{\locations}^{\degree+1} \times \card{\aalphabet}
\times 2^{(\degree+1) 
\times m \times (\beta(\degree+1))^{k_0}}
\right)
\end{equation}
}

\paragraph*{Overall complexity}
Taking into account the cost to generate the
satisfiable types (see Equation~(\ref{equation-sattypes})),
Büchi tree automaton construction (see Equation~(\ref{equation-bta-construction})), 
and nonemptiness checking (see Equation~(\ref{equation-nonemptiness})),
the overall time complexity belongs to:

{\footnotesize
\begin{equation}
\label{equation-overall-complexity}
\mathcal{O}\left(
2^{\apolynomial_1(m \times (\beta(\degree+1))^{k_0})}
+ 
\apolynomial_2(\card{\locations}^{\degree+1} \times \card{\aalphabet} \times 2^{(\degree+1) \times m \times 
(\beta(\degree+1))^{k_0}}) \times \card{\transitions} \times 
\left(\maxconstraintsize{\aautomaton} + \degree \times m \times (\beta(\degree+1))^{k_0} 
\right)\right).
\end{equation}
}

\begin{thm}
  \label{theorem-tgca}
  Let \(\acdomain\) be a concrete domain satisfying (C1), (C2) and (C3.1).
  The nonemptiness problem for TGCA over 
 \(\acdomain\) is in \exptime.
\end{thm}

As a slight digression,  \exptime-hardness  already holds
with the concrete domain $\pair{\adomain}{=}$ with infinite domain $\adomain$
as a consequence of~\cite[Appendix~B.1]{Demri&Quaas25}.
If we assume (C3.$k$) for some $k \geq 2$ instead of (C3.1)
in Theorem~\ref{theorem-tgca}, we get $k$-\exptime-membership.

\section{Constraint Automata for Checking Consistency}
\label{section-encoding-consistency}

Below, we reduce the consistency problem for $\ALCO(\acdomain)$
to the nonemptiness problem for TGCA over $\acdomain$. 
Correctness of the reduction only requires that the concrete domain $\acdomain$
satisfies the condition (C4). As shown in Section~\ref{section-tgca},
the translation from TGCA to BTA requires the satisfaction of (C1) and
$\csp{\acdomain}$ is decidable. In order to get \exptime-membership of
the consistency problem for $\ALCO(\acdomain)$, the condition
(C2) and (C3.1) are further assumed.

At this point, it is worth emphasizing that using the automata-based approach with
ABoxes
is not so common in the literature.
Indeed, several works dedicated to the consistency problem for description logics with concrete domains
consider only TBoxes in ontologies,
see e.g.~\cite{Baaderetal03bis,Lutz&Milicic07,CarapelleTurhan16,Labai&Ortiz&Simkus20},
most probably because this is much easier to conceive,
but with nominals, assertions can be encoded by GCIs.
There is a notable exception with recent works~\cite{Borgwardt&DeBortoli&Koopmann24,Baaderetal25}.
Indeed, handling ABoxes means that the interpretations satisfying
ontologies should rather be viewed as forests and not only as simple trees.
Fortunately, forests can be encoded as trees as soon as it is possible
to manage the global information about the different trees of the forest
in a tree-like manner. Presently, the global information should include
which individual names belong to the interpretation of subconcepts,
but also to have access to the  values returned by the concrete features
for the distinguished
elements
of the interpretation domain
interpreting the individual names. Hopefully, this can be performed
thanks to additional registers in the constraint automata.
The need for some global information within the automata-based approach
in logical formalisms having syntactic objects such as
nominals or individual names  can be found in the
works~\cite[Section~6.3]{Baaderetal03bis} and~\cite{Demri&Sattler02}.

\subsection{Preliminaries}
\label{section-automaton-preliminaries}
Given an ontology \(\aontology = \pair{\atbox}{\aabox}\),
we shall build a TGCA
\[
\aautomaton = (\locations, \aalphabet, \degree, \beta,  \locations_{in}, \transitions, F)
\]
over \(\acdomain\) such that
\(
\aontology
\)
is consistent iff $\alang(\aautomaton) \neq \emptyset$.
This section is dedicated to prepare the definition of such a reduction.
More specifically, we briefly explain why the ontology $\aontology$ can be
assumed to be in some normal form, and
we introduce several notions
that are helpful to design the constraint automaton $\aautomaton$. 
Below, we introduce the notions of concept types (similar to Hintikka sets),
and (local/global/contextual) abstractions that are finite pieces
of information about elements in  some interpretation domain
$\aidomain$. The construction of  $\aautomaton$
is finalised in Section~\ref{section-construction}
and the proof for correctness 
is postponed to Section~\ref{section-correctness-main-reduction}. 

\paragraph*{Normalisation.}
A concept is in \defstyle{negation normal form} (NNF)
if the negation operator for concepts 
occurs only
in front of
concept names or nominals. 
Every concept is logically
equivalent to a concept in NNF (and there is a simple
effective construction running in quadratic time).
In particular,
$\dlneg (\cdrestriction{\aquantifier}{v_1: \arp_1, \dots, v_k: \arp_k}{\acons})$
is logically equivalent to
$\cdrestriction{\overline{\aquantifier}}{v_1: \arp_1, \dots, v_k: \arp_k}{\neg \acons}$,
where $\aquantifier \in \set{\exists, \forall}$,
$\overline{\forall} = \exists$ and $\overline{\exists} = \forall$.
This is valid since the constraints in CD-restrictions
are {\em closed under negations}.  
Also, every GCI \(\aconcept \sqsubseteq \aconceptbis\) is 
equivalent to \(\top \sqsubseteq \lnot \aconcept \sqcup \aconceptbis\).
Hence, we assume a fixed ontology 
\(\aontology = \pair{\atbox}{\aabox}\) such that
all the concepts are in NNF and the GCIs are of the form
$\top \sqsubseteq \aconcept$. 
This normalisation is standard and harmless for complexity.

We write
$\subconcepts{\aontology}$
to denote the set of
NNF subconcepts
occurring in the ontology $\aontology$ 
augmented with the concepts $\anominal$ and $\dlneg  \anominal$
for all individual names $\aindname$ occurring
in $\aontology$ in assertions and nominals (details omitted). 
So,  all the elements of $\subconcepts{\aontology}$
are also in NNF and $\card{\subconcepts{\aontology}}$ is linear in the size of $\aontology$.
We introduce the following sets:
\begin{itemize}
\item \(\individualnames(\aontology) \egdef \{\aindname_0,\dots,\aindname_{\eta-1}\}\) is the
set of 
    individual names occurring in assertions/nominals in $\aontology$,
\item \(\cfeatures(\aontology) \egdef \{\acfeature_1,\dots,\acfeature_\alpha\}\) is the set of concrete 
    features
    occurring in concepts
    in \(\subconcepts{\aontology}\)
    and,
\item 
    \(\rolenames(\aontology) \egdef \{\arolename_1,\dots,\arolename_\kappa\}\)
    is
    the set of role names 
    occurring
    in \(\subconcepts{\aontology}\).
\end{itemize}
    
\paragraph*{Abstractions.}
Consistency of the ontology $\aontology$ asks for the existence
of an interpretation $\ainter$ such that $\ainter \models \aontology$.
In the automata-based approach,
this reduces to the existence of an infinite data tree $\adatatree$
such that $\adatatree \in \alang(\aautomaton)$.
Such a witness tree $\adatatree$  represents an interpretation
$\ainter$ (nodes in $\interval{0}{\degree-1}^*$ roughly correspond to elements of $\aidomain$),
so $\alang(\aautomaton)$ is designed to accept such representations.
An accepting run on $\adatatree$
thus ensures that $\adatatree$, viewed as an $\ALCO(\acdomain)$ interpretation,
satisfies $\aontology$.
In an accepting run $\arun$ on $\adatatree$, each node $\anode$ is labelled by a transition
$(\alocation, \aletter, \acons, \alocation_0, \ldots, \alocation_{\degree-1})$,
where the location $\alocation$ contains some
 \emph{finite symbolic abstraction} about the node $\anode$ 
 viewed as an element from the interpretation domain. This is
 where {\em local, global and contextual abstractions} enter into play.

 Given an element $\aind \in \aidomain$ for some
 interpretation $\ainter$, the \defstyle{local abstraction} of the element $\aind$
 in $\ainter$, written $\localabstraction{\aind}{\ainter}$, 
 is  a triple $\triple{\actype}{\symblinks}{\aactvector}$ defined as follows. 
 \begin{itemize}
 \item $\actype$ is the \defstyle{concept type} and
   $\actype \egdef \set{\aconcept \in \subconcepts{\aontology}
   \mid \aind \in \inter{\aconcept}}$,
 \item $\aactvector$ is the \defstyle{activity vector}
   in $\set{\perp,\top}^{\alpha}$
   such that $\aactvector[i] = \top$ $\equivdef$
   $\inter{\acfeature_i}(\aind)$ is defined, for all $i \in \interval{1}{\alpha}$
(recall that concrete features
are interpreted by partial functions),
 \item $\symblinks$ is the set of \defstyle{symbolic links}
   in $\rolenames(\aontology) \times \individualnames(\aontology)$
   such that
   $\pair{\arolename}{\aindname} \in \symblinks$
   $\equivdef$
   $\pair{\aind}{\inter{\aindname}} \in \inter{\arolename}$. 
 \end{itemize}
 It is not immediate that this notion of local abstraction suffices,
 but we shall show that it does.
 Observe that the notion of local abstraction is parameterised by ontologies
 but this is not emphasized in the notation. 
 Of course, in the forthcoming TGCA $\aautomaton$, we also 
 make use of the registers to encode the concrete features from interpretations
 of $\ALCO(\acdomain)$.
 By contrast, the \defstyle{global abstractions} act as persistent 
 memories relevant for the whole interpretation.
 Unsurprisingly, they assign to each individual
 name in $\individualnames(\aontology)$ a local abstraction;
 due to a technical reason, 
 we can restrict these to just the concept type and the activity vector.
Since runs of the one-way automaton $\aautomaton$ cannot revisit nodes,
the global abstraction enables
to recover the local abstraction
of the individual names from $\individualnames(\aontology)$
from  any node. 
We use a standard notion of \defstyle{types} to represent locally consistent 
sets of subconcepts,
similarly to Hintikka sets to prove completeness
of propositional logic for tableaux-style calculi,
see also~\cite[Section~5.1.2]{Baaderetal17}. 
The set of \defstyle{concept types}, 
denoted $\ctypes{\aontology}$, 
consists of all subsets \(\actype \subseteq \subconcepts{\aontology}\) such that
\begin{enumerate}
\item \(\bot \notin \actype\),
\item for all concept names \(\aconceptname \in \ctypes{\aontology}\), if \(\aconceptname \in \actype\) 
    then \(\dlneg \aconceptname \notin \actype\),
 \item for all $j \in \interval{0}{\eta-1}$,
   either $\{ \hspace{-0.02in} \aindname_j \hspace{-0.02in}\}
   \in \actype$
   or $\dlneg \{ \hspace{-0.02in} \aindname_j \hspace{-0.02in}\}
   \in \actype$ (but never both),
 \item if \(\aconcept \sqcap \aconceptbis \in \actype\), then \(\aconcept \in \actype\) and \(\aconceptbis \in \actype\),
 \item 
    if \(\aconcept \sqcup \aconceptbis \in \actype\), then \(\aconcept \in \actype\) or \(\aconceptbis \in \actype\),
  \item for every GCI
    \(\top \sqsubseteq \aconcept \in \atbox\) (recall \(\atbox\) is the TBox), we have \(\aconcept \in \actype\).
\end{enumerate}
An \defstyle{anonymous} concept type $\actype$
is such that
$\set{\dlneg \{ \hspace{-0.02in} \aindname_0 \hspace{-0.02in}\},
  \ldots, \dlneg \{ \hspace{-0.02in} \aindname_{\eta-1} \hspace{-0.02in}\}}
\subseteq \actype$.
An $\aindname$-type $\actype$ is such that
$\anominal \in \actype$ and for
all $\anominalbis$ with $\aindnamebis \in \individualnames(\aontology)$
distinct from $\aindname$,
we have $\neg \anominalbis \in \actype$
(see e.g.~the named states in~\cite[Section~5]{Demri&Sattler02} and the named
types in~\cite[Section~4]{Borgwardt&DeBortoli&Koopmann24}). 

An \defstyle{activity vector} $\aactvector$
is a tuple in \(\{\top, \bot\}^\alpha\)
where the \(i\)-th component $\aactvector[i]$ indicates whether the  concrete feature
$\acfeature_i$ (from $\cfeatures(\aontology)$)
is defined (\(\top\)) or not (\(\bot\)).
The set of activity vectors is written $\act$.
Since the interpretation of each concrete feature is a partial function, 
only the registers corresponding to positions where $\aactvector[i] = \top$ 
represent meaningful feature values.
While the data tree assigns values to all registers uniformly, 
entries with $\aactvector[i] = \bot$ are 
ignored when encoding $\ALCO(\acdomain)$ interpretations.
A \defstyle{symbolic link} is a pair
$\pair{\arolename}{\aindname} \in
\rolenames(\aontology) 
\times \individualnames(\aontology)$
intended to represent that
the current element is related to the
interpretation of
the individual name \(\aindname\) via
the interpretation of the role name \(\arolename\).
The set \(\slinks\) of symbolic link sets is the powerset
\(\powerset{\rolenames(\aontology) \times \individualnames(\aontology)}\).
A graphical representation is given below. The dashed arrow from
\(\aindname_0\) to \(\aindname_1\), labelled by \(\arolename_2\), represents the
symbolic link induced by the ABox assertion
\(
\rassertion{\aindname_0}{\aindname_1}{\arolename_2} \in \aabox .
\)
The second dashed arrow, labelled by \(\arolename_1\), illustrates another
possible symbolic link from an anonymous element to the element representing
\(\aindname_0\).

\begin{center}
    \scalebox{1}{      
\begin{tikzpicture}[node distance=1.4cm, bend angle=10,
                    level 1/.style={sibling distance=1.5cm},
                    level 2/.style={sibling distance=1cm},
                    level 3/.style={sibling distance=0.8cm},
                    level distance=1.5cm]

  \node[state] (root) {$\varepsilon$}
   child { node[state] (aone) {$\aindname_0$}
           edge from parent node[above left] {}
   }
   child { node[state] (atwo) {$\aindname_1$}
     child { node[state]  (a) {}
       child { node[state] {}
           edge from parent node[above left] {}
   }
   child { node[state] {}
           edge from parent node[above left] {}
   }
   child { node[state] {}
           edge from parent node[above left] {}
   }
   child { node[state]  {}
           edge from parent node[above left] {}
   }
           edge from parent node[above left] {}
   }
   child { node[state] {}
           edge from parent node[above left] {}
   }
   child { node[state] {}
           edge from parent node[above left] {}
   }
   child { node[state]  {}
           edge from parent node[above left] {}
   }
           edge from parent node[above left] {}
   }
   child { node[state] {$\square$}
           edge from parent node[above left] {}
   }
   child { node[state] (lastsquare)  {$\square$}
           edge from parent node[above left] {}
   }
   ;
   \path [->,dashed] (a)  edge[bend left=45]  node[sloped,above]  {$\arolename_1$} (aone);
   \path [->,dashed] (aone)  edge[bend left=20]  node[sloped,below]  {$\arolename_2$} (atwo);
   
\end{tikzpicture}
    }
\end{center}

A \defstyle{local abstraction} $\labs$ is a triple in $\ctypes{\aontology} \times \slinks \times \act$.
A \defstyle{global abstraction} $\gabs$ is a function
\(
\gabs : \individualnames(\aontology) \to \ctypes{\aontology} \times \act
\).
A location of the forthcoming automaton $\aautomaton$ is structured
in such a way that it contains a local abstraction $\labs$ (related
to the associated node) and a global abstraction $\gabs$ that records
a fixed knowledge about the individual names. Such pairs
are called \defstyle{contextual abstractions} $\cabs= \pair{\gabs}{\labs}$.

\paragraph*{Branching degree $\degree$.}
We need  to determine
a branching degree $\degree$ that allows us to satisfy all kinds
of existential restrictions.
Fortunately, the value $\degree$ shall be polynomial in the size
of $\aontology$. 
To do so, we introduce the following values.

\begin{itemize}
\item \(N_{\mathsf{ex}} \egdef \card{\set{\exrest{\arolename}{\aconcept} \mid
    \exrest{\arolename}{\aconcept} \in \subconcepts{\aontology}}}\).

\item \(N_{\mathsf{cd}} \egdef
  \card{
    \set{\cdrestriction{\exists}{v_1: \arp_1, \dots, v_k: \arp_k}{\acons}
      \mid \cdrestriction{\exists}{v_1: \arp_1, \dots, v_k: \arp_k}{\acons}
       \in \subconcepts{\aontology}
  }}\).

\item \(N_{\mathsf{var}} \egdef 
  \max (\set{0} \cup \set{k \mid
    \cdrestriction{\exists}{v_1: \arp_1, \dots, v_k: \arp_k}{\acons}
       \in \subconcepts{\aontology}})
  \).
\end{itemize}

The rationale to define $\degree$ is based on the following
requirements. For each $\exrest{\arolename}{\aconcept} \in \subconcepts{\aontology}$,
one needs one direction to have a witness individual.
Similarly, for each $\cdrestriction{\exists}{v_1: \arp_1, \dots, v_k: \arp_k}{\acons}
\in \subconcepts{\aontology}$, 
$k$ directions are sufficient to get $k$ witness values in $\adomain$.
In addition, the tree interpretation property for
$\ALCO(\acdomain)$ is based on the representation
of a forest such that each of its trees corresponds
to the unfolding of the interpretation from some individual name
in $\individualnames(\aontology)$. 
Since this forest is encoded below the root of the data tree accepted by
\(\aautomaton\), the degree must also be at least \(\eta=\card{\individualnames(\aontology)}\). 
The unused directions are sent to a sink location, represented by \(\square\). 
This is illustrated below with \(\eta = 2\) and \(\degree = 4\).

\begin{center}
    \scalebox{0.8}{      
\begin{tikzpicture}[node distance=1.4cm, bend angle=10,
                    level 1/.style={sibling distance=1.5cm},
                    level 2/.style={sibling distance=2cm},
                    level 3/.style={sibling distance=1cm},
                    level distance=1.5cm]

  \node[state] (root) {$\varepsilon$}
   child { node[state]  {$\aindname_0$}
           edge from parent node[above left] {}
   }
   child { node[state] {$\aindname_1$}
           edge from parent node[above left] {}
   }
   child { node[state] {$\square$}
           edge from parent node[above left] {}
   }
   child { node[state] (a) {$\square$}
           edge from parent node[above left] {}
         }
   ;

\end{tikzpicture}
    }
\end{center}

So, it makes sense to define $\degree$ as the value
\(\max\big\{N_{\mathsf{ex}} + N_{\mathsf{cd}} \times N_{\mathsf{var}}, 
\card{\individualnames(\aontology)} \big\}\); the
forthcoming soundness proof
of Lemma~\ref{lemma-encoding-soundness}
shall confirm this formally.

\paragraph*{Correspondence between directions, role names
  and restrictions.}
Once the degree $\degree$ is defined, one needs to establish
a discipline to put in correspondence directions in $\interval{0}{\degree-1}$,
role names
and restrictions. To do so, below, we define the maps $\lambda$ and
$dir$. Indeed, a standard approach consists in booking directions
in $\interval{0}{\degree-1}$ for each role name in $\rolenames(\aontology)$.
This is needed because the data trees have no labels on edges. Another
technique used in~\cite[Section~3.2]{Baader09} adds to the local
abstractions the role name of the incoming edge (unique thanks to the
tree structure). 
Herein, the map $\lambda$ takes care of
assigning directions to existential restrictions
and to role paths in existential CD-restrictions. By contrast,
the map $dir$ is {\em uniquely} determined by
$\lambda$ and returns a set of directions among $\interval{0}{\degree-1}$
to each role name. Such a discipline is required
to construct the transitions in $\aautomaton$ as one needs
to agree on the directions that are relevant for getting
witnesses of a given restriction.
The extension to functional role names would be then pretty smooth, see
Section~\ref{section-functional-role-names}.
We introduce the following sets.

\begin{itemize}
\item \(\mathcal{E}_r \egdef
  \{\exrest{\arolename_{i_1}}{\aconcept_1}, \ldots,
  \exrest{\arolename_{i_{N_{\mathsf{ex}}}}}{\aconcept_{N_{\mathsf{ex}}}}
  \}\) contains all the existential restrictions from $\subconcepts{\aontology}$.

\item \(\mathcal{E}_{\mathsf{cd}} \egdef
  \{\aecdconcept_1, \ldots, \aecdconcept_{N_{\mathsf{cd}}}\}\)
  contains all the existential CD-restrictions from $\subconcepts{\aontology}$.
\end{itemize}

We define the labeling domain as:
\(
\mathcal{B} \egdef
\mathcal{E}_r \cup \{ (\aecdconcept_i, j)
\mid 1 \le i \le N_{\mathsf{cd}},\ 1 \le j \le N_{\mathsf{var}} \}.
\)
\iftoggle{versionlong}{

\begin{defi}[Branch labeling function]
  The global labeling function is the injective map
  \(
    \lambda : \mathcal{B} \hookrightarrow \interval{0}{\degree-1}
  \)
  defined as follows:
  \begin{itemize}
  \itemsep 0cm
  \item \(\lambda(\exrest{\arolename_{i_k}}{\aconcept_k}) \egdef k - 1\),
    for \(1 \le k \le N_{\mathsf{ex}}\),
  \item \(\lambda(\aecdconcept_i, j) \egdef
    N_{\mathsf{ex}} + (i{-}1) \times N_{\mathsf{var}} + (j{-}1)\),
    for \(1 \le i \le N_{\mathsf{cd}}\) and \(1 \le j \le N_{\mathsf{var}}\).
  \end{itemize}
  Thus, \(\lambda\) maps \(\mathcal{B}\) bijectively onto its image. Directions in
  \(\interval{0}{\degree-1} \setminus \operatorname{im}(\lambda)\), if any, are unused.
\end{defi}

The definition of the map $dir$
that is uniquely
determined by $\lambda$
is provided below. 

\begin{defi}[Role directions]
Let \(\arolename \in \rolenames(\aontology)\). The set of
directions associated with the role name \(\arolename\) is defined as:
\[
\dir{\arolename} \egdef \left\{\, i \in \interval{0}{\degree{-}1} \;\middle|\;
\begin{array}{l}
\lambda^{-1}(i) = \exrest{\arolename}{\aconcept} \text{ for some } \aconcept, \text{ or} \\
\lambda^{-1}(i) = (\aecdconcept, j) \text{ where } \aecdconcept =
\cdrestriction{\exists}{v_1{:}\arolepath_1, \dots, v_k{:}\arolepath_k}{\acons}, \\
\quad\text{with } 1 \le j \le k \text{ and } \arolepath_j = \arolename \cdot \acfeature
\end{array}
\,\right\}.
\]
\end{defi}
}{
 The global labeling function \(\lambda : \mathcal{B} \to
  \interval{0}{\degree-1}\), that is designed as a bijection, 
  is defined as:
  
  \begin{itemize}

\item \(\lambda(\exrest{\arolename_{i_k}}{\aconcept_k}) \egdef k - 1\),
  for \(1 \le k \le N_{\mathsf{ex}}\),
\item \(\lambda(\aecdconcept_i, j) \egdef N_{\mathsf{ex}} + (i{-}1) \times
  N_{\mathsf{var}} + (j{-}1)\), for \(1 \le i \le N_{\mathsf{cd}},\ 1 \le j \le N_{\mathsf{var}}\).
  \end{itemize}
  
  The definition of the map $dir$
is provided below. 
Let \(\arolename \in \rolenames(\aontology)\).
The set of
directions associated with the role name \(\arolename\) is defined as:

\[
\dir{\arolename} \egdef \left\{\, i \in \interval{0}{\degree{-}1} \;\middle|\;
\begin{array}{l}
\lambda^{-1}(i) = \exrest{\arolename}{\aconcept} \text{ for some } \aconcept, \text{ or} \\
\lambda^{-1}(i) = (\aecdconcept, j) \text{ where } \aecdconcept =
\cdrestriction{\exists}{v_1{:}\arolepath_1, \dots, v_k{:}\arolepath_k}{\acons}, \\
\quad\text{with } 1 \le j \le k \text{ and } \arolepath_j = \arolename \cdot \acfeature
\end{array}
\,\right\}.
\]
}

\paragraph*{Dimension $\beta$ and conventions about the registers.} 
The forthcoming automaton $\aautomaton$ accepts
infinite data trees $\adatatree: \interval{0}{\degree-1}
\to \aalphabet \times \adomain^{\beta}$ and it remains
to define
the dimension $\beta$ before moving to
the final steps of the construction in Section~\ref{section-construction}.
Since $\cfeatures(\aontology)$ contains $\alpha$ concrete features, 
we require at least $\alpha$ registers.
Though global abstractions are persistent memories, this is not
sufficient to handle the values returned by the
concrete features for the individual names. This is useful
to consider the satisfaction of CD-restrictions requiring
as witness individuals those interpreting
some individual name in~$\individualnames(\aontology)$.
That is why, we augment the dimension by
$\card{\individualnames(\aontology)} \times
\alpha$ so that it is possible to carry all over the runs
the values of the concrete features for such individuals.
Consequently, we set
$\beta \egdef (\eta +1) \times \alpha$. 
To use the registers in $\registers{\beta}{\degree}$,
with $\beta$ and $\degree$ defined above, 
we follow conventions that happen to be handy and that diverge slightly
from the general notations introduced in Section~\ref{section-tgca}. 
The designated registers $\aregister_1, \ldots, \aregister_{\beta}$
referring to values at the current node shall be represented by
\[
\aregister_1, \ldots, \aregister_{\alpha},
\aregister_{\aindname_0, 1}, \ldots,\aregister_{\aindname_0,\alpha},
\ldots,
\aregister_{\aindname_{\eta-1}, 1}, \ldots,\aregister_{\aindname_{\eta-1},\alpha} .
\]
Hence, implicitly $\aregister_j$ refers to the value of the concrete feature
$\acfeature_j$, if any. Similarly, $\aregister_{\aindname_{\ell}, j}$
refers to the value of the concrete feature $\acfeature_j$ for the interpretation
of the individual name $\aindname_{\ell}$.
Furthermore, we use $\aregister_j^i$ (resp.
$\aregister_{\aindname_{\ell}, j}^i$) to refer to the value
of $\acfeature_j$ for the $i$th child
(resp. for the  individual name $\aindname_{\ell}$).
However, by construction of $\aautomaton$, we require
that $\aregister_{\aindname_{\ell}, j}$ is enforced to be equal to
$\aregister_{\aindname_{\ell}, j}^i$ as such registers are dedicated
to store global values.
\iftoggle{versionlong}{
\begin{figure}[t]
  \centering
  \begin{tikzpicture}[scale=1, every node/.style={scale=1}]
    \node[draw, thick, minimum width=2cm, minimum height=1cm] (n) at (0,0) {\textbf{Node \(\anode\)}};
    \node[above=0.1cm of n] (nreg) {\(x_1, \dots, x_\alpha\)};

    \foreach \i in {0,1,2} {
      \node[draw, thick, minimum width=1.6cm, minimum height=0.8cm] (c\i) at (-3 + 3*\i, -2.5) {\textbf{Child \(\anode \cdot \i\)}};
      \node[below=0.1cm of c\i] (creg\i) {\(x_{1}^{\i}, \dots, x_{\alpha}^{\i}\)};
      \draw[->, thick] (n) -- (c\i);
    }

    \foreach \i/\name in {0/\aindname,1/\aindnamebis} {
      \node[draw, thick, minimum width=1.8cm, minimum height=0.8cm, fill=blue!5] (ind\i) at (4.2, 1.8 - 1.2*\i) {\(\name\)};
      \node[right=0.1cm of ind\i] (indreg\i) {\(x_{\name,1}, \dots, x_{\name,\alpha}\)};
      \draw[->, thick, dashed] (n) -- (ind\i);
    }

    \foreach \i in {0,1,2} {
      \node[below=0.1cm of creg\i] (copylabel\i) {\footnotesize Copies of \(x_{a,j}, x_{b,j}\)};
      \node[below=0.15cm of copylabel\i] (copyvars\i) {\footnotesize \(x_{a,j}^{\i},\ x_{b,j}^{\i}\)};
    }

  \end{tikzpicture}
  \caption{Accessible registers and variable names at node \(\anode\).
    }
  \label{fig:variables-and-registers}
\end{figure}
Figure~\ref{fig:variables-and-registers} illustrates these conventions.
}

\subsection{Definition of the Constraint Automaton \texorpdfstring{$\aautomaton$}{}}
\label{section-construction}

Now, we define the  TGCA \(\aautomaton = (\locations, \aalphabet, \degree, \beta, 
\locations_{in}, \transitions, F)\).
Each non-root node 
\( \anode \in \interval{0}{\degree-1}^+\) in a run of the 
automaton may represent an element of the interpretation domain, but this is not
systematic. 
A node  represents a domain element if its location is
not the designated {\em sink location}—a special marker
to identify {\em inactive} nodes. 
In the active case, the location 
encodes a contextual abstraction as previously defined.
By contrast, the root node $\varepsilon$
does not correspond to any domain element. 
It serves as a technical device to assemble a forest-shaped model into a single data tree:
its immediate children within \(\interval{0}{\eta-1}\)
encode the interpretations 
of the individual names in the ABox
(recall we can safely assume (UNA)). 
The transitions are designed to enforce the semantical properties
of the description logic $\ALCO(\acdomain)$
based on the contextual abstractions.
They ensure that
the restrictions are satisfied via directions in $\interval{0}{\degree-1}$
and  via symbolic links; that the global abstraction remains consistent across the run;
and that the tree structure 
corresponds to a forest-shaped interpretation of the ontology.
Below, we present the definition
of the constraint automaton $\aautomaton$ based on 
the preliminary definitions in Section~\ref{section-automaton-preliminaries}.

\paragraph*{Set of locations.} 
The set $\locations$ is defined as
the set $\cabsset \cup \set{\rootlocation, \square}$
and $F \egdef \locations$,
where

\begin{itemize}

\item \(\cabsset \egdef \gabsset \times \labsset\) is the set 
  of \defstyle{contextual abstractions}, where \(\gabsset\) 
  is the set of \defstyle{global abstractions} and \(\labsset\) the set 
  of \defstyle{local abstractions},
 
\item \(\rootnode\) is a distinguished location used only at the root (\defstyle{root location})
  and $\locations_{in} \egdef \set{\rootlocation}$,
\item \(\square\) is a \defstyle{sink location} used to label nodes
  that represent no domain elements.
  It propagates systematically over the whole structure. 
\end{itemize}

\paragraph*{Alphabet.}
The alphabet \(\aalphabet\) is  \(\{\aletter\}\); 
the automaton $\aautomaton$ does not rely on input symbols—only the structure 
of the tree and the tuples in $\adomain^{\beta}$ matter.

\paragraph*{Transition relation.}
Before providing the formal definition, let us briefly explain its main
principles.
Each non-root node in a run is either associated with a 
contextual abstraction representing a candidate element of the interpretation domain,
or is labelled with the sink location \(\square\).
If a node is associated with a contextual abstraction, 
its concept type  encodes the concepts it must belong to in any 
corresponding interpretation: if a concept appears in the concept type of a
node,  then the interpreted element is required to satisfy that concept
and this creates obligations (see the conditions for general transitions).
Role names are interpreted with the help of two kinds of links:
\begin{itemize}
\item[(i)] \emph{direct links} to
the children, where each direction in $\interval{0}{\degree-1}$
is possibly attached to a role name and,
\item[(ii)] \emph{symbolic links} connecting a node 
to the interpretation of individual names from the ABox. These links are used to 
satisfy CD-restrictions, existential restrictions and value restrictions.
\end{itemize}

Constraints are evaluated using register values drawn from the current 
node and its children (as determined by \(\lambda\)). 
The current node stores its own concrete feature values via
the registers \(\aregister_j\)
and also the concrete feature values for the interpretation of the individual names
via the registers 
\(\aregister_{\aindname, j}\) as part of the global information. 
These latter values stored in registers for the interpretation of the individual names
must remain
constant during the run, and this
is explicitly enforced 
by the transition relation with the help of the constraints. Activity vectors determine
which concrete feature values are defined and 
guarantee that only the defined values are referred to in the constraints.
The global abstraction—common to all nodes labelled
by non-sink locations—must remain constant too.

\cut{
\iftoggle{versionlong}{
\begin{itemize}
  \item \textbf{Padding transitions} (a unique \(\atransition_{\square}\)) map \(\square\) 
    to itself on all branches, maintaining the tree structure without semantic impact.
  \item \textbf{Root transitions} initialize the run at \(\varepsilon\), assigning to each 
    successor a contextual abstraction that encodes the interpretation of an individual name from the ABox.
  \item \textbf{General transitions} apply at
   nodes labelled by contextual 
   abstractions. They enforce the satisfaction of subconcepts in types
   and propagate the global information. 
\end{itemize}

Here come the formal definitions:
}
}
Formally,
the transition relation $\transitions$ is a subset of
\(\locations \times \aalphabet \times 
\treeconstraints{\beta, \degree} \times \locations^\degree\)
and contains three kinds of transitions. 
\begin{description}
\item[\textbf{(a) Padding transition}]
There is a unique padding transition \(
\atransition_{\square} \egdef (\square, \aletter, \top, \square, \dots, \square)
\).
\item[\textbf{(b) Root transitions}]
  From the root location $\rootlocation$, the automaton $\aautomaton$ performs an 
  initialization step by assigning contextual abstractions to the first 
  $\eta$ children, each of which represents the 
  interpretation of an individual name from the ABox. Each such contextual 
  abstraction is of the form $(\gabs_i, (\actype_i, \symblinks_i, \aactvector_i))$.
  For each individual name $\aindname_i$, the local abstraction assigned to 
  the $i$-th child must match the corresponding entry in the global abstraction: 
  specifically, the concept type and activity vector must coincide with 
  $\gabs(\aindname_i)$. In addition, the value of each concrete feature 
  $\acfeature_j$ for  $\aindname_i$ must be consistently stored in the register 
  $\aregister_{\aindname_i, j}$ at each node, which is treated as a constant 
  throughout the run. This value originates from the local register $\aregister_j$ 
  of the $i$-th child, 
  which represents the interpretation of $\aindname_i$. Therefore, the consistency 
  condition $\aregister_{\aindname_i, j} = \aregister_j$ must hold at that node.  
  These requirements 
  together form the \defstyle{global information consistency check}.
  Beyond this, the local abstraction must correctly reflect the ABox assertions 
  involving $\aindname_i$.

  Formally, root transitions are of the form
  \[
  (\rootlocation, \aletter, \acons_{\mathsf{root}}, \cabs_0, \ldots, \cabs_{\eta-1}, \square, \ldots, \square),
  \] 
  where each \(\cabs_i = (\gabs_i, (\actype_i, \symblinks_i, \aactvector_i))\) 
  is a contextual abstraction and $\actype_i$ is an
  $\aindname_i$-type.
  The constraint \(\acons_{\mathsf{root}}\) is equal to the expression
  \[
  \bigwedge_{i,i' \in \interval{0}{\eta-1}, j \in \interval{1}{\alpha}, \aactvector_{\aindname_i}[j] = \top} 
  \aregister_{j}^i = \aregister_{\aindname_i, j}^{i'}.
  \] 
  \iftoggle{versionlong}{
  Indeed, the constraint \(\acons_{\mathsf{root}}\) enforces the following equalities.
  \begin{itemize}
    \item For all \(0 \le i < \eta\) and all \(1 \le j \le \alpha\), we have:
      \(
      \aregister_j^i = \aregister_{\aindname_{i}, j}^i,
      \)
      ensuring that the register for the \(j\)th concrete feature of individual name
      \(\aindname_{i}\) matches its local value at the node \(i\)
      (we require $\aactvector_{\aindname_i}[j] = \top$);
    \item For all \(0 \le i_1, i_2 < \eta\), all \(\aindname \in 
      \individualnames(\aontology)\), and all \(1 \le j \le \alpha\), we require:
      \(
      \aregister_{\aindname, j}^{i_1} = \aregister_{\aindname, j}^{i_2},
      \)
      ensuring that the registers for each individual name are 
      globally consistent across all the children
      (we require $\aactvector_{\aindname_{i_1}}[j] = \top$
      and $\aactvector_{\aindname_{i_2}}[j] = \top$).
  \end{itemize}
}

\iftoggle{versionlong}{
  The transition must also satisfy the following structural conditions.
  \begin{enumerate}
\item \(\gabs_0 = \cdots = \gabs_{\eta-1}\).
    \item For each \(0 \le i < \eta\), the local abstraction matches 
      the global one: \((\actype_i, \aactvector_i) = \gabs(\aindname_{i})\),
    \item \(\actype_i\) contains all concepts asserted for \(\aindname_{i}\) in the ABox:
      \(
      \set{\aconcept \mid \cassertion{\aindname_{i}}{\aconcept} \in \aabox} \subseteq \actype_i,
      \)
    \item \(\symblinks_i\) is compatible with the ABox $\aabox$:
      \(
      \set{(\arolename, \aindnamebis) \mid \rassertion{\aindname_{i}}{\aindnamebis}{\arolename} \in \aabox}
      \subseteq \symblinks_i.
      \)
  \end{enumerate}
}{
The transition must also satisfy the following structural conditions: 
\(\gabs_0 = \cdots = \gabs_{\eta-1}\) and 
for all $i \in \interval{0}{\eta-1}$,
\begin{center}
  \((\actype_i, \aactvector_i) = \gabs(\aindname_{i})\),
  \(
      \set{\aconcept \mid \cassertion{\aindname_{i}}{\aconcept} \in \aabox} \subseteq \actype_i,
      \) and
      \(
      \set{(\arolename, \aindnamebis) \mid \rassertion{\aindname_{i}}{\aindnamebis}{\arolename} \in \aabox}
      \subseteq \symblinks_i
      \).
\end{center} 
}
  From now on, for each individual name \(\aindname \in \individualnames(\aontology)\),
  we let \(\gabs(\aindname) = (\actype_{\aindname}, \aactvector_{\aindname})\), where 
  \(\actype_{\aindname}\) denotes the concept type and \(\aactvector_{\aindname}\) 
  the activity vector associated to \(\aindname\) via the global abstraction \(\gabs\). 
  This notation is well-defined due to the global information consistency check 
  performed by the root transition.

\item[\textbf{(c) General transitions}]
  From a location
\(\cabs = (\gabs, (\actype, \symblinks, \aactvector))\), 
we need to find witness individuals for  the existential  restrictions 
and CD–restrictions occurring in
the concept type
\(\actype\). 
Each  witness can be realized in two ways: either through a  child, 
determined via the branch labeling function \(\lambda\), 
or via a symbolic link in \(\symblinks\) pointing to
an individual name.
As for value restrictions, the transition must verify that 
\emph{all} possible witnesses satisfy the required concept or constraint. 
For universal CD–restrictions, the constraint must 
hold across all applicable combinations of registers extracted from 
those directions and links.
General transitions  also maintain some global information
and are of the form 
\(
\atransition = (\cabs, \aletter, \acons_\atransition, \alocation_0, \dots, \alocation_{\degree-1})
\),
where $\cabs = (\globalinfo, (\actype, \symblinks, \aactvector))$
satisfies the conditions below.
\begin{enumerate}
\item \textbf{Existential restrictions.}
  For all \(\exrest{\arolename}{\aconcept} \in \actype\),
  there is $\pair{\arolename}{\aindname}
  \in \symblinks$ such that $\aconcept \in \actype_{\aindname}$
  or \(i = \lambda(\exrest{\arolename}{\aconcept})\),  
  \(\alocation_i\) is of the form
  \((\globalinfo_i, (\actype_i, \symblinks_i, \aactvector_i))\)
  and $\aconcept \in \actype_i$. 

\item \textbf{Value restrictions.}
  For all \(\varest{\arolename}{\aconcept} \in \actype\),
  for all \(i \in \dir{\arolename}\) with \(\alocation_i\) of the form
  \((\globalinfo_i, (\actype_i, \symblinks_i, \aactvector_i))\),
  we have \(\aconcept \in \actype_i\)
  and  for all \((\arolename, \aindname) \in \symblinks\),
  we have \(\aconcept \in \actype_\aindname\).  

\item \textbf{CD-restrictions.}  
  Let \(\mathrm{CD}_\exists(\actype)\) and \(\mathrm{CD}_\forall(\actype)\) be respectively
  the sets of existential and universal 
CD-restrictions in \(\actype\). 
Let \(\aconcept = \cdrestriction{\aquantifier}{v_1: \arp_1, \dots, v_k: \arp_k}{\acons(v_1,\dots,v_k)}\) 
be such a CD-restriction with
quantifier
\(\aquantifier \in \{\exists, \forall\}\). 
For each role path
\(\arp_j\), we define a
candidate set $\asetter_j$ of
registers
as follows:

\begin{itemize}

\item If \(\arp_j = \acfeature_l\), then
  if $\aactvector[l] = \top$, then $\asetter_j \egdef  \set{\aregister_l}$,
  otherwise $\asetter_j \egdef \emptyset$.
  \cut{
  \(
  \asetter_j := 
  \begin{cases}
    \{\aregister_l\}, & \text{if } \aactvector[l] = \top, \\
    \emptyset, & \text{otherwise}.
  \end{cases}
  \)
}
  \item If \(\arp_j = \arolename \cdot \acfeature_l\), then \(\asetter_j\) consists of the values of 
    the feature \(\acfeature_l\) 
    taken from nodes reachable via the role name  \(\arolename\), provided the corresponding feature is defined.
\cut{
\begin{tightcenter}
{\small 
$\hspace*{-0.1in} \asetter_j \egdef \set{ \aregister_{l}^{i} \mid i \in \dir{\arolename},\ 
\alocation_i = (\globalinfo_i, (\actype_i, \symblinks_i, \aactvector_i)),
  \aactvector_i[l] = \top} 
         \cup \set{ \aregister_{\aindname,l} \mid (\arolename, \aindname) 
         \in \symblinks,\ \aactvector_\aindname[l] = \top}.$}
\end{tightcenter}
}
{\small 
  \[
  \asetter_j \egdef \set{ \aregister_{l}^{i} \mid i \in \dir{\arolename},\ 
\alocation_i = (\globalinfo_i, (\actype_i, \symblinks_i, \aactvector_i)),
  \aactvector_i[l] = \top} 
         \cup \set{ \aregister_{\aindname,l} \mid (\arolename, \aindname) 
           \in \symblinks,\ \aactvector_\aindname[l] = \top}.
         \]
         }
\end{itemize}

Let \(\asetter \egdef \asetter_1 \times \cdots \times \asetter_k\).
\iftoggle{versionlong}{
We define the constraint:
\[
\acons_{\aconcept} \egdef
\begin{cases}
  \bigvee_{(\aregisterbis_1,\dots,\aregisterbis_k) \in \asetter}
  \acons(\aregisterbis_1,\dots,\aregisterbis_k), & \text{if } \aconcept
  \in \mathrm{CD}_\exists(\actype), \\[1mm]
  \bigwedge_{(\aregisterbis_1,\dots,\aregisterbis_k) \in \asetter}
  \acons(\aregisterbis_1,\dots,\aregisterbis_k), & \text{if } \aconcept
  \in \mathrm{CD}_\forall(\actype).
\end{cases}
\]
}{
If $\aconcept
\in \mathrm{CD}_\exists(\actype)$, then
$\acons_{\aconcept} \egdef \bigvee_{(\aregisterbis_1,\dots,\aregisterbis_k) \in \asetter}
\acons(\aregisterbis_1,\dots,\aregisterbis_k)$,
otherwise $\acons_{\aconcept} \egdef \bigwedge_{(\aregisterbis_1,\dots,\aregisterbis_k) \in \asetter}
\acons(\aregisterbis_1,\dots,\aregisterbis_k)$.
}
By convention, the empty disjunction is $\perp$
and the empty conjunction $\top$.

The constraint $\acons_\atransition$ is defined below, where
the second part expresses the consistency of the registers dedicated to individual names.
\[
\acons_\atransition \egdef
\overbrace{
  \bigwedge_{\aconcept \in \mathrm{CD}_\exists(\actype) \cup \mathrm{CD}_\forall(\actype)}
  \acons_\aconcept
}^{\mbox{\small dedication to CD-restrictions}}
\wedge
\overbrace{
  \bigwedge_{\substack{i \in \interval{0}{\degree-1}  \\ \alocation_i \neq \square}} 
\bigwedge_{j = 1}^\alpha \ \ \ 
\bigwedge_{j' \in \interval{0}{\eta-1}, \ \aactvector_{\aindname_{j'}}[j] = \top} 
\aregister_{\aindname_{j'},j} = \aregister_{\aindname_{j'},j}^i
}^{\mbox{\small propagation of global values}}.
\]

The constraint
\(\acons_\atransition\) is uniquely determined by the
sequence $\alocation, \alocation_0, \ldots, \alocation_{\degree-1}$.
\cut{
Note that \(\acons_\atransition\) is uniquely determined by the source location \(\alocation\) and 
its children \(\alocation_0, \dots, \alocation_{\degree-1}\). 
This is because the set of CD–restrictions
to be enforced, the mapping of register values, the directions \(\dir{-}\), and the active symbolic links
are all encoded in the contextual abstractions (i.e., \(\alocation\) and each \(\alocation_i\)).
As a result, once the locations are fixed, the construction of the associated constraint
\(\acons_\atransition\) is fully determined.
}

\item \textbf{Global abstraction preservation.}  
For each \(i < \degree\) with \(\alocation_i \neq \square\), we require \(\globalinfo_i = \globalinfo\).

\item \textbf{Inactive children.}  For
  $i \in \interval{0}{\degree-1}$,
  if $\lambda^{-1}(i) = \exrest{\arolename}{\aconcept}$ and $\lambda^{-1}(i) \not \in \actype$
  or $\lambda^{-1}(i) =
  \pair{\cdrestriction{\exists}{v_1: \arolepath_1, \ldots, v_k: \arolepath_k}{\acons}}{j}$ and
 ( $j > k$ or $\cdrestriction{\exists}{v_1: \arolepath_1, \ldots, v_k: \arolepath_k}{\acons}
  \not \in \actype$), then $\alocation_i = \square$.

\item \textbf{Anonymity.} For $i < \degree$ such that
  \(\alocation_i\) is of the form
  \((\globalinfo_i, (\actype_i, \symblinks_i, \aactvector_i))\),
  $\actype_i$ is an anonymous concept type. 
\end{enumerate}
Such general transitions are reminiscent of the augmented types
in~\cite[Section~3]{Borgwardt&DeBortoli&Koopmann24}. 
\cut{
The first conjunct, handling CD–restrictions, may be of exponential size in \(|\aontology|\)
because for each existential or universal CD–restriction of arity k, the constraint
\(\acons_\aconcept\) involves a disjunction (or conjunction) over all k-tuples of candidate
registers. Since the number of such tuples grows exponentially in k, and there may be
multiple CD–restrictions in the concept type, the overall size is exponential in the worst case.

In contrast, the second conjunct which ensures consistency of registers for individual names across the children,
is a conjunction over all individual names and register indices, and thus has size polynomial
in the size of the ontology.
}
\end{description}

\subsection{Correctness of the Reduction}
\label{section-correctness-main-reduction}
It remains to establish that
\(\alang(\aautomaton) \neq \emptyset\)  iff
the ontology \(\aontology\) is consistent.  
Forthcoming Lemma~\ref{lemma-encoding-soundness} states the left-to-right
direction and its proof amounts to design an interpretation
from a data tree in \(\alang(\aautomaton)\) with its accepting run.
Elements of the interpretation domain are non-root nodes that are not labelled
by the sink location, interpretation of the concept names can be read from
the concept types, and the interpretation of the concrete features are read
from values in the tuples from $\adomain^{\beta}$ in the data tree.
Requirements on the transitions complete the analysis  to
guarantee that an interpretation satisfying the ontology
is indeed produced.
\begin{lem}[Soundness]
\label{lemma-encoding-soundness}
  \(\alang(\aautomaton) \neq \emptyset\) implies \(\aontology\) is consistent.
\end{lem}

\begin{proof}
 Assume that \(\alang(\aautomaton) \neq \emptyset\).
 There exists a data tree \(\adatatree : \interval{0}{\degree-1}^* \to
 \aalphabet \times \adomain^\beta\) and an accepting run
 \(\arun : \interval{0}{\degree-1}^* \to \transitions\) on $\adatatree$ such that 
    for every node \(\anode\),
    \(\adatatree(\anode) = \pair{\aletter_{\anode}}{\atuple_{\anode}}\)
    and \(\arun(\anode) = (\alocation_{\anode}, \aletter_{\anode},
                    \acons_{\anode}, \alocation_{\anode \cdot 0}, \dots,
                    \alocation_{\anode \cdot (\degree-1)})\).
                    Sometimes, we use the fact that the node $\anode$ is labelled by a location
                    (and not necessarily
                    by a transition) and we mean $\alocation_{\anode}$. 
The core idea of the proof is to extract an interpretation $\ainter$ from the
accepting run of the automaton, using its structural components (concept types,
values for the concrete features from the data tree $\adatatree$, symbolic links)
to define the interpretation domain
and the interpretations of concept names and role names.
Then, we verify that this interpretation satisfies all the GCIs and assertions
of the ontology.
    
    We construct the $\ALCO(\acdomain)$ interpretation
    $\ainter = (\aidomain, \inter{\cdot})$ from $\adatatree$ and $\arun$ as follows.
    \begin{itemize}
    \item The interpretation domain is
    \(
    \aidomain \egdef \{ \anode \in \interval{0}{\degree-1}^+
    \mid \alocation_{\anode} \neq \square \}
    \). 
    The root $\varepsilon$ never belongs to $\aidomain$ but it serves
    to connect the interpretations of the individual names to their 
    subtrees. Additionally, the nodes labelled with the sink location \(\square\)
    are excluded from the interpretation domain as these nodes are considered
    as inactive. 

    By default, if $\alocation_{\anode} \not \in \set{\rootlocation, \square}$,
    then $\alocation_{\anode}$ is of the form
    $\pair{\gabs}{\triple{\actype_{\anode}}{\symblinks_{\anode}}{\aactvector_{\anode}}}$. 
  \item For each $i \in \interval{0}{\eta-1}$, \(\inter{\aindname_i} \egdef i\).
    In the run $\arun$, the location assigned to the node $i$ (interpreting
    $\aindname_i$)
     is 
     \((\gabs, (\actype_{\aindname_i}, \symblinks_{\aindname_i}, \aactvector_{\aindname_i}))\) 
     with \((\actype_{\aindname_i}, act_{\aindname_i}) = \gabs(\aindname_i)\) by definition
     of the root transitions.

    \item For each concept name \(\aconceptname\), define
    \(
    \inter{\aconceptname} \egdef \{ \anode \in \aidomain \mid \aconceptname \in \actype_{\anode} \}.
    \)

    \item For each role name \(\arolename \in \rolenames(\aontology)\), define
      \[   
      \inter{\arolename} \egdef\;
      \overbrace{
    \left\{\, (\anode, \anode \cdot i) \in \aidomain \times \aidomain\;\middle|\;
    i \in \dir{\arolename}\right\}
    }^{\mbox{from the tree structure}}
    \;\cup\;
    \overbrace{
      \left\{\, (\anode, \inter{\aindname})  \in \aidomain \times \aidomain
      \;\middle|\; (\arolename, \aindname) \in \symblinks_{\anode} \,\right\}
      }^{\mbox{from symbolic links}}.
    \]

  \item For each concrete feature \(\acfeature_j \in \cfeatures(\aontology)\),
    we define the partial 
    function \(\inter{\acfeature_j} : \aidomain \to \adomain\) such that 
    \(
    \inter{\acfeature_j}(\anode) \egdef \atuple_{\anode}[j]
    \) if $\aactvector_{\anode}[j] = \top$,  
    otherwise $\inter{\acfeature_j}(\anode)$ is undefined. 
    \end{itemize}
    Observe that $\ainter$ satisfies (UNA) w.r.t.
    $\set{\aindname_0, \ldots, \aindname_{\eta-1}}$. 
    
    Showing  $\ainter \models \aontology$ boils down to 
    establish the  \defstyle{type soundness property}:
    for every \(\anode \in \aidomain\),
    and for every concept \(\aconcept \in \subconcepts{\aontology}\), 
    if \(\aconcept \in \actype_{\anode}\), then \(\anode \in \inter{\aconcept}\),
    assuming that the location $\alocation_{\anode}$ labelling the node  $\anode$ is of the form
    \((\gabs, (\actype_{\anode}, \symblinks_{\anode}, \aactvector_{\anode}))\). 
    The proof is by structural induction. 
    
    \begin{description}
    \item[\textbf{Case:} $\aconcept = \aconceptname$ (concept name)]  
    By the definition of $\inter{\aconceptname}$, we have 
    \(
    \aconceptname \in \actype_{\anode}
    \)
    implies \(\anode \in \inter{\aconceptname}\). 

  \item[\textbf{Case:} $\aconcept = \dlneg \aconceptname$, where $\aconceptname$ is a concept
    name]  This is one of the two cases with the negation operator as the complex
    concepts are assumed to be in NNF. 
    Since $\actype_{\anode}$ is a concept type, we have 
    \(
    \dlneg \aconceptname \in \actype_{\anode}
    \)
    implies
    \(\aconceptname \notin \actype_{\anode}\).
    By definition of $\ainter$, $\anode \notin \inter{\aconceptname}$, hence 
    \(\anode \in \inter{(\dlneg \aconceptname)}\).

  \item[\textbf{Case:} $\aconcept = \{ \hspace{-0.02in} \aindname_j \hspace{-0.02in}\}$]
    If $\aconcept \in \actype_{\anode}$, then necessarily
    $\anode = j$ by construction of the transitions, and therefore
    $\anode \in \inter{\aconcept}$ by definition of $\ainter$.

  \item[\textbf{Case:} $\aconcept = \dlneg \{ \hspace{-0.02in} \aindname_j \hspace{-0.02in}\}$]
    If $\aconcept \in \actype_{\anode}$, then necessarily
    $\anode \neq j$ by construction of the transitions, and therefore
    $\anode \not \in \inter{\aconcept}$ by definition of $\ainter$.

    \item[\textbf{Case:} $\aconcept = \top$]  
    Trivially, $\inter{\top} = \aidomain$, so $\anode \in \inter{\top}$.

    \item[\textbf{Case:} $\aconcept = \bot$]  
    This case cannot arise since $\bot \notin \actype_{\anode}$ by definition of concept types.

    \item[\textbf{Case:} $\aconcept = \aconceptbis \dlwedge \aconceptter$]  
      From $\aconcept \in \actype_{\anode}$ and $\actype_{\anode}$ is a
      concept type, we conclude 
    \(\aconceptbis \in \actype_{\anode}\) and \(\aconceptter \in \actype_{\anode}.\)
    By the induction hypothesis,
    both $\anode \in \inter{\aconceptbis}$ and $\anode \in \inter{\aconceptter}$, 
    hence $\anode \in \inter{(\aconceptbis \sqcap \aconceptter)}$
    by the semantics of $\ALCO(\acdomain)$.

    \item[\textbf{Case:} $\aconcept = \aconceptbis \dlvee \aconceptter$]  
    If $\aconcept \in \actype_{\anode}$, then at least one of 
    \(\aconceptbis\) or \(\aconceptter\) belongs to $\actype_{\anode}$.
    By the induction hypothesis,
    $\anode \in \inter{\aconceptbis}$ or $\anode \in \inter{\aconceptter}$,
    hence $\anode \in \inter{(\aconceptbis \sqcup \aconceptter)}$.

  \item[\textbf{Case:} $\aconcept = \exrest{\arolename}{\aconceptbis}$]
    Suppose that $\aconcept \in \actype_{\anode}$.
    By the definition of general transitions  for $\aautomaton$
    and more specifically
    the part about existential restrictions, 
    there is $\pair{\arolename}{\aindname}
  \in \symblinks_{\anode}$ such that $\aconceptbis \in \actype_{\aindname}$
  or \(i = \lambda(\exrest{\arolename}{\aconceptbis})\),  
  \(\alocation_{\anode \cdot i}\) is of the form
  \((\globalinfo_i, (\actype_i, \symblinks_i, \aactvector_i))\)
  and $\aconceptbis \in \actype_i$.
  By the induction hypothesis,
  $\inter{\aindname} \in \inter{\aconceptbis}$
  or
  $\anode \cdot i \in \inter{\aconceptbis}$.
  Moreover, by definition of $\inter{\arolename}$,
  $\pair{\arolename}{\aindname}
  \in \symblinks_{\anode}$ implies
  $\pair{\anode}{\inter{\aindname}} \in \inter{\arolename}$
  and $\pair{\anode}{\anode \cdot i} \in \inter{\arolename}$
  whenever $i \in dir(\arolename)$ and $\alocation_{\anode \cdot i}$
  is not the sink location.
  Consequently, there is $\aind \in \aidomain$
  such that
  $\pair{\anode}{\aind} \in \inter{\arolename}$
  and $\aind \in \inter{\aconceptbis}$, whence 
  $\anode \in \inter{\aconcept}$. 

    \item[\textbf{Case:} $\aconcept = \varest{\arolename}{\aconceptbis}$]  
      Assume $\aconcept \in \actype_{\anode}$
      and $\pair{\anode}{\aind} \in \inter{\arolename}$.
      We distinguish two cases depending how the edge
      $\pair{\anode}{\aind}$ is included in $\inter{\arolename}$
      (parent-child edge versus a symbolic link). 
      \begin{description}

      \item[Subcase $\pair{\anode}{\anode \cdot i} \in \inter{\arolename}$,
        $\alocation_{\anode \cdot i} \neq \square$
        and $i \in dir(\arolename)$]
      By the definition of general transitions for $\aautomaton$ and more specifically
      the part about value restrictions,
      if \(\alocation_{\anode \cdot i}\) is of the form
      \((\globalinfo_i, (\actype_i, \symblinks_i, \aactvector_i))\),
      then  \(\aconceptbis \in \actype_i\).
      By the induction hypothesis, we have
      $\anode \cdot i \in \inter{\aconceptbis}$.

    \item[Subcase $\pair{\arolename}{\aindname} \in \symblinks_{\anode}$]
    By the definition of general transitions  for $\aautomaton$ and more specifically
    the part about value restrictions,
    we have $\aconceptbis \in \actype_{\aindname}$.
    By the induction hypothesis, we can conclude 
    $\inter{\aindname} \in \inter{\aconceptbis}$. 
      
      \end{description}
      In conclusion, for all $\pair{\anode}{\aind} \in \inter{\arolename}$,
      we have $\aind \in \inter{\aconceptbis}$, whence
      $\anode \in \inter{\aconcept}$. 

\item[\textbf{Case:} $\aconcept = \cdrestriction{\exists}{v_1: \arp_1, \dots, v_k: \arp_k}{\acons}$]
Assume that $\aconcept \in \actype_{\anode}$. 
By the definition of general transitions for $\aautomaton$ and more specifically
the part about the constraints,
$\acons_{\anode}$ has a conjunct $\acons_{\aconcept}$
of the form
\[
\acons_{\aconcept}
=
\bigvee_{(\aregisterbis_1,\dots,\aregisterbis_k) \in \asetter_1 \times \cdots \times \asetter_k}
  \acons(\aregisterbis_1,\dots,\aregisterbis_k).
  \]

Let \(\avaluation_{\anode} : \registers{\beta}{\degree} \to
  \adomain\) be the valuation defined by
  \(
  \avaluation_{\anode}(\aregister_j) = \atuple_n[j] \text{ and }
  \avaluation_{\anode}(\aregister_j^i) = \atuple_{\anode \cdot i}[j],
  \)
  for all \(1 \le j \le \beta,\ 0 \le i < \degree\),
  see the definition of runs in Section~\ref{section-tgca}.
  Since $\arun$ is a run, we have
  \(\avaluation_{\anode} \models \acons_{\anode}\)
  and therefore there is
  $(\aregisterbis_1,\dots,\aregisterbis_k) \in
  \asetter_1 \times \cdots \times \asetter_k$ such that
  $\avaluation_{\anode} \models
  \acons(\aregisterbis_1,\dots,\aregisterbis_k)$.
  Now, by construction of the $\asetter_j$'s
  and by the interpretation of the concrete features in
  $\ainter$, 
  one can show that for all $j$,
  \begin{equation}
  \label{equation-rpj-to-zj}
  \inter{\arp_j}(\anode) =
  \set{\avaluation_{\anode}(\aregisterbis)
    \mid \aregisterbis \in \asetter_j}.
  \end{equation}
  In a way,~(\ref{equation-rpj-to-zj}) states that the definition
  of $\asetter_j$ is correct. 
  Consequently,
  \[
    [v_1 \mapsto \avaluation_{\anode}(\aregisterbis_1), \dots,
      v_k \mapsto \avaluation_{\anode}(\aregisterbis_k)] \models \acons
  \]
  and
  $
  (\avaluation_{\anode}(\aregisterbis_1), \ldots,
  \avaluation_{\anode}(\aregisterbis_k)) \in
  \inter{\arp_1}(\anode) \times \cdots \times
  \inter{\arp_k}(\anode)
  $. 
  By the semantics of $\ALCO(\acdomain)$, we get
  $\anode \in \inter{\aconcept}$.
  It remains to prove~(\ref{equation-rpj-to-zj}).

  First, assume that $\arp_j = \acfeature_{\ell}$.
  If $\aactvector_{\anode}[\ell] = \perp$, then
  $\asetter_j = \emptyset$ by definition of $\asetter_j$
  and $\inter{\acfeature_{\ell}}(\anode)$ is undefined by
  definition of $\inter{\acfeature_{\ell}}$.
  In the case $\aactvector_{\anode}[\ell] = \top$, then
  $\asetter_j = \set{\aregister_{\ell}}$ by definition of $\asetter_j$
  and $\inter{\acfeature_{\ell}}(\anode)$ is
  equal to $\atuple_{\anode}[\ell]$ by definition of
  $\inter{\acfeature_{\ell}}$.
  Since $\avaluation_{\anode}(\aregister_{\ell})
  = \atuple_{\anode}[\ell]$ by definition of
  $\avaluation_{\anode}$, for both cases
  we get 
  $\inter{\arp_j}(\anode) =
  \set{\avaluation_{\anode}(\aregisterbis)
    \mid \aregisterbis \in \asetter_j}$. 

  Second, assume that $\arp_j = \arolename \cdot
  \acfeature_{\ell}$.

  \begin{enumerate}

  \item If $\pair{\arolename}{\aindname} \in \symblinks_{\anode}$
    and $\aactvector_{\aindname}[\ell] = \top$, then
    $\aregister_{\aindname, \ell} \in \asetter_j$.
    By definition of $\inter{\arolename}$ and
    $\inter{\acfeature_{\ell}}$,
    we have $\pair{\anode}{\inter{\aindname}} \in
    \inter{\arolename}$ and $\inter{\acfeature_{\ell}}(\inter{\aindname})$
    is defined. Consequently,
    $\acfeature_{\ell}(\inter{\aindname}) \in \inter{\arp_j}(\anode)$.
    If $\aindname = \aindname_j$ for some $j \in \interval{0}{\eta-1}$,
    then $\acfeature_{\ell}(\inter{\aindname})$
    is equal to $\avaluation_{\anode}(\aregister_{\aindname_j, \ell})$
    because the value of the register remains constant
    all over the run $\arun$.
    Hence, $\avaluation_{\anode}(\aregister_{\aindname_j, \ell})\in
    \inter{\arp_j}(\anode)$.

  \item If $i \in dir(\arolename)$, $\alocation_{\anode \cdot i}
    \neq \square$ and $\aactvector_{\anode \cdot i}[\ell] = \top$,
    then $\aregister_{\ell}^i \in \asetter_j$.
    By definition of $\inter{\arolename}$ and
    $\inter{\acfeature_{\ell}}$,
    we have $\pair{\anode}{\anode \cdot i} \in
    \inter{\arolename}$ and $\inter{\acfeature_{\ell}}(\anode \cdot i)$
    is defined.
    Consequently,  $\inter{\acfeature_{\ell}}(\anode \cdot i)
    \in \inter{\arp_j}(\anode)$. 
    Since $\avaluation_{\anode}(\aregister_{\ell}^i)
    = \avaluation_{\anode \cdot i}(\aregister_{\ell})$ by definition
    of the valuations
    and $\inter{\acfeature_{\ell}}(\anode \cdot i)
    = \avaluation_{\anode \cdot i}(\aregister_{\ell})$
    by definition of $\inter{\acfeature_{\ell}}$, we get 
    $\avaluation_{\anode}(\aregister_{\ell}^i)\in
    \inter{\arp_j}(\anode)$.

  \item Now, we prove the other inclusion.  Let $\avalue \in \inter{\arp_j}(\anode)$. There is
    $\aind$ such that $\pair{\anode}{\aind} \in \inter{\arolename}$
    and $\inter{\acfeature_{\ell}}(\aind) = \avalue$.
    We distinguish two cases depending how the edge
    $\pair{\anode}{\aind}$  is included in $\inter{\arolename}$
    (parent-child edge versus a symbolic link).
    \begin{description}
    \item[Subcase $\pair{\arolename}{\aindname} \in \symblinks_{\anode}$
      and $\inter{\aindname} = \aind$]
      Since $\inter{\acfeature_{\ell}}(\aind)$ is defined,
      by definition
      of $\inter{\acfeature_{\ell}}$, $\aactvector_{\aindname}[\ell] = \top$.
      By definition of $\asetter_j$,
      $\aregister_{\aindname, \ell} \in \asetter_j$.
      Moreover, $\avaluation_{\anode}(\aregister_{\aindname, \ell})
      = \avalue$, because such values are propagated all over the
      run. Consequently,
      $\avaluation_{\anode}(\aregister_{\aindname, \ell}) \in 
      \set{\avaluation_{\anode}(\aregisterbis)
        \mid \aregisterbis \in \asetter_j}$.
    \item[Subcase $\pair{\anode}{\anode \cdot i} \in \inter{\arolename}$,
        $\alocation_{\anode \cdot i} \neq \square$
      and $i \in dir(\arolename)$]
      Since $\inter{\acfeature_{\ell}}(\anode \cdot i)$ is defined,
      by definition
      of $\inter{\acfeature_{\ell}}$, $\aactvector_{\anode \cdot i}[\ell]
      = \top$.
      By definition of $\asetter_j$,
      $\aregister_{\ell}^{i} \in \asetter_j$.
      Moreover, $\avaluation_{\anode}(\aregister_{\ell}^i)
      = \avaluation_{\anode \cdot i}(\aregister_{\ell})
      = \inter{\acfeature_{\ell}}(\anode \cdot i)$,
      by definition of the valuations and
      $\inter{\acfeature_{\ell}}$.
      Consequently, 
      $\avaluation_{\anode}(\aregister_{\ell}^i) \in 
      \set{\avaluation_{\anode}(\aregisterbis)
        \mid \aregisterbis \in \asetter_j}$.   
    \end{description}
  \end{enumerate}

\item[\textbf{Case:}
  $\aconcept = \cdrestriction{\forall}{v_1: \arp_1, \dots, v_k: \arp_k}{\acons}$]  

Assume that $\aconcept \in \actype_{\anode}$. 
By the definition of general transitions for
$\aautomaton$ and more specifically
the part about the constraints,
$\acons_{\anode}$ has a conjunct $\acons_{\aconcept}$
of the form
\[
\acons_{\aconcept}
=
\bigwedge_{(\aregisterbis_1,\dots,\aregisterbis_k) \in \asetter_1 \times \cdots \times \asetter_k}
  \acons(\aregisterbis_1,\dots,\aregisterbis_k).
  \]
Let \(\avaluation_{\anode} : \registers{\beta}{\degree} \to
\adomain\) be the valuation defined as in the previous
case. 
  Since $\arun$ is a run, we have
  \(\avaluation_{\anode} \models \acons_{\anode}\)
  and therefore for all 
  $(\aregisterbis_1,\dots,\aregisterbis_k) \in
  \asetter_1 \times \cdots \times \asetter_k$,
  we have 
$\avaluation_{\anode} \models
  \acons(\aregisterbis_1,\dots,\aregisterbis_k)$.
  Now, by construction of the $\asetter_j$'s
  and by the interpretation of the concrete features in
  $\ainter$,  for all $j$, we have seen that (\ref{equation-rpj-to-zj})
  holds true.
  Consequently,
  for all $(\avalue_1, \ldots, \avalue_k)
  \in \inter{\arp_1}(\anode) \times \cdots \times
  \inter{\arp_k}(\anode)$,
  we have 
  $[v_1 \mapsto \avalue_1, \dots,
    v_k \mapsto \avalue_k] \models \acons$.
  By the semantics of $\ALCO(\acdomain)$, we get
  $\anode \in \inter{\aconcept}$.

    \end{description}

    We have shown that for all \(\anode \in \aidomain\) and for all 
    \(\aconcept \in \subconcepts{\aontology}\), if \(\aconcept \in \actype_{\anode}\), 
    then \(\anode \in \inter{\aconcept}\).
    It remains to verify that the interpretation \(\ainter\) satisfies the 
    ontology \(\aontology= (\atbox, \aabox)\).
    \begin{description}
        \item[\textbf{TBox satisfaction}]  
        Recall that every GCI in $\atbox$ is of the form
        \(\top \sqsubseteq \aconcept\), where \(\aconcept\) is in negation normal form (NNF).  
        For every node \(\anode \in \aidomain\), the 
        concept type \(\actype_{\anode}\) belongs 
        to \(\ctypes{\aontology}\), which ensures that
        for all 
        \(\top \sqsubseteq \aconcept \in \atbox\),
        we have $\aconcept \in \actype_{\anode}$.
        By the type soundness property, this implies 
        \(\anode \in \inter{\aconcept}\), i.e., 
        \(\inter{\top} = \aidomain \subseteq \inter{\aconcept}\).  
        Thus, \(\ainter \models \top \sqsubseteq \aconcept\) for all
        GCIs $\top \sqsubseteq \aconcept$ in $\atbox$, whence
        \(\ainter \models \atbox\).

        \item[\textbf{ABox satisfaction}]  
          Let \(\cassertion{\aindname}{\aconcept} \in \aabox\) be a
          concept assertion. Then, by construction of 
        the global information function \(\globalinfo\), we have:
        \(\aconcept \in \actype_{\aindname}\) (see the requirements for root transitions).
        By the type soundness property, it follows that
        \(\inter{\aindname} \in \inter{\aconcept}\), 
        so \(\ainter \models \cassertion{\aindname}{\aconcept}\).

        Similarly, for a role assertion
        \(\rassertion{\aindname}{\aindnamebis}{\arolename} \in \aabox\),
        the requirements on the root transitions
        guarantee that 
        \((\arolename, \aindnamebis) \in \symblinks_{\aindname}\).
        According to the definition 
        of the interpretation of role names, this ensures that:
        \((\inter{\aindname}, \inter{\aindnamebis}) \in \inter{\arolename},\)
        so \(\ainter \models \rassertion{\aindname}{\aindnamebis}{\arolename}\).

    \end{description}
    Thus, the interpretation \(\ainter\) satisfies
    all the GCIs in $\atbox$ and all the assertions in $\aabox$, 
    so it is a model of the ontology \(\aontology\). 
\end{proof}

\begin{lem}[Completeness]
\label{lemma-encoding-completeness}
\(\aontology\) is consistent
with an interpretation satisfying (UNA) implies \(\alang(\aautomaton) \neq \emptyset\).
\end{lem}

  \begin{proof}
    Suppose the ontology \(\aontology = (\atbox, \aabox)\) is consistent, 
    and let \(\ainter = (\aidomain, \inter{\cdot})\) be an interpretation
    such that $\ainter \models \aontology$ and $\ainter$ satisfies (UNA).
    It is worth noting that assuming (UNA) is not required
    for the developments below, but we make such an assumption
    to be able at a later stage
    to obtain that consistency under (UNA) reduces to 
    nonemptiness for constraint automata. 
    Below, we define a data tree $\adatatree: \interval{0}{\degree-1}^* \to
    \aalphabet \times \adomain^{\beta}$,
    a partial function $\amapbis: \interval{0}{\degree-1}^* \rightharpoonup \aidomain$,
    a map
    $\arun^{\dag}: \interval{0}{\degree-1}^* \to
    \locations$,
    and a map $\arun: \interval{0}{\degree-1}^* \to
    \transitions$ satisfying the following properties.
    \begin{description}
    \item[(P1)] The map $\arun$ is an accepting run on $\adatatree$ and consequently,
      $\adatatree \in \alang(\aautomaton)$ and $\alang(\aautomaton) \neq \emptyset$.
    \item[(P2)] For all nodes $\anode \in \interval{0}{\degree-1}^*$,
      $\amapbis(\anode)$ is defined iff
      $\anode$ is not the root node $\varepsilon$
      and $\arun^{\dag}(\anode)$ is different from $\square$.
    \item[(P3)] For all $\anode \in \interval{0}{\degree-1}^*$,
      $\arun(\anode)$  is of the form
      $(\arun^{\dag}(\anode), \aletter, \acons,
      \arun^{\dag}(\anode \cdot 0), \ldots, \arun^{\dag}(\anode \cdot (\degree-1)))$.
    \item[(P4)] For all $\anode \in \interval{0}{\degree-1}^*$ with $\amapbis(\anode)$ defined
      and $j \in \interval{1}{\alpha}$,
      if $\adatatree(\anode) = \pair{\aletter}{\atuple}$
      and $\inter{\acfeature_j}(\amapbis(\anode))$ is defined,
      then $\atuple[j] = \inter{\acfeature_j}(\amapbis(\anode))$.
    \item[(P5)] For all $\anode \in \interval{0}{\degree-1}^+$
      such that $\arun^{\dag}(\anode)$ is of the
      form $(\globalinfo, (\actype, \symblinks, \aactvector))$,
      we have $\localabstraction{\amapbis(\anode)}{\ainter} = (\actype, \symblinks, \aactvector)$. 
    \end{description}
    By construction of the constraint automaton $\aautomaton$,
    for all the locations $\alocation, \alocation_0, \ldots, \alocation_{\degree-1}$,
    there is at most one constraint $\acons$ such that
    $(\alocation, \aletter, \acons, \alocation_0, \ldots, \alocation_{\degree-1})$
    is a transition in $\transitions$. That is why, the map $\arun$
    is uniquely determined by the map $\arun^{\dag}$, actually we only need
    to verify that indeed $\arun$ is a total function in order to satisfy
    the property (P3). The satisfaction of the property (P4) can be easily realised once
    the map $\amapbis$ is defined.
    Similarly, the satisfaction of the property (P5) can be also easily realised
    once the map $\amapbis$ is defined, but one needs to check that
    indeed  $\localabstraction{\amapbis(\anode)}{\ainter}$ is a local abstraction,
    which is not difficult to verify (omitted). 

    We define $\adatatree$, $\amapbis$ and $\arun^{\dag}$ inductively
    on the depth of the nodes. Before handling the base case with the root node
    $\varepsilon$, let us define the global abstraction
    $\globalinfo$ that is present all over the run.
    For all $j \in \interval{0}{\eta-1}$,
    $\globalinfo(\aindname_j) \egdef \pair{\actype_{\aindname_j}}{\aactvector_{\aindname_j}}$
    with $\localabstraction{\inter{\aindname_j}}{\ainter}
    = \triple{\actype_{\aindname_j}}{\symblinks_{\aindname_j}}{\aactvector_{\aindname_j}}$. 

    For the base case, $\amapbis(\varepsilon)$ is undefined and
    $\arun^{\dag}(\varepsilon) \egdef \rootlocation$. For all $j \in
    \interval{0}{\eta-1}$,
    $\amapbis(j) \egdef \inter{\aindname_j}$ and
    $\arun^{\dag}(j) \egdef \pair{\globalinfo}{\localabstraction{\inter{\aindname_j}}{\ainter}}$.
      For all $j \in \interval{\eta}{\degree-1}$,
      $\amapbis(j)$ is undefined and
      $\arun^{\dag}(j) = \square$.

      Let us handle now the induction step for the nodes of depth at least two.
      \proofsubparagraph{Induction hypothesis}
      For all nodes \(\anode\) at depth at most \(D \geq 1\),
      $\arun^{\dag}(\anode)$ is defined, and 
      if $\amapbis(\anode)$ is defined and
      $\arun^{\dag}(\anode) = (\globalinfo, (\actype_{\anode}, \symblinks_{\anode}, \aactvector_{\anode}))$,
      then
      $(\actype_{\anode}, \symblinks_{\anode}, \aactvector_{\anode})
      = \localabstraction{\amapbis(\anode)}{\ainter}
      $.

      \proofsubparagraph{Inductive construction}
      Let \(i \in \interval{0}{\degree-1}\) such that
      the depth of $\anode \cdot i$ is at most $D +1$,
      and $\anode \cdot i$ is the smallest element in
      $\interval{0}{\degree-1}^{+}$ with respect to
      the lexicographical ordering such that $\arun^{\dag}(\anode \cdot i)$ is not yet defined.
      If $\arun^{\dag}(\anode) = \square$, then
      $\amapbis(\anode \cdot i)$ is undefined 
      and $\arun^{\dag}(\anode \cdot i) \egdef \square$. Otherwise, we proceed
      by a case analysis.

      \begin{itemize}

      \item If \(\lambda^{-1}(i) = \exrest{\arolename}{\aconcept}\)
        and $\exrest{\arolename}{\aconcept} \not \in \actype_{\anode}$,
        then $\amapbis(\anode \cdot i)$ is undefined 
        and $\arun^{\dag}(\anode \cdot i) \egdef \square$.

      \item If \(\lambda^{-1}(i) = \exrest{\arolename}{\aconcept}\)
        and $\exrest{\arolename}{\aconcept} \in \actype_{\anode}$,
        then $\amapbis(\anode) \in \inter{(\exrest{\arolename}{\aconcept})}$ by the induction hypothesis.
        There is \(\aind \in \aidomain\) such that
        \(\pair{\amapbis(\anode)}{\aind} \in \inter{\arolename}\) and 
        \(\aind \in \inter{\aconcept}\).
        If there is such an individual $\aind$ such that $\aind \in \set{\inter{\aindname_0}, \ldots,
          \inter{\aindname_{\eta -1}}}$, then $\amapbis(\anode \cdot i)$ is undefined 
        and $\arun^{\dag}(\anode \cdot i) \egdef \square$ (the witness is already accounted for 
    via a symbolic link).
        Otherwise, we pick an arbitrary witness $\aind$,
        and $\amapbis(\anode \cdot i) \egdef \aind$ and
        $\arun^{\dag}(\anode \cdot i) \egdef \pair{\globalinfo}{\localabstraction{\aind}{\ainter}}$. 
        
      \item  If \(\lambda^{-1}(i) =
        \pair{\aconcept^{\star}
        }{j}\)
        and ($\aconcept^{\star} \not \in \actype_{\anode}$ or $j > k$) with $\aconcept^{\star}
        = \cdrestriction{\exists}{v_1: \arolepath_1, \ldots, v_k: \arolepath_k}{\acons}$,
       then $\amapbis(\anode \cdot i)$ is undefined 
        and $\arun^{\dag}(\anode \cdot i) \egdef \square$.

      \item If \(\lambda^{-1}(i) = \pair{\aconcept^{\star}
        }{j}\),
        $\aconcept^{\star} \in \actype_{\anode}$ and $j \leq k$
        with $\aconcept^{\star}
        = \cdrestriction{\exists}{v_1: \arolepath_1, \ldots, v_k: \arolepath_k}{\acons}$,
        then $\amapbis(\anode) \in \inter{(\aconcept^{\star})}$ by the induction hypothesis.
        By the semantics of $\ALCO(\acdomain)$,
        there is a tuple \((\avalue_1, \dots, \avalue_k) \in \inter{\arp_1} \times \cdots \times
        \inter{\arp_k}\) such that
        \[
          [v_1 \mapsto \avalue_1, \ldots, v_k \mapsto \avalue_k]
          \models \acons. 
          \]
          We pick an arbitrary such a tuple. 
          For all $j \in \interval{1}{k}$ such that
          $\arp_j$ is of length two of the form
          $\arolename_j' \cdot \acfeature'_j$,
          there is $\aind_j \in \aidomain$ such that
          \(\pair{\amapbis(\anode)}{\aind_j} \in \inter{(\arolename'_j)}\) and
          $\inter{\acfeature_j'}(\aind_j) = \avalue_j$.
          For all $i'$ such that
          \(\lambda^{-1}(i') = \pair{\cdrestriction{\exists}{v_1:
              \arolepath_1, \ldots, v_k: \arolepath_k}{\acons}}{j}\) for such $j$'s
          and $\aind_j \not \in \set{\inter{\aindname_0}, \ldots,
          \inter{\aindname_{\eta -1}}}$,
          $\amapbis(\anode \cdot i') \egdef \aind_j$ and 
          $\arun^{\dag}(\anode \cdot i') \egdef \pair{\globalinfo}{\localabstraction{\aind_j}{\ainter}}$.
          For all the other $i'$, $\amapbis(\anode \cdot i')$ is undefined 
          and $\arun^{\dag}(\anode \cdot i') \egdef \square$.
          The case $i' = i$ has been necessarily handled.
          We need to go through all the indices $i'$ (and not to consider $i$ alone)
          because the same tuple $(\avalue_1, \dots, \avalue_k)$ needs to be considered.

    \end{itemize}

      By the construction of the constraint automaton $\aautomaton$, and in particular
      the part about constraints (see Section~\ref{section-construction}), for
      all $\anode \in \interval{0}{\degree-1}^*$,
      there is exactly one constraint $\acons_{\anode}$
      such that 
      $(\arun^{\dag}(\anode), \aletter, \acons_{\anode},
      \arun^{\dag}(\anode \cdot 0), \ldots, \arun^{\dag}(\anode \cdot (\degree-1)))$
      is a transition of $\aautomaton$ and this is precisely
      how $\arun(\anode)$ is defined.
      This holds mainly because we took care to propagate the
      location $\square$ as soon as it appears on a node. 
     
      \proofsubparagraph{Data tree $\adatatree$}
     It remains to define the 
    data tree \(\adatatree :
    \interval{0}{\degree-1}^* \to \aalphabet \times \adomain^\beta\)
    such that for all node $\anode$,
    \(\adatatree(\anode) \egdef (\aletter, \atuple_{\anode})\) with $\atuple_{\anode}$
    defined as follows ($\aletter$ is the unique letter of the alphabet). 
    \begin{itemize}
    \item $\adatatree(\varepsilon) \egdef \pair{\aletter}{\atuple}$
      for some arbitrary tuple $\atuple$. 
    \item For all $j \in \interval{1}{\alpha}$,
      if $\inter{\acfeature_j}(\amapbis(\anode))$ is undefined,
      then $\atuple_{\anode}[j]$ is arbitrary. Otherwise, 
      $\atuple_{\anode}[j] \egdef \inter{\acfeature_j}(\amapbis(\anode))$.
    \item For all $j \in \interval{1}{\alpha}$ and $i \in \interval{0}{\eta-1}$,
      if $\inter{\acfeature_j}(\inter{\aindname_i})$ is undefined,
      then $\atuple_{\anode}[\alpha \cdot (i+1) + j]$ is arbitrary.
      Otherwise, $\atuple_{\anode}[\alpha \cdot (i+1) + j] \egdef
      \inter{\acfeature_j}(\inter{\aindname_i})$.
      We recall here that  the registers $\aregister_1, \ldots, \aregister_{\beta}$
      in $\aautomaton$ are also represented by the sequence
$
\aregister_1, \ldots, \aregister_{\alpha},
\aregister_{\aindname_0, 1}, \ldots,\aregister_{\aindname_0,\alpha},
\ldots,
\aregister_{\aindname_{\eta-1}, 1}, \ldots,\aregister_{\aindname_{\eta-1},\alpha}
$, which explains the index $\alpha \cdot (i+1) + j$ above.
    \end{itemize}

    \proofsubparagraph{The map $\arun$: an accepting run on $\adatatree$} It remains to show that $\arun$ is an
    accepting run on the data tree $\adatatree$. We have to show that
    indeed (a) $\arun$ is a map from $\interval{0}{\degree-1}^*$ to $\transitions$,
    (b) $\arun^{\dag}(\varepsilon) \in \locations_{in}$,
    (c) $\arun$ is accepting and the condition (ii) for runs holds.
    The satisfaction of (b) is immediate from the base case above
    and (c) is straightforward because $F = \locations$.

    In order to check the property (a), we go through all the values $\arun(\anode)$ and check
    that it satisfies the conditions for defining transitions in Section~\ref{section-construction}.
    This checking does not require the data values from $\adatatree$, unlike the verification of the
    condition (ii). Furthermore, if the conditions
    on $\arun^{\dag}(\anode), \arun^{\dag}(\anode \cdot 0), \ldots,\arun^{\dag}(\anode \cdot (\degree-1))$
    are satisfied, then the definition of $\acons_{\anode}$ is uniquely determined.
    Note also that all the conditions about the global abstraction
    $\globalinfo$ are naturally satisfied because $\globalinfo$ is
    the only global abstraction involved in the definition of $\arun^{\dag}$. 

    \begin{itemize}
    \item If $\arun^{\dag}(\anode) = \square$, then
      we have seen that
      $\arun^{\dag}(\anode \cdot 0) = \cdots =\arun^{\dag}(\anode \cdot (\degree-1)) = \square$
      by definition of $\arun^{\dag}$,
      which allows us to use the padding transition $\atransition_{\square}$.

    \item We have $\arun(\varepsilon)
      = (\rootlocation, \aletter, \acons_{\mathsf{root}},
      \pair{\gabs}{\localabstraction{\inter{\aindname_0}}{\ainter}}, \ldots,
      \pair{\gabs}{\localabstraction{\inter{\aindname_{\eta-1}}}{\ainter}},
      \square, \ldots, \square)$, which guarantees that all the requirements
      for the root transitions are satisfied. The details are omitted.
      By way of example, assuming that
      $\localabstraction{\inter{\aindname_j}}{\ainter}
      = \triple{\actype_j}{\symblinks_j}{\aactvector_j}$,
      if $\cassertion{\aindname_j}{\aconcept} \in \aabox$,
      then
      $\inter{\aindname_j} \in \inter{\aconcept}$ ($\ainter$
      satisfies $\aontology$)
      and therefore $\aconcept \in \actype_j$.
      The other requirements can be shown similarly.

    \item If $\arun^{\dag}(\anode)$ is
      of the form
      $\pair{\globalinfo}{\triple{\actype_{\anode}}{\symblinks_{\anode}}{\aactvector_{\anode}}}$,
      then let us check that
      the conditions
    on $\arun^{\dag}(\anode), \arun^{\dag}(\anode \cdot 0), \ldots,\arun^{\dag}(\anode \cdot (\degree-1))$
    are satisfied.
    \begin{description}
    \item[Requirements from existential restrictions]
      Let $\exrest{\arolename}{\aconcept} \in \actype_{\anode}$ with
      $i = \lambda(\exrest{\arolename}{\aconcept})$.
      We have seen that either $\arun^{\dag}(\anode \cdot i) =
      \pair{\globalinfo}{\triple{\actype_{\anode \cdot i}}{\symblinks_{\anode \cdot i}}{
          \aactvector_{\anode \cdot i}}}$ and $\aconcept \in
      \actype_{\anode \cdot i}$
      or there is some $j$ such that $\globalinfo(\aindname_j) =
      \pair{\actype_j}{\aactvector_j}$ and $\aconcept \in \actype_j$.
      This is sufficient to satisfy these requirements.

    \item[Requirements from value restrictions] Similar to the previous checking. 
    \item[Requirements from global abstraction preservation]
      Immediate since $\globalinfo$ is
    the only global abstraction involved in the definition of $\arun^{\dag}$.
  \item[Requirements from inactive children] For each $i \in \interval{0}{\degree-1}$,
    the requirements are satisfied and this can be read from
    the cases
    (\(\lambda^{-1}(i) = \exrest{\arolename}{\aconcept}\)
    and $\exrest{\arolename}{\aconcept} \not \in \actype_{\anode}$)
    or (\(\lambda^{-1}(i) =
        \pair{\cdrestriction{\exists}{v_1: \arolepath_1, \ldots, v_k: \arolepath_k}{\acons}}{j}\)
        and ($\lambda^{-1}(i) \not \in \actype_{\anode}$ or $j > k$)).
      \item[Requirements about anonymity]
        Let $i < \degree$ such that
        $\alocation_i = \pair{\gabs}{\triple{\actype_i}{\symblinks_i}{\aactvector_i}}$.
        By construction of $\amapbis$, the map $\amapbis$ for the children
        never returns a value in $\set{\inter{\aindname_0}, \ldots,\inter{\aindname_{\eta-1}}}$.  
        Therefore, $\neg \anominal \in \actype_i$
        for all $\aindname \in \individualnames(\aontology)$. 
    \end{description}
    
    \end{itemize}

   It remains to check the condition (ii), which is recalled below. 
   \begin{romanenumerate}
   \item[(ii)] For all nodes $\anode \in \interval{0}{\degree-1}^*$
     with $\adatatree(\anode) = \pair{\aletter}{\atuple_{\anode}}$,
     we define the valuation 
    \(\avaluation_{\anode} : \registers{\beta}{\degree} \to
  \adomain\) such that
  \(
  \avaluation_{\anode}(\aregister_j) = \atuple_n[j] \text{ and }
  \avaluation_{\anode}(\aregister_j^i) = \atuple_{\anode \cdot i}[j],
  \)
  for all \(1 \le j \le \beta,\ 0 \le i < \degree\). Then
  \(\avaluation_{\anode} \models \acons_{\anode}\), where $\acons_{\anode}$
  is the constraint of the transition $\arun(\anode)$.
   \end{romanenumerate}
   If $\arun^{\dag}(\anode) = \square$, then $\acons_{\anode} = \top$ and
   the property obviously holds.
   If $\arun^{\dag}(\anode) = \rootlocation$ (i.e. $\anode = \varepsilon$),
   then we have to establish that
   \[
   \avaluation_{\varepsilon} \models \bigwedge_{i,i' \in \interval{0}{\eta-1}, j \in \interval{1}{\alpha},
     \aactvector_{\aindname_i}[j] = \top} 
   \aregister_{j}^i = \aregister_{\aindname_i, j}^{i'} .
   \]
   Let $i,i' \in \interval{0}{\eta-1}$ and $j \in \interval{1}{\alpha}$
   such that $\aactvector_{\aindname_i}[j] = \top$.
   By definition of $\avaluation_{\varepsilon}$,
   $\avaluation_{\varepsilon}(\aregister_{j}^i)
   = \atuple_{i}[j]$. Since $\aactvector_{\aindname_i}[j] = \top$,
   by definition of $\atuple_{i}$,
   $\atuple_{i}[j] = \inter{\acfeature_j}(\inter{\aindname_i})$.
   By definition of $\avaluation_{\varepsilon}$,
   $\avaluation_{\varepsilon}(\aregister_{\aindname_i, j}^{i'})
   = \atuple_{i'}[\alpha \times (i+1) +j]$.
   Since $\aactvector_{\aindname_i}[j] = \top$, by definition of $\atuple_{i'}$
   $\atuple_{i'}[\alpha \times (i+1) +j] = \inter{\acfeature_j}(\inter{\aindname_i})$,
   which allows to conclude that $\avaluation_{\varepsilon}$
   satisfies the above-mentioned generalised conjunction.

   If $\arun^{\dag}(\anode)$ is of the
   form $\pair{\globalinfo}{\triple{\actype_{\anode}}{\symblinks_{\anode}}{\aactvector_{\anode}}}$,
   with 
   $\arun(\anode) = 
   (\arun^{\dag}(\anode), \aletter, \acons_{\anode},
   \arun^{\dag}(\anode \cdot 0), \ldots, \arun^{\dag}(\anode \cdot (\degree-1)))$, then
   $\acons_{\anode}$ is equal to
   \[
   \acons_\atransition =
\left(\bigwedge_{\aconcept \in \mathrm{CD}_\exists(\actype_{\anode}) \cup \mathrm{CD}_\forall(\actype_{\anode})}
\acons_\aconcept\right)
\wedge
\left(\bigwedge_{\substack{i \in \interval{0}{\degree-1}  \\ \arun^{\dag}(\anode \cdot i) \neq \square}} 
\bigwedge_{j = 1}^\alpha \ \ 
\bigwedge_{j' \in \interval{0}{\eta-1}, \ \aactvector_{\aindname_{j'}}[j] = \top} 
\aregister_{\aindname_{j'},j} = \aregister_{\aindname_{j'},j}^i
\right).
\]
Let us show that $\avaluation_{\anode} \models \acons_\atransition$.
\begin{itemize}
\item Let $i \in \interval{0}{\degree-1}$ such that $\arun^{\dag}(\anode \cdot i) \neq \square$,
  $j \in \interval{1}{\alpha}$, and $j' \in \interval{0}{\eta-1}$ such that $\aactvector_{\aindname_{j'}}[j] = \top$.
  By definition of $\avaluation_{\anode}$,
  $\avaluation_{\anode}(\aregister_{\aindname_{j'},j}) =
  \atuple_{\anode}[\alpha (j'+1) +j]$.
  By definition of $\atuple_{\anode}$,
  since  $\aactvector_{\aindname_{j'}}[j] = \top$,
  $\atuple_{\anode}[\alpha (j'+1) +j]$ is equal to
  $\inter{\acfeature_j}(\inter{\aindname_{j'}})$.
  Similarly,
  $\avaluation_{\anode}(\aregister_{\aindname_{j'},j}^i) =
  \atuple_{\anode \cdot i}[\alpha (j'+1) +j]$
  by definition of $\avaluation_{\anode}$.
  By definition of $\atuple_{\anode \cdot i}$,
  since  $\aactvector_{\aindname_{j'}}[j] = \top$,
  $\atuple_{\anode \cdot i}[\alpha (j'+1) +j]$ is also equal to
  $\inter{\acfeature_j}(\inter{\aindname_{j'}})$.

\item Let $\aconcept \in \mathrm{CD}_\exists(\actype_{\anode})$ with
  $\aconcept = \cdrestriction{\exists}{v_1: \arp_1, \dots, v_k: \arp_k}{\acons}$.
  Let us show that
  \[
  \avaluation_{\anode} \models
  \bigvee_{(\aregisterbis_1,\dots,\aregisterbis_k) \in \asetter_1 \times \cdots \times \asetter_k}
  \acons(\aregisterbis_1,\dots,\aregisterbis_k) .
  \]
  We have seen earlier that we have picked
  a tuple \((\avalue_1, \dots, \avalue_k) \in \inter{\arp_1} \times \cdots \times
        \inter{\arp_k}\) such that
        \[
          [v_1 \mapsto \avalue_1, \ldots, v_k \mapsto \avalue_k]
          \models \acons,
          \]
  from which $\arun^{\dag}(\anode \cdot i)$ is defined for all $i \in \interval{0}{\degree-1}$. 
 For all $j \in \interval{1}{k}$ such that
          $\arp_j$ is of length two of the form
          $\arolename_j' \cdot \acfeature'_j$,
          there is $\aind_j \in \aidomain$ such that
          \(\pair{\amapbis(\anode)}{\aind_j} \in \inter{(\arolename'_j)}\) and
          $\inter{\acfeature_j'}(\aind_j) = \avalue_j$.
          For all $i'$ such that
          \(\lambda^{-1}(i') = \pair{\cdrestriction{\exists}{v_1:
              \arolepath_1, \ldots, v_k: \arolepath_k}{\acons}}{j}\) for such $j$'s
          and $\aind_j \not \in \set{\inter{\aindname_0}, \ldots,
          \inter{\aindname_{\eta -1}}}$,
          $\amapbis(\anode \cdot i') \egdef \aind_j$ and 
          $\arun^{\dag}(\anode \cdot i') \egdef \pair{\globalinfo}{\localabstraction{\aind_j}{\ainter}}$.
          For all the other $i'$, $\amapbis(\anode \cdot i')$ is undefined 
          and $\arun^{\dag}(\anode \cdot i') \egdef \square$.
          Let us define a tuple $(\aregisterbis_1, \ldots, \aregisterbis_k)
          \in  \asetter_1 \times \cdots \times \asetter_k$
          such that $(\avaluation_{\anode}(\aregisterbis_1), \ldots,
          \avaluation_{\anode}(\aregisterbis_k)) = (\avalue_1, \dots, \avalue_k)$,
          which shall allow us to conclude that
          \[\avaluation_{\anode} \models
  \bigvee_{(\aregisterbis_1,\dots,\aregisterbis_k) \in \asetter_1 \times \cdots \times \asetter_k}
  \acons(\aregisterbis_1,\dots,\aregisterbis_k) .
  \] 

  Let $\gamma \in \interval{1}{k}$.
  If $\arp_{\gamma} = \acfeature_{\ell}$ for some $\ell$, then
  let us take $\aregisterbis_{\gamma} = \aregister_{\ell}$
  and $\avaluation_{\anode}(\aregister_{\ell}) = \atuple_{\anode}[\ell]$
  by definition of $\avaluation_{\anode}$.
  Moreover, $\avalue_{\gamma} = \inter{\acfeature_{\ell}}(\amapbis(\anode))$
  by the semantics of $\ALCO(\acdomain)$.
    By definition of $\atuple_{\anode}[\ell]$, we have
    $\atuple_{\anode}[\ell] = \inter{\acfeature_{\ell}}(\amapbis(\anode))$.
  Consequently, $\avaluation_{\anode}(\aregisterbis_{\gamma}) = \avalue_{\gamma}$. 
  Finally, since $\inter{\acfeature_{\ell}}(\amapbis(\anode))$ is defined,
  $\aactvector_{\anode}[\ell] = \top$ and
  therefore $\aregisterbis_{\gamma} \in \asetter_{\gamma}$.

  If $\arp_{\gamma} = \arolename_{\ell'} \cdot \acfeature_{\ell}$
  for some $\ell, \ell'$, then we perform a case analysis to define
  $\aregisterbis_{\gamma}$.
  If $\aind_{\gamma} \in \set{\inter{\aindname_0}, \ldots,\inter{\aindname_{\eta-1}}}$,
  then let us take  $\aregisterbis_{\gamma} = \aregister_{\aindname_{H},\ell}$
  for some $H$ such that $\aind_{\gamma} = \inter{\aindname_{H}}$. 
  Moreover, $\avalue_{\gamma} = \inter{\acfeature_{\ell}}(\inter{\aindname_H})$
  by the semantics of $\ALCO(\acdomain)$.
  By definition of $\avaluation_{\anode}$,
  $\avaluation_{\anode}(\aregister_{\aindname_{H},\ell}) = \atuple_{\anode}[\alpha \times (H+1) + \ell]$.
  By definition of $\atuple_{\anode}$,
  $\atuple_{\anode}[\alpha \times (H+1) + \ell] =
  \inter{\acfeature_{\ell}}(\inter{\aindname_H})$.
  Consequently, $\avaluation_{\anode}(\aregisterbis_{\gamma}) = \avalue_{\gamma}$. 
  Since  \(\pair{\amapbis(\anode)}{\aind_{\gamma}} \in \inter{(\arolename_{\ell'})}\),
  $\inter{\acfeature_{\ell}}(\aind_{\gamma})$ is defined
  and $\aactvector_{\aindname_{H}}[\ell]=\top$,
  we get $\pair{\arolename_{\ell'}}{\aindname_{H}} \in \symblinks_{\anode}$
  and therefore $\aregisterbis_{\gamma} \in \asetter_{\gamma}$.

  In the case $\aind_{\gamma} \not \in \set{\inter{\aindname_0}, \ldots,\inter{\aindname_{\eta-1}}}$,
  $\aregisterbis_{\gamma} = \aregister_{\ell}^i$ with
  $\lambda^{-1}(i) = \pair{\aecdconcept}{\gamma}$. 
  Moreover, $\avalue_{\gamma} = \inter{\acfeature_{\ell}}(\aind_{\gamma})$ by definition of
  $\avalue_{\gamma}$ and
  $\amapbis(\anode \cdot i) = \aind_{\gamma}$ by definition of $\amapbis$.
  We have $\avaluation_{\anode}(\aregister_{\ell}^i) = \atuple_{\anode \cdot i}[\ell]$
  by definition of $\avaluation_{\anode}$
  and $\atuple_{\anode \cdot i}[\ell] = \inter{\acfeature_{\ell}}(\amapbis(\anode \cdot i))$
  by definition of $\atuple_{\anode \cdot i}$. Consequently,
  $\avalue_{\gamma} = \avaluation_{\anode}(\aregisterbis_{\gamma})$.
  Since $\amapbis(\anode \cdot i)$ is defined,
  $\arun^{\dag}(\anode \cdot i) \neq \square$,
  $i \in dir(\arolename_{\ell'})$ and $\aactvector_{\anode \cdot i}[\ell] = \top$,
  we get $\aregisterbis_{\gamma} \in \asetter_{\gamma}$. 
  This completes the case 
  $\aconcept = \cdrestriction{\exists}{v_1: \arp_1, \dots, v_k: \arp_k}{\acons}$
  and the proof of 
   $\avaluation_{\anode} \models
  \bigvee_{(\aregisterbis_1,\dots,\aregisterbis_k) \in \asetter_1 \times \cdots \times \asetter_k}
  \acons(\aregisterbis_1,\dots,\aregisterbis_k)$.

\item Let $\aconcept \in \mathrm{CD}_\forall(\actype_{\anode})$ with
  $\aconcept = \cdrestriction{\forall}{v_1: \arp_1, \dots, v_k: \arp_k}{\acons}$.
  In order to show that $\avaluation_{\anode} \models
  \bigwedge_{(\aregisterbis_1,\dots,\aregisterbis_k) \in \asetter_1 \times \cdots \times \asetter_k}
  \acons(\aregisterbis_1,\dots,\aregisterbis_k)$, we proceed similarly to the
  previous case by performing a case analysis on the elements in
  the $\asetter_{\gamma}$'s. The details are omitted herein. \qedhere
\end{itemize}  
\end{proof}

  \subsection{Final Complexity Analysis}
The size of the TGCA 
\(\aautomaton\) 
is bounded according to the following analysis. 
\begin{itemize}
\item \textbf{Locations.}
The set of locations \(\locations\) consists of:
\begin{itemize}
\item the special root location \(\rootlocation\) and the sink location $\square$,
\item all contextual abstractions \((\gabs, \labs)\) with
      \(\labs = (\actype, \symblinks, \aactvector)\) and
      \(\gabs : \individualnames(\aontology) \to \ctypes{\aontology} \times \act\).
\end{itemize}
The number of concept types is at most \(2^{\card{\subconcepts{\aontology}}}\).
The number of sets of symbolic links is at most
$2^{\card{\rolenames(\aontology)} \cdot \card{\individualnames(\aontology)}}$.
The number of activity vectors is equal to
$2^{\alpha}$ with \(\alpha = \card{\cfeatures(\aontology)}\). 
Consequently, we have 
\[
|\locations| \in \mathcal{O}\left( 
\underbrace{
  (2^{(|\subconcepts{\aontology}| + \alpha)})^{\card{\individualnames(\aontology)}}}_{\text{global abstraction}} 
  \cdot 
  \underbrace{2^{\card{\subconcepts{\aontology}}}}_{\text{local concept type}} 
  \cdot 
  \underbrace{2^{\alpha}}_{\text{local activity vector}} 
  \cdot 
  \underbrace{2^{\card{\rolenames(\aontology)} \cdot \card{\individualnames(\aontology)}}}_{\text{local symbolic links}} 
\right)
\]
which is exponential in \(\size{\aontology}\).
So there exists a polynomial \(\apolynomial\) such that
\(
\card{\locations} \in \mathcal{O}(2^{\apolynomial(\size{\aontology})}).
\)

\item \textbf{Transitions.}
  Every transition is determined by a tuple of locations
  \((\alocation, \alocation_0, \dots, \alocation_{\degree - 1})\)
and a unique constraint \(\acons\) computed from these locations as explained in
Section~\ref{section-construction}. Hence the number of possible transitions is at most:
\(
\card{\transitions} \leq \card{\locations}^{\degree + 1}
\). Recall that $\card{\aalphabet} =1$. 

\item \textbf{Constraint size.}
Each constraint \(\acons_\atransition\) is the conjunction of:
\begin{itemize}
\itemsep 0 cm 
\item CD–constraints \(\acons_\aconcept\) for all \(\aconcept \in \mathrm{CD}_\exists(\actype) \cup \mathrm{CD}_\forall(\actype)\).
For a CD–restriction of arity $k$, the corresponding constraint ranges over at most
\(\card{\registers{\beta}{\degree}}^k\)
register tuples, hence is of size exponential in $k$. Since the value $k$ is bounded
by a constant depending on the ontology, and the number of CD–restrictions is
linear in \(\size{\aontology}\),
this part has size at most exponential in \(\size{\aontology}\).
If the arity is fixed, this part is polynomial.

\item Register equalities \(\aregister_{\aindname,j} = \aregister_{\aindname,j}^i\) for
each $i < \degree$, \(\aindname \in \individualnames(\aontology)\), and $1 \le j \le \alpha$,
resulting in at most \(\degree \cdot \card{\individualnames(\aontology)} \cdot \alpha\)
equalities.
This part is polynomial in \(\size{\aontology}\) ($\degree$, $\eta$, $\alpha$ are polynomial
in $\size{\aontology}$).
\end{itemize}
Thus, the total size of \(\acons_\atransition\) is exponential in
\(\size{\aontology}\) in the worst case and therefore
$\maxconstraintsize{\aautomaton}$ is exponential
in $\size{\aontology}$.

\item \textbf{Branching degree and register count.}
  The degree $\degree$ is bounded by the number of existential restrictions and
  the number of variables in CD–restrictions, and
  \(\beta = \alpha (1 + \card{\individualnames(\aontology)})\).
  Both are polynomial in \(\size{\aontology}\),
as established in Section~\ref{section-automaton-preliminaries}.
\end{itemize}


\cut{
\subsection{Final Complexity Analysis}
\iftoggle{versionlong}{\input{complexity-analysis-tgca-a}
}{
The construction of the TGCA $\aautomaton$
can be done in exponential-time in \(\size{\aontology}\), 
since all components—locations, transitions, and 
constraints—can be generated and assembled in exponential time, 
By
substituting these parameters into the general complexity expression for
the nonemptiness check of TGCA
}
}
This yields the main result below.
\begin{thm} \label{theorem-main} 
The consistency problem for $\ALCO(\acdomain)$ is in $k$-\exptime if
$\acdomain$ satisfies the conditions (C1), (C2), (C3.k) and (C4), for any
$k \geq 1$.
\end{thm}

As a corollary of the combination of Lemma~\ref{lemma-encoding-soundness}
and Lemma~\ref{lemma-encoding-completeness}, if $\aontology$ is consistent, then
it admits a forest-like interpretation satisfying it with branching degree bounded
by
\[
\max\big\{N_{\mathsf{ex}} + N_{\mathsf{cd}} \times N_{\mathsf{var}},\ 
\card{\individualnames(\aontology)} \big\}.
\]
This provides an alternative
to the tree unfolding of the interpretations. Besides, the concrete domain
$\acdomain = (\Zed, (+_k)_{k \in \Nat})$ for which the consistency problem
$\ALC(\acdomain)$  is shown undecidable in~\cite[Proposition~1]{Rydval22}, has an
infinite set of predicate symbols but does not satisfy the notion of amalgamation
used in Section~\ref{section-concrete-domains} (see also~\cite[Example~3]{Rydval22}). 
\section{Extensions and Variants}
\label{section-extensions}

In this section, we consider the additional ingredients described in
Section~\ref{section-beyond-alc}
and we show how our automata-based approach
can be smoothly adapted, preserving the \exptime-membership.
By way of example, below we consider inverse roles,
functional role names and 
constraint assertions.
Section~\ref{section-integers} is dedicated to the set of integers with the linear ordering
(not satisfying (C1)) but for which we can use results from~\cite{Demri&Quaas25}. 

\subsection{Adding Functional Role Names}
\label{section-functional-role-names} 

We extend the logic \(\ALCO(\acdomain)\) with functional role names to
obtain \(\ALCOF(\acdomain)\), see Section~\ref{section-beyond-alc}.
The main trick to adapt what is done for \(\ALCO(\acdomain)\)
is to put new requirements on the map $\lambda$ and to give up the fact that
$\lambda$ is an injective map.

Before providing more details, we need to resolve a question that was
also relevant for \(\ALCO(\acdomain)\).
By adapting developments in Section~\ref{section-consistency-problem},
we can restrict ourselves to interpretations satisfying (UNA).
However, the equivalence relation  on $\set{\aindname_0, \ldots, \aindname_{\eta-1}}$
such that $\aindname \equiv \aindnamebis$
iff $\inter{\aindname} = \inter{\aindnamebis}$ is this time a bit more constrained.
Indeed, ($\rassertion{\aindname}{\aindnamebis}{\arolename} \in \aabox$
and
$\rassertion{\aindname}{\aindnameter}{\arolename} \in \aabox$
with a functional role name $\arolename$) implies 
$\aindnamebis \equiv \aindnameter$.
Consequently, it is enough to consider quotient relations satisfying these
additional constraints, and then to restrict ourselves to interpretations
satisfying (UNA) with respect to the resulting equivalence classes.

In order to construct the automaton $\aautomaton$, we follow  the construction
from Section~\ref{section-construction} except for the requirements below and
the ones about the map $\lambda$.
We assume that the (arbitrary or functional) role names occurring
in $\aontology$ are still written $\set{\arolename_1, \ldots, \arolename_{\kappa}}$
and that we are able to distinguish the functional ones. 
For the description logic $\ALCO(\acdomain)$, $\lambda$ is defined as an
{\em injective} map, which is no longer the case to handle functional role names. 
Assume that $\arolename$
is a functional role name and, $\arolename', \arolename''$ are distinct arbitrary
role names (possibly functional but not necessarily).
We require the satisfaction of the following conditions. 
\begin{itemize}
\item For all $\exrest{\arolename}{\aconcept},\exrest{\arolename}{\aconcept'}
  \in \subconcepts{\aontology}$, $\lambda(\exrest{\arolename}{\aconcept})
  = \lambda(\exrest{\arolename}{\aconcept'})$.
\item For all $\exrest{\arolename'}{\aconcept'},\exrest{\arolename''}{\aconcept''} \in
  \subconcepts{\aontology}$, we have
  $\lambda(\exrest{\arolename'}{\aconcept'})
  \neq \lambda(\exrest{\arolename''}{\aconcept''})$.
\item For all
  $\cdrestriction{\exists}{v_1: \arp_1, \dots, v_k: \arp_k}{\acons},
   \cdrestriction{\exists}{v_1': \arp_1', \dots, v_{k'}': \arp_{k'}'}{\acons'},
  \in\subconcepts{\aontology}$, for all
  $i,i'$ with $\arp_i = \arolename \cdot \acfeature$ and
  $\arp_{i'}' = \arolename \cdot \acfeature'$,
  \[
  \lambda(\pair{\cdrestriction{\exists}{v_1: \arp_1, \dots, v_k: \arp_k}{\acons}}{i})
  =
  \lambda(\pair{\cdrestriction{\exists}{v_1': \arp_1', \dots,
      v_{k'}': \arp_{k'}'}{\acons'}}{i'}).
  \]
\item A few more conditions are needed
that roughly
  state that elements of the labeling domain $\mathcal{B}$
  with the same functional role name lead to the same
  direction, and other elements of the labeling domain
  lead to distinct directions. Here they are.
  \begin{itemize}
  \item For all
  $\cdrestriction{\exists}{v_1: \arp_1, \dots, v_k: \arp_k}{\acons},
   \cdrestriction{\exists}{v_1': \arp_1', \dots, v_{k'}': \arp_{k'}'}{\acons'}
  \in\subconcepts{\aontology}$, for all
  $i,i'$ with $\arp_i = \arolename' \cdot \acfeature'$ and
  $\arp_{i'}' = \arolename'' \cdot \acfeature''$,
  \[
  \lambda(\pair{\cdrestriction{\exists}{v_1: \arp_1, \dots, v_k: \arp_k}{\acons}}{i})
  \neq
  \lambda(\pair{\cdrestriction{\exists}{v_1': \arp_1', \dots,
      v_{k'}': \arp_{k'}'}{\acons'}}{i'}).
  \]
  \item  For all $\exrest{\arolename'}{\aconcept'},
  \cdrestriction{\exists}{v_1: \arp_1, \dots, v_k: \arp_k}{\acons} \in \subconcepts{\aontology}$
  with $\arp_i = \arolename'' \cdot \acfeature''$ for some $i$,
  we have
  \[
  \lambda(
  \exrest{\arolename'}{\aconcept'}
  )
  \neq
  \lambda(\pair{\cdrestriction{\exists}{v_1: \arp_1, \dots,
      v_{k'}: \arp_{k'}}{\acons}}{i}).
  \]
  \item For all distinct $\exrest{\arolename'}{\aconcept'},\exrest{\arolename'}{\aconcept''} \in
  \subconcepts{\aontology}$, we have
  $\lambda(\exrest{\arolename'}{\aconcept'})
  \neq \lambda(\exrest{\arolename'}{\aconcept''})$ if $\arolename'$ is not functional.
  \item For all
  $\cdrestriction{\exists}{v_1: \arp_1, \dots, v_k: \arp_k}{\acons},
   \cdrestriction{\exists}{v_1': \arp_1', \dots, v_{k'}': \arp_{k'}'}{\acons'}
  \in\subconcepts{\aontology}$, for all
  $i,i'$ with $\arp_i = \arolename' \cdot \acfeature'$ and
  $\arp_{i'}' = \arolename'' \cdot \acfeature''$,
  \[
  \lambda(\pair{\cdrestriction{\exists}{v_1: \arp_1, \dots, v_k: \arp_k}{\acons}}{i})
  \neq
  \lambda(\pair{\cdrestriction{\exists}{v_1': \arp_1', \dots,
      v_{k'}': \arp_{k'}'}{\acons'}}{i'}).
  \]
  \item  For all $\exrest{\arolename}{\aconcept'},
  \cdrestriction{\exists}{v_1: \arp_1, \dots, v_k: \arp_k}{\acons} \in \subconcepts{\aontology}$
  with $\arp_i = \arolename \cdot \acfeature$ for some $i$,
  we have
  \[
  \lambda(
  \exrest{\arolename}{\aconcept'}
  )
  =
  \lambda(\pair{\cdrestriction{\exists}{v_1: \arp_1, \dots,
      v_{k'}: \arp_{k'}}{\acons}}{i}).
  \]
  \end{itemize}
\end{itemize}

The map $dir$ is uniquely determined by $\lambda$, even if $\lambda$
is constrained. 
Such a map $\lambda$ can be easily defined and the construction
in Section~\ref{section-construction} is adapted as follows. 
The set of general transitions in $\aautomaton$ is defined as for the description
logic $\ALCO(\acdomain)$ (see Section~\ref{section-construction})
except that the map $\lambda$ satisfies the properties
stated earlier. We select from such general transitions a subset
as we need to ensure that for any functional role name $\arolename$,
it is not possible to have a symbolic link with $\arolename$ and
a direction $i$ such that $i \in dir(\arolename)$, and the location of
the $i$th child is different from $\square$. This can be made precise as follows.
For any transition of the form
\[
((\globalinfo, (\actype, \symblinks, \aactvector)), \aletter, \acons,
 \alocation_0, \ldots,\alocation_{\degree-1}),
 \]
 if $\alocation_i \neq \square$ and $i \in dir(\arolename)$ for some functional
 role name $\arolename$, then there is no symbolic link
 of the form $\pair{\arolename}{\aindname}$ in $\symblinks$. 
 Furthermore, we require that there are no $\pair{\arolename}{\aindname},
 \pair{\arolename}{\aindname'} \in \symblinks$ such that $\arolename$ is functional
 and $\aindname$ is distinct from $\aindname'$. 
 
 The proof of Lemma~\ref{lemma-encoding-soundness} can be easily adapted for
 $\ALCOF(\acdomain)$ by observing that for every functional role name $\arolename$,
 the binary relation $\inter{\arolename}$ is indeed weakly functional. This is a consequence
 of the requirements on $\lambda$ and of the above requirement for general transitions.
 Similarly, the proof of Lemma~\ref{lemma-encoding-completeness} can also be adapted
 by noting that if $\ainter$ satisfies $\aontology$ with the
 functional role names interpreted by weakly functional binary relations,
 then the construction of $\adatatree$ and $\arun$ leads to an accepting run
 on $\adatatree$. The above requirement on general transitions shall be satisfied
 and the constraints on the map $\lambda$ are not an obstacle due
 to the interpretations of the functional role names.

 Complexity-wise, $\degree$ and $\beta$ are unchanged and the new requirements
 on the general transitions can be checked in polynomial time. This allows us to preserve
 the complexity upper bounds and to get Theorem~\ref{theorem-ALCOF} below.

 \begin{thm} \label{theorem-ALCOF} 
The consistency problem for $\ALCOF(\acdomain)$ is in $k$-\exptime 
if $\acdomain$ satisfies the conditions (C1), (C2), (C3.k) and (C4),
for any $k \geq 1$. 
\end{thm}

Role paths with sequences of role names of length
$> 1$
are allowed in~\cite{Lutz&Milicic07}, if made only of functional role names.
We believe we could handle it easily.

\subsection{Adding Constraint Assertions}
\label{section-constraint-assertions}

Beyond concept and role assertions, we can extend the ABox to
include  assertions involving concrete features
and  expressing relationships between concrete feature values of individual names.
Such assertions are called 
\defstyle{constraint assertions}, and have the  form
$
\acons(\acfeature_1(\aindname_1), \dots, \acfeature_k(\aindname_k))
$, 
where
$\acons(v_1, \ldots, v_k)$ is an arbitrary $\acdomain$-constraint.
An interpretation \(\ainter\) satisfies the constraint assertion
$
\acons(\acfeature_1(\aindname_1), \dots, \acfeature_k(\aindname_k))
$ if
\(
\acdomain \models \acons
(\inter{\acfeature_1}(\inter{\aindname_1}), \dots, \inter{\acfeature_k}(\inter{\aindname_k})).
\)

The constraint assertions of
the form
$
P(\acfeature_1(\aindname_1), \dots, \acfeature_k(\aindname_k))
$ for some  \(k\)-ary predicate  \(P\)
are called {\em predicate} assertions in the literature,
see e.g.~\cite{Borgwardt&DeBortoli&Koopmann24}. 
The $\acfeature_i$'s are concrete features and the $\aindname_j$'s
are individual names. Such assertions can be handled in~\cite[Section~4]{Borgwardt&DeBortoli&Koopmann24}.  
In particular, if the concrete domain \(\acdomain\) is equipped with a family of unary 
predicates \(=_{\avalue}\) interpreted by the singleton set \(\{\avalue\}\), 
then we can treat the special case of a \defstyle{feature assertion}—that is, a statement 
of the form \(\acfeature(\aindname)=\avalue\),
which asserts that the concrete 
feature \(\acfeature\) takes the value \(\avalue\) on the interpretation of the individual
name \(\aindname\).

\paragraph*{Encoding into the automaton.}
To handle constraint assertions in the construction
of $\aautomaton$, we add to the 
unique constraint $\acons_{\mathsf{root}}$ occurring only
in the root transitions, 
constraints
$\acons(\aregister_{\aindname_{i_1},j_1}, \dots, \aregister_{\aindname_{i_k},j_k})$.
Furthermore, for all $\ell \in \interval{1}{k}$,
we require that $\aactvector_{\aindname_{i_{\ell}}}[j_{\ell}]$ is equal to $\top$
in order to ensure that
\(\acfeature_{j_{\ell}}\) is defined for each individual name \(\aindname_{i_{\ell}}\).

\begin{thm} \label{theorem-predicate-assertions} 
  The consistency problem for $\ALCO(\acdomain)$ augmented with constraint assertions
  is in $k$-\exptime if $\acdomain$ satisfies the conditions (C1), (C2), (C3.k) and (C4),
  for any $k \geq 1$. 
\end{thm}

Note that nothing prevents us from using constraint assertions 
and functional role names in the same logic while preserving the complexity upper bound.

\paragraph*{Main differences with~\cite{Borgwardt&DeBortoli&Koopmann24}.}
Unlike the approach in~\cite{Borgwardt&DeBortoli&Koopmann24}, our framework does not
systematically exclude concrete domains with infinitely many 
predicate symbols such as $\triple{\Rat}{<}{(=_{q})_{q \in \Rat}}$
by relying on conditions (C1), (C2), (C3.1), and (C4). 
This makes it possible to directly express arbitrary singleton equality predicates as feature assertions,
without requiring 
the {\em homogeneity} condition on the concrete domain as done
for~\cite[Theorem 11]{Borgwardt&DeBortoli&Koopmann24}.
However, this assumes that $\acdomain$ includes unary predicates $=_{\avalue}$ with $\avalue \in \adomain$
and satisfies the appropriate conditions (Ci). 

 In~\cite{Borgwardt&DeBortoli&Koopmann24}, homogeneity plays a crucial role in the reduction:
 it ensures the existence of 
 an isomorphism between substructures of the concrete domain that preserves the interpretation of
 feature values, 
 thereby allowing singleton predicates to be simulated via isomorphic renaming. In contrast, our
 construction supports 
 such predicates natively through the infinite relational signature of the concrete
 domain but the conditions still apply.

\cut{
Because of lack of space, Appendix~\ref{appendix-functional-role-names}
(resp. Appendix~\ref{appendix-constraint-assertions})
is dedicated to functional role names (resp. to constraint assertions)
but we state their main results.

\begin{thm}
  The consistency problem for $\ALCO(\acdomain)$ augmented either
  with constraint assertions
  or with functional role names
  is in $k$-\exptime if $\acdomain$ satisfies the conditions (C1), (C2), (C3.k) and (C4),
  for any $k \geq 1$. 
\end{thm}

}

\subsection{Adding Inverse Roles}
\label{seciton-inverse-roles}
Now, let us consider  in more detail \(\ALCI(\acdomain)\),
the extension of \(\ALC(\acdomain)\) with inverse roles,
see Section~\ref{section-beyond-alc}.
The roles $\arole$
in \(\ALCI(\acdomain)\) are either role names $\arolename$ or
their inverses $\inverse{\arolename}$.
Role assertions involve
only
role names because inverse role names
add no expressivity in such assertions.
By convention, we write
$\rassertion{\aindname}{\aindnamebis}{\arolename^{-}} \in \aabox$ to denote
that the role assertion 
 $\rassertion{\aindnamebis}{\aindname}{\arolename}$ belongs to $\aabox$. 
Observe that nominals are not
present in \(\ALCI(\acdomain)\) and the handling of \(\ALCI(\acdomain)\) with nominals
remains an {\em open problem}.
Let us show how to deal with \(\ALCI(\acdomain)\).
Given a role $\arole$, we write $\inverse{\arole}$ to
denote its inverse, i.e. $\inverse{\arole} \egdef \arolename^{-}$
if $\arole = \arolename$ and $\inverse{\arole} \egdef \arolename$
if $\arole = \arolename^{-}$. 
Given an ontology $\aontology$ for \(\ALCI(\acdomain)\),
we write $\roles{\aontology}$ to
denote the set $\set{\arole_1, \ldots, \arole_{\kappa}}$ of roles
occurring in $\aontology$.
In the definition of $\subconcepts{\aontology}$, we
``remove'' the parts about nominals. 
To extend the automaton construction
to
\(\ALCI(\acdomain)\), the transitions for $\aautomaton$
need to encode access to new witnesses by considering
also
{\em the edge} to the parent node.

\paragraph*{Parent abstractions and new contextual abstractions.}
A \defstyle{parent abstraction} $\pabs$ is either a pair
$\pair{\gamma}{\amap}$
with $\gamma \in \interval{0}{\eta-1}$
and $\amap: \individualnames(\aontology) \to \act$
(encoding unknown parent, incoming direction and activity vectors of the
individual names)
or a triple
\[
\triple{\arole^{\uparrow}}{\actype^{\uparrow}}{\aactvector^{\uparrow}}
\in \roles{\aontology} \times \ctypes{\aontology} \times \act .
\]
As expected, $\arole^{\uparrow}$ encodes the role to reach a node from its parent node,
$\actype^{\uparrow}$ is the parent's concept type and
$\aactvector^{\uparrow}$ is the  parent's activity vector.
\cut{
The pairs $\pair{\gamma}{\amap}$ are not exactly
some pieces of information about the parent node, and this is mainly used
for nodes related to individual names: we find it convenient
to formalise it this way.
}
The set $\cabsset$ of contextual abstractions is now defined
as the product $\pabsset \times \labsset$,
where $\pabsset$ denotes the set of parent abstractions.
In the transitions of the forthcoming automaton $\aautomaton$,
parent abstractions are forwarded from the parent node to the children nodes. 
Contextual abstractions no longer include global abstractions
since we gave up the nominals (but the pair
$\pair{\gamma}{\amap}$ contains a bit of global information).
Similarly, the local abstractions have no more
symbolic links. 

\paragraph*{Automaton construction for $\ALCI(\acdomain)$.}
We define the  TGCA \(\aautomaton = 
(\locations, \aalphabet, \degree, \beta, \locations_{in}, 
\transitions, F)\)  from some ontology \(\aontology\) as follows,
extending the construction in Section~\ref{section-construction}.
As a rule of thumb, we are concise for the parts similar to
$\ALCO(\acdomain)$ and we briefly comment the new parts related
to inverse roles. Nevertheless, we provide full definitions and
proofs since changes occur everywhere and we wish to make it easy
to check the correctness  from previous developments
for $\ALCO(\acdomain)$. 

\begin{itemize}
\item $\degree \egdef \max \set{N_{\mathsf{ex}} + N_{\mathsf{cd}} \times N_{\mathsf{var}}, 
  \card{\individualnames(\aontology)}}$, where $N_{\mathsf{ex}}$
  is the number of existential restrictions in $\subconcepts{\aontology}$.
  To  support the reasoning on the concrete feature
  values of  the parent nodes, the new  registers
  $\aregister_1^\uparrow, \ldots, \aregister_{\alpha}^\uparrow$
  are dedicated to the parent nodes.
  Consequently, $\beta \egdef \alpha \times (\eta +2)$.
  It is worth noting at this stage that the $\beta \times \eta$
  registers devoted to the $\eta$ individual names are useful
  only at the very beginning of the runs. 
Unlike what happens in the automata
  for $\ALCO(\acdomain)$, the registers dedicated to
  the individual names are not maintained constant
  along the run but only used
  at the nodes $0, \ldots, \eta-1$ (labelled by some location
  of the form $\pair{\pair{\gamma}{\amap}}{\labs}$). 
\item $\locations \egdef \cabsset \cup \set{\rootlocation, \square}$
  (with the suitable definition for $\cabsset$),
  $F \egdef \locations$, $\locations_{in} \egdef \set{\rootlocation}$
  and $\aalphabet \egdef \set{\aletter}$. 
\item The
  transition
  relation $\transitions$ is a subset of
\(
\locations \times \aalphabet \times 
\treeconstraints{\beta, \degree} \times \locations^\degree
\).
As for \(\ALCO(\acdomain)\), we distinguish three kinds of transitions
but the general transitions have now two categories.
There is still a unique padding transition
$\atransition_\square \egdef (\square, \aletter, \top, \square, \dots, \square)$.
In the  root transitions
\[
(\rootlocation, \aletter, \acons_{\mathsf{root}}, \cabs_0, 
\ldots, \cabs_{\eta-1}, \square, \ldots, \square),
\]
each $\cabs_{i}$ is of the form
$(\pair{i}{\amap_i}, \pair{\actype_i}{\aactvector_i})$
with the same requirements as for $\ALCO(\acdomain)$
(except that there is no more global abstractions)
 but we require that $\amap_0 = \cdots = \amap_{\eta-1}$. 
 Furthermore, $\acons_{\mathsf{root}}$ keeps the same value.
There is nevertheless an additional requirement due to the absence of
global abstractions: for all $\rassertion{\aindname_i}{\aindname_{i'}}{\arolename}
\in \aabox$, for all $\varest{\arolename}{\aconcept} \in \actype_i$
and $\varest{\arolename^{-}}{\aconcept'} \in \actype_{i'}$,
we have $\aconcept \in \actype_{i'}$ and $\aconcept' \in \actype_{i}$.
\iftoggle{versionlong}{
The presence of inverse role names in 
\(\ALCI(\acdomain)\) introduces two distinct directions for satisfying
the restrictions: either via a child node or
via the parent, when the incoming role matches the inverse of
the required role.
For the nodes $\gamma \in \interval{0}{\eta-1}$, the two distinct
directions are to the child node and to  nodes in $\interval{0}{\eta-1}$.
\cut{
\begin{itemize}
 \item \emph{Forward:} via a child node.
 \item \emph{Backward:} via the parent, when the incoming role matches the inverse of
   the required role.
  \end{itemize}
 }
  The first case is already in the construction for
   \(\ALCO(\acdomain)\).
   To handle inverse role names soundly, any value
   restriction \(\varest{\arole}{\aconcept}\) 
   satisfied in a child or the node representing the individual names via
   symbolic link must be reflected 
   in the parent node if reached via \(\inverse{\arole}\).

}{}
The main changes are for the general transitions
  \(
  \atransition = (\cabs, \aletter, \acons_\atransition,
  \alocation_0, \dots, \alocation_{\degree-1})
  \)
 satisfying the following conditions
   with $\cabs =
   \pair{\pabs}{
     \pair{\actype}{\aactvector}}$.
   
  \begin{description}
 
  \item[(Existential restrictions)]  For all \(\exrest{\arole}{\aconcept} \in \actype\),
  \(i = \lambda(\exrest{\arole}{\aconcept})\),  
  \(\alocation_i\) of the form
  \((\pabs_i, (\actype_i,  \aactvector_i))\)
  and $\aconcept \in \actype_i$ or
  $\inverse{\arole} = \arole^{\uparrow}$ and
  $\aconcept \in \actype^{\uparrow}$
  if
  $\pabs =  \triple{\arole^{\uparrow}}{\actype^{\uparrow}}{\aactvector^{\uparrow}}$ (new  option).

  \item[(Value restrictions)]
   For all \(\varest{\arole}{\aconcept} \in \actype\),
   for all \(i \in \dir{\arole}\) with
   \(\alocation_i = (\pabs_i, (\actype_i,  \aactvector_i))\),
  we have \(\aconcept \in \actype_i\), 
  and if $\inverse{\arole} = \arole^{\uparrow}$, then  $\aconcept \in \actype^{\uparrow}$
  if $\pabs =  \triple{\arole^{\uparrow}}{\actype^{\uparrow}}{\aactvector^{\uparrow}}$
  (new  requirement).

\item[(Back-propagation)]
  If
  \(\alocation_i = (\pabs_i,  (\actype_i, \aactvector_i))\),
  $\varest{\arole}{\aconcept} \in \actype_i$ and
  $i \in dir(\inverse{\arole})$, then
  $\aconcept \in \actype$.  
  This is a new condition due to inverse roles. 

\item[(CD-restrictions)]
  Let \(\mathrm{CD}_\exists(\actype)\) and \(\mathrm{CD}_\forall(\actype)\) be
  respectively the sets of existential and universal 
CD-restrictions in \(\actype\). 
Let \(\aconcept = \cdrestriction{\aquantifier}{v_1: \arp_1, \dots, v_k: \arp_k}{\acons}\) 
be such a CD-restriction with quantifier \(\aquantifier \in \{\exists, \forall\}\). 
For each role path \(\arp_j\), we define
\(\asetter_j \subseteq 
\registers{\beta}{\degree}\) as follows.
\begin{itemize}
\item If \(\arp_j = \acfeature_l\), then
  if $ \aactvector[l] = \top$, then 
  $\asetter_j \egdef \{\aregister_l\}$
  else $\asetter_j \egdef \emptyset$. 
\item If \(\arp_j = \arole \cdot \acfeature_l\), then \(\asetter_j\)
  is defined as follows.
  If $\pabs =  \triple{\arole^{\uparrow}}{\actype^{\uparrow}}{\aactvector^{\uparrow}}$, then
  \[
  \asetter_j \egdef  \left\{ \aregister_{l}^{i} \mid i \in \dir{\arole},\ 
  \alocation_i = (\pabs_i,  (\actype_i, \aactvector_i)),\
  \aactvector_i[l] = \top \right\} 
  \cup
  \left\{
        \aregister_l^{\uparrow} \;\middle|\;
            \text{\(\arole^{\uparrow} = \inverse{\arole}\),}
          \; \aactvector^{\uparrow}[l] = \top
          \right\}.
\]
\cut{
\[
{\small
  \begin{aligned}
  \asetter_j \egdef \; & \left\{ \aregister_{l}^{i} \mid i \in \dir{\arole},\ 
  \alocation_i = (\pabs_i,  (\actype_i, \aactvector_i)),\
  \aactvector_i[l] = \top \right\} 
  \cup
  \left\{
        \aregister_l^{\uparrow} \;\middle|\;
            \text{\(\arole^{\uparrow} = \inverse{\arole}\),}
          \; \aactvector^{\uparrow}[l] = \top
          \right\}
  \end{aligned}
}.
\]
}
Otherwise $\pabs$ is of the form  $\pair{\gamma}{\amap}$, we need  to take advantage
of the registers dedicated to individual names:
\cut{
$\asetter_j \egdef \;  \left\{ \aregister_{l}^{i} \mid i \in \dir{\arole},\ 
  \alocation_i = (\pabs_i,  (\actype_i, \aactvector_i)),\
  \aactvector_i[l] = \top \right\} 
  \cup 
   \set{\aregister_{\aindname_{\gamma'},l} \mid
    \rassertion{\aindname_{\gamma}}{\aindname_{\gamma'}}{\arole}
    \in \aabox,
    \amap(\aindname_{\gamma'})[l] = \top
   }$.
}
\[
\begin{aligned}
  \asetter_j \egdef \; & \left\{ \aregister_{l}^{i} \mid i \in \dir{\arole},\ 
  \alocation_i = (\pabs_i,  (\actype_i, \aactvector_i)),\
  \aactvector_i[l] = \top \right\} 
  \cup \\ 
  & \set{\aregister_{\aindname_{\gamma'},l} \mid
    \rassertion{\aindname_{\gamma}}{\aindname_{\gamma'}}{\arole}
    \in \aabox,
    \amap(\aindname_{\gamma'})[l] = \top
  }.
\end{aligned}
\]
\end{itemize}
Let
$\asetter \egdef \prod \asetter_j$. 
If $\aconcept
\in \mathrm{CD}_\exists(\actype)$,
then
$\acons_{\aconcept} \egdef 
\bigvee_{(\aregisterbis_1,\dots,\aregisterbis_k) \in \asetter}
\acons(\aregisterbis_1,\dots,\aregisterbis_k)$,
otherwise
$\acons_{\aconcept} \egdef 
\bigwedge_{(\aregisterbis_1,\dots,\aregisterbis_k) \in \asetter}
\acons(\aregisterbis_1,\dots,\aregisterbis_k)$.
The constraint $\acons_\atransition$ is defined below, where the
new part
expresses the consistency of the registers dedicated
to the parent node (\(\aregister_j^{\uparrow \ i}\) denotes the register
at the \(i\)th child for the parent's value of 
\(\acfeature_j\)).
\[
\acons_\atransition =
\bigwedge_{\aconcept \in \mathrm{CD}_\exists(\actype) \cup \mathrm{CD}_\forall(\actype)}
\acons_\aconcept
\wedge
\bigwedge_{\substack{i \in \interval{0}{\degree-1} \\ \alocation_i \neq \square}}
\ \ 
\bigwedge_{j \in \interval{1}{\alpha}, \aactvector[j]=\top}
\aregister_j^{\uparrow \ i} = \aregister_j.
\]

\item[(Inactive children)] Same condition as for $\ALCO(\acdomain)$. 
\cut{
  For all $i \in \interval{0}{\degree-1}$,
  if $\lambda^{-1}(i) = \exrest{\arole}{\aconcept}$ and $\lambda^{-1}(i) \not \in \actype$
  or $\lambda^{-1}(i) =
  \pair{\cdrestriction{\exists}{v_1: \arolepath_1, \ldots, v_k: \arolepath_k}{\acons}}{j}$ and
 ( $j > k$ or $\cdrestriction{\exists}{v_1: \arolepath_1, \ldots, v_k: \arolepath_k}{\acons}
  \not \in \actype$), then $\alocation_i = \square$.
}

  \item[(Parent information propagation)] 
    For each $i$ such that \(\alocation_i\) is a contextual abstraction
    $(\triple{\arole^{\uparrow}_i}{\actype^{\uparrow}_i}{\aactvector^{\uparrow}_i},
    (\actype_i,  \aactvector_i))\), we require
    $i \in dir(\arole_i^{\uparrow})$ and $\pair{\actype}{\aactvector}
    = \pair{\actype^{\uparrow}_i}{\aactvector^{\uparrow}_i}$.
    This is a new condition. 
\end{description}
\end{itemize}
Soundness is shown as for
Lemma~\ref{lemma-encoding-soundness-alci}
by  extending  the interpretation
of role names,
which requires to verify more conditions.

\begin{lem}[Soundness] 
\label{lemma-encoding-soundness-alci}
  \(\alang(\aautomaton) \neq \emptyset\) implies \(\aontology\) is consistent.
\end{lem}

\begin{proof} We slightly adapt the proof of Lemma~\ref{lemma-encoding-soundness}
  but we provide the details to allow the reader to perform a quick check.
  Nevertheless, there are slight changes everywhere. 
 Assume that \(\alang(\aautomaton) \neq \emptyset\).
 There exist a data tree \(\adatatree : \interval{0}{\degree-1}^* \to
 \aalphabet \times \adomain^\beta\) and an accepting run
 \(\arun : \interval{0}{\degree-1}^* \to \transitions\) on $\adatatree$ such that 
    for every node \(\anode\),
    \(\adatatree(\anode) = \pair{\aletter_{\anode}}{\atuple_{\anode}}\)
    and \(\arun(\anode) = (\alocation_{\anode}, \aletter_{\anode},
                    \acons_{\anode}, \alocation_{\anode \cdot 0}, \dots,
                    \alocation_{\anode \cdot (\degree-1)})\).
                    Sometimes, we use the fact that $\anode$ is labelled by a location
                    (and not necessarily
                    by a transition) and we mean $\alocation_{\anode}$. 
    We construct the $\ALCI(\acdomain)$ interpretation
    $\ainter = (\aidomain, \inter{\cdot})$ from $\adatatree$ and $\arun$ as follows.
    \begin{itemize}
    \item The interpretation domain is
    \(
    \aidomain \egdef \{ \anode \in \interval{0}{\degree-1}^+
    \mid \alocation_{\anode} \neq \square \}
    \). 

  \item For each $i \in \interval{0}{\eta-1}$, \(\inter{\aindname_i} \egdef i\).
    In the run $\arun$, the location assigned to the node $i$ (interpreting
    $\aindname_i$)
     is 
     \((\pair{i}{\amap}, (\actype_{\aindname_i},  \aactvector_{\aindname_i}))\) 
      by definition
     of the root transitions.

    \item For each concept name \(\aconceptname\), define
    \(
    \inter{\aconceptname} \egdef \{ \anode \in \aidomain \mid \aconceptname \in \actype_{\anode} \}.
    \)

    \item For each role name \(\arolename \in \rolenames(\aontology)\), define
      \[
\begin{aligned}
      \inter{\arolename} \egdef\; &
    \left\{\, (\anode, \anode \cdot i) \in \aidomain \times \aidomain\;\middle|\;
    i \in \dir{\arolename}\right\}
    \;\cup\; \\
    & \underbrace{
    \left\{\, \pair{\anode \cdot i}{\anode}  \in \aidomain \times \aidomain\;\middle|\;
    i \in \dir{\arolename^{-}}\right\}
    }_{\mbox{from the tree structure (new)}} \cup
    \set{
      \pair{\gamma}{\gamma'} \mid
      \rassertion{\aindname_{\gamma}}{\aindname_{\gamma'}}{\arolename}
     \in \aabox \ {\rm and} \ \gamma,\gamma' \in \interval{0}{\eta-1}      
    }
\end{aligned}
    \]
  \item For each concrete feature \(\acfeature_j \in \cfeatures(\aontology)\),
    we define the partial 
    function \(\inter{\acfeature_j} : \aidomain \to \adomain\) such that 
    \(
    \inter{\acfeature_j}(\anode) \egdef \atuple_{\anode}[j]
    \) if $\aactvector_{\anode}[j] = \top$,  
    otherwise $\inter{\acfeature_j}(\anode)$ is undefined. 
    \end{itemize}
    Observe that $\ainter$ satisfies (UNA) w.r.t.
    $\set{\aindname_0, \ldots, \aindname_{\eta-1}}$. 
    
    Showing  $\ainter \models \aontology$ boils down to 
    establish the  \defstyle{type soundness property}:
    for every \(\anode \in \aidomain\),
    and for every concept \(\aconcept \in \subconcepts{\aontology}\), 
    if \(\aconcept \in \actype_{\anode}\), then \(\anode \in \inter{\aconcept}\),
    assuming that the location $\alocation_{\anode}$ labelling the node  $\anode$ is of the form
    \((\pabs, 
    (\actype_{\anode},\aactvector_{\anode}))\).
    The parent abstract $\pabs$ is either a triple
    $\triple{\arole^{\uparrow}}{\actype^{\uparrow}}{\aactvector^{\uparrow}}$ or some
    pair $\pair{\gamma}{\amap}$ with $\gamma \in \interval{0}{\eta-1}$. 
    The proof is by structural induction and we treat only the cases that differ
    from the proof of Lemma~\ref{lemma-encoding-soundness}.

    \begin{description}
    
  \item[\textbf{Case:} $\aconcept = \exrest{\arole}{\aconceptbis}$]
    Suppose that $\aconcept \in \actype_{\anode}$.
    By the definition of general transitions  for $\aautomaton$
    and more specifically
    the part about existential restrictions, 
     \(i = \lambda(\exrest{\arole}{\aconceptbis})\),  
  \(\alocation_{\anode \cdot i}\) is of the form
  \((\pabs_i, (\actype_i,  \aactvector_i))\)
  and $\aconceptbis \in \actype_i$
  or $\inverse{\arole} = \arole^{\uparrow}$ and
  $\aconceptbis \in \actype^{\uparrow}$
  if $\pabs$ is a triple containing $\actype^{\uparrow}$.
  By the induction hypothesis,
  $\anode \cdot i \in \inter{\aconceptbis}$
  or $\anodebis \in \inter{\aconceptbis}$ ($\anodebis$ is the parent of $\anode$).
  Moreover, by definition of $\inter{\arole}$,
   $\pair{\anode}{\anode \cdot i} \in \inter{\arole}$
  whenever $i \in dir(\arole)$ and $\alocation_{\anode \cdot i}$
  is not the sink location.
  Moreover, if $\inverse{\arole} = \arole^{\uparrow}$ and
  $\aconcept \in \actype^{\uparrow}$, then
  by the induction hypothesis
  $\anodebis \in \inter{\aconceptbis}$
  and $\pair{\anode}{\anodebis} \in \inter{\arole}$. 
  Consequently, there is $\aind \in \aidomain$
  such that
  $\pair{\anode}{\aind} \in \inter{\arole}$
  and $\aind \in \inter{\aconceptbis}$, whence 
  $\anode \in \inter{\aconcept}$.

    \item[\textbf{Case:} $\aconcept = \varest{\arole}{\aconceptbis}$]  
      Assume $\aconcept \in \actype_{\anode}$
      and $\pair{\anode}{\aind} \in \inter{\arole}$.
      We distinguish three cases depending how the edge
      $\pair{\anode}{\aind}$ is included in $\inter{\arole}$.
      \begin{description}

      \item[Subcase $\pair{\anode}{\anode \cdot i} \in \inter{\arole}$,
        $\alocation_{\anode \cdot i} \neq \square$
        and $i \in dir(\arole)$]
      By the definition of general transitions for $\aautomaton$ and more specifically
      the part about value restrictions,
      if \(\alocation_{\anode \cdot i}\) is of the form
      \((\pabs_i, (\actype_i, \aactvector_i))\),
      then  \(\aconceptbis \in \actype_i\).
      By the induction hypothesis, we have
      $\anode \cdot i \in \inter{\aconceptbis}$.

  \item[Subcase  $\pabs = \triple{\arole^{\uparrow}}{\actype^{\uparrow}}{\aactvector^{\uparrow}}$ and $\inverse{\arole} = \arole^{\uparrow}$]
     By the definition of general transitions for $\aautomaton$ and more specifically
     the part about value restrictions, $\aconceptbis \in \actype^{\uparrow}$.
     By the induction hypothesis, $\anodebis \in \inter{\aconceptbis}$,
     where $\anodebis$ is the parent of $\anode$.

   \item[Subcase $\anode = \gamma$ and $\aind = \gamma'$]
     This entails that $\rassertion{\aindname_{\gamma}}{\aindname_{\gamma'}}{\arole}
     \in \aabox$ by the definition of $\inter{\arole}$
     (remember we have a convention to talk about
     $\arole$). By the requirements for the root transitions,
     $\aconceptbis \in \actype_{\gamma'}$ and therefore by the induction
     hypothesis, $\gamma' \in \inter{\aconceptbis}$.

      \end{description}
      
      In conclusion, for all $\pair{\anode}{\aind} \in \inter{\arole}$,
      we have $\aind \in \inter{\aconceptbis}$, whence
      $\anode \in \inter{\aconcept}$.

    \item[\textbf{Case:} $\aconcept = \cdrestriction{\exists}{v_1: \arp_1, \dots, v_k: \arp_k}{\acons}$ and
      $\pabs_{\anode}$ not of the form $\pair{\gamma}{\amap}$]
Assume that $\aconcept \in \actype_{\anode}$. 
By the definition of general transitions for $\aautomaton$ and more specifically
the part about the constraints,
$\acons_{\anode}$ has a conjunct $\acons_{\aconcept}$
of the form
$\bigvee_{(\aregisterbis_1,\dots,\aregisterbis_k) \in \asetter}
  \acons(\aregisterbis_1,\dots,\aregisterbis_k)$. 
Let \(\avaluation_{\anode} : \registers{\beta}{\degree} \to
  \adomain\) be the valuation defined by
  \(
  \avaluation_{\anode}(\aregister_j) = \atuple_{\anode}[j]\),
  \(
  \avaluation_{\anode}(\aregister_j^{\uparrow}) = \atuple_{\anodebis}[j]\)
  for $\anodebis$ the parent of $\anode$,
  and
  \(
  \avaluation_{\anode}(\aregister_j^i) = \atuple_{\anode \cdot i}[j],
  \)
  for all \(1 \le j \le \beta,\ 0 \le i < \degree\),
  see the definition of runs in Section~\ref{section-tgca}.
  Since $\arun$ is a run, we have
  \(\avaluation_{\anode} \models \acons_{\anode}\)
  and therefore there is
  $(\aregisterbis_1,\dots,\aregisterbis_k) \in
  \asetter$ such that
  $\avaluation_{\anode} \models
  \acons(\aregisterbis_1,\dots,\aregisterbis_k)$.
  Now, by construction of the $\asetter_j$'s
  and by the interpretation of the concrete features in
  $\ainter$, 
  one can show that for all $j$,
  \begin{equation}
  \label{equation-rpj-to-zj-bis}
  \inter{\arp_j}(\anode) =
  \set{\avaluation_{\anode}(\aregisterbis)
    \mid \aregisterbis \in \asetter_j}.
  \end{equation}
  Consequently,
  $[v_1 \mapsto \avaluation_{\anode}(\aregisterbis_1), \dots,
    v_k \mapsto \avaluation_{\anode}(\aregisterbis_k)] \models \acons$
  and $(\avaluation_{\anode}(\aregisterbis_1), \ldots,
  \avaluation_{\anode}(\aregisterbis_k)) \in
  \inter{\arp_1}(\anode) \times \cdots \times
      \inter{\arp_k}(\anode)$.
  By the semantics of $\ALCI(\acdomain)$, we get
  $\anode \in \inter{\aconcept}$.
  It remains to prove~(\ref{equation-rpj-to-zj-bis}).

  First, assume that $\arp_j = \acfeature_{\ell}$.
  If $\aactvector_{\anode}[\ell] = \perp$, then
  $\asetter_j = \emptyset$ by definition of $\asetter_j$
  and $\inter{\acfeature_{\ell}}(\anode)$ is undefined by
  definition of $\inter{\acfeature_{\ell}}$.
  In the case $\aactvector_{\anode}[\ell] = \top$, then
  $\asetter_j = \set{\aregister_{\ell}}$ by definition of $\asetter_j$
  and $\inter{\acfeature_{\ell}}(\anode)$ is
  equal to $\atuple_{\anode}[\ell]$ by definition of
  $\inter{\acfeature_{\ell}}$.
  Since $\avaluation_{\anode}(\aregister_{\ell})
  = \atuple_{\anode}[\ell]$ by definition of
  $\avaluation_{\anode}$, for both cases
  we get 
  $\inter{\arp_j}(\anode) =
  \set{\avaluation_{\anode}(\aregisterbis)
    \mid \aregisterbis \in \asetter_j}$. 

  Second, assume that $\arp_j = \arole \cdot
  \acfeature_{\ell}$.
  \begin{enumerate}

  \item If $i \in dir(\arole)$, $\alocation_{\anode \cdot i}
    \neq \square$ and $\aactvector_{\anode \cdot i}[\ell] = \top$,
    then $\aregister_{\ell}^i \in \asetter_j$.
    By definition of $\inter{\arole}$ and
    $\inter{\acfeature_{\ell}}$,
    we have $\pair{\anode}{\anode \cdot i} \in
    \inter{\arole}$ and $\inter{\acfeature_{\ell}}(\anode \cdot i)$
    is defined.
    Consequently,  $\inter{\acfeature_{\ell}}(\anode \cdot i)
    \in \inter{\arp_j}(\anode)$. 
    Since $\avaluation_{\anode}(\aregister_{\ell}^i)
    = \avaluation_{\anode \cdot i}(\aregister_{\ell})$ by definition
    of the valuations
    and $\inter{\acfeature_{\ell}}(\anode \cdot i)
    = \avaluation_{\anode \cdot i}(\aregister_{\ell})$
    by definition of $\inter{\acfeature_{\ell}}$, we get 
    $\avaluation_{\anode}(\aregister_{\ell}^i)\in
    \inter{\arp_j}(\anode)$.

  \item If \(\arole^{\uparrow} = \inverse{\arole}\)
    and $\aactvector^{\uparrow}[l] = \top$, then
    $\aregister_l^{\uparrow} \in \asetter_j$.
     By definition of $\inter{\arole}$ and
    $\inter{\acfeature_{\ell}}$,
    we have $\pair{\anode}{\anodebis} \in
    \inter{\arole}$ for $\anodebis$ parent of $\anode$,
    and $\inter{\acfeature_{\ell}}(\anodebis)$
    is defined. Consequently, $\avaluation_{\anode}(\aregister_{\ell}^{\uparrow})\in
    \inter{\arp_j}(\anode)$.

  \item Now, we prove the other inclusion.
    Let $\avalue \in \inter{\arp_j}(\anode)$. There is
    $\aind$ such that $\pair{\anode}{\aind} \in \inter{\arole}$
    and $\inter{\acfeature_{\ell}}(\aind) = \avalue$.
    We distinguish two cases depending how the edge
    $\pair{\anode}{\aind}$  is included in $\inter{\arole}$:
    the second
    one is new and about the parent nodes. 
    \begin{description}
    \item[Subcase $\pair{\anode}{\anode \cdot i} \in \inter{\arole}$,
        $\alocation_{\anode \cdot i} \neq \square$
      and $i \in dir(\arole)$]
      Since $\inter{\acfeature_{\ell}}(\anode \cdot i)$ is defined,
      by definition
      of $\inter{\acfeature_{\ell}}$, $\aactvector_{\anode \cdot i}[\ell]
      = \top$.
      By definition of $\asetter_j$,
      $\aregister_{\ell}^{i} \in \asetter_j$.
      Moreover, $\avaluation_{\anode}(\aregister_{\ell}^i)
      = \avaluation_{\anode \cdot i}(\aregister_{\ell})
      = \inter{\acfeature_{\ell}}(\anode \cdot i)$,
      by definition of the valuations and
      $\inter{\acfeature_{\ell}}$.
      Consequently, 
      $\avaluation_{\anode}(\aregister_{\ell}^i) \in 
      \set{\avaluation_{\anode}(\aregisterbis)
        \mid \aregisterbis \in \asetter_j}$.
    \item[Subcase  $\pair{\anode}{\anodebis} \in \inter{\arole}$
      for the parent $\anodebis$]
       Since $\inter{\acfeature_{\ell}}(\anodebis)$ is defined,
      by definition
      of $\inter{\acfeature_{\ell}}$, $\aactvector_{\anodebis}[\ell]
      = \top$.
      Moreover, since $\pair{\anode}{\anodebis} \in \inter{\arole}$,
      by definition of $\inter{\arole}$, this means
      $\arole^{\uparrow} = \inverse{\arole}$.
      By definition of $\asetter_j$,
      $\aregister_{\ell}^{\uparrow} \in \asetter_j$.
       Moreover, $\avaluation_{\anode}(\aregister_{\ell}^{\uparrow})
      = \avaluation_{\anodebis}(\aregister_{\ell})
      = \inter{\acfeature_{\ell}}(\anodebis)$,
      by definition of the valuations and
      $\inter{\acfeature_{\ell}}$.
      Consequently, 
      $\avaluation_{\anode}(\aregister_{\ell}^{\uparrow}) \in 
      \set{\avaluation_{\anode}(\aregisterbis)
        \mid \aregisterbis \in \asetter_j}$.
    \end{description}
  \end{enumerate}

\item[\textbf{Case:} $\aconcept = \cdrestriction{\exists}{v_1: \arp_1, \dots, v_k: \arp_k}{\acons}$ and
  $\pabs_{\anode} = \pair{\gamma}{\amap}$]
  This is similar to the previous case but we provide the details
  for the sake of exhaustivity. Remember there is a case analysis
  for defining the $\asetter_j$'s. 
  
  Assume that $\aconcept \in \actype_{\anode}$. 
The constraint
$\acons_{\anode}$ has a conjunct $\acons_{\aconcept}$
of the form
\[
\bigvee_{(\aregisterbis_1,\dots,\aregisterbis_k) \in \asetter}
\acons(\aregisterbis_1,\dots,\aregisterbis_k).
\]
Herein,
the distinguished registers $\aregister_1, \ldots, \aregister_{\beta}$
referring to values at the current node shall be represented by
$
\aregister_1, \ldots, \aregister_{\alpha},
\aregister_1^{\uparrow}, \ldots, \aregister_{\alpha}^{\uparrow}, 
\aregister_{\aindname_0, 1}, \ldots,\aregister_{\aindname_0,\alpha},
\ldots,
\aregister_{\aindname_{\eta-1}, 1}, \ldots,\aregister_{\aindname_{\eta-1},\alpha}
$.
Let \(\avaluation_{\anode} : \registers{\beta}{\degree} \to
  \adomain\) be the valuation defined by
  \(
  \avaluation_{\anode}(\aregister_j) = \atuple_{\anode}[j]\),
  \(
  \avaluation_{\anode}(\aregister_j^{\uparrow}) = \atuple_{\anodebis}[j]\)
  for $\anodebis$ the parent of $\anode$,
  and
  \(
  \avaluation_{\anode}(\aregister_j^i) = \atuple_{\anode \cdot i}[j],
  \)
  for all \(1 \le j \le \beta,\ 0 \le i < \degree\),
  see the definition of runs in Section~\ref{section-tgca}.
  Since $\arun$ is a run, we have
  \(\avaluation_{\anode} \models \acons_{\anode}\)
  and therefore there is
  $(\aregisterbis_1,\dots,\aregisterbis_k) \in
  \asetter$ such that
  $\avaluation_{\anode} \models
  \acons(\aregisterbis_1,\dots,\aregisterbis_k)$.
  Now, by construction of the $\asetter_j$'s
  and by the interpretation of the concrete features in
  $\ainter$, 
  one can show that for all $j$,
  \begin{equation}
  \label{equation-rpj-to-zj-ter}
  \inter{\arp_j}(\anode) =
  \set{\avaluation_{\anode}(\aregisterbis)
    \mid \aregisterbis \in \asetter_j}.
  \end{equation}
Consequently,
  $[v_1 \mapsto \avaluation_{\anode}(\aregisterbis_1), \dots,
    v_k \mapsto \avaluation_{\anode}(\aregisterbis_k)] \models \acons$
  and $(\avaluation_{\anode}(\aregisterbis_1), \ldots,
  \avaluation_{\anode}(\aregisterbis_k)) \in
  \inter{\arp_1}(\anode) \times \cdots \times
      \inter{\arp_k}(\anode)$.
  By the semantics of $\ALCI(\acdomain)$, we get
  $\anode \in \inter{\aconcept}$.
  It remains to prove~(\ref{equation-rpj-to-zj-ter}).
  The case $\arp_j = \acfeature_{\ell}$ is handled exactly as the previous one,
  so it is omitted. Now, assume that $\arp_j = \arole \cdot
  \acfeature_{\ell}$.
  \begin{enumerate}
   \item If $i \in dir(\arole)$, $\alocation_{\anode \cdot i}
    \neq \square$ and $\aactvector_{\anode \cdot i}[\ell] = \top$,
    then $\aregister_{\ell}^i \in \asetter_j$.
    This case is handled exactly as the previous one
    and we can conclude $\avaluation_{\anode}(\aregister_{\ell}^i)\in
    \inter{\arp_j}(\anode)$. 
  \item  If $\rassertion{\aindname_{\gamma}}{\aindname_{\gamma'}}{\arole} \in \aabox$
    and $\amap(\aindname_{\gamma'})[\ell] = \top$, then
    $\aregister_{\aindname_{\gamma'}, \ell} \in \asetter_j$.
    By definition of $\inter{\arole}$ and
    $\inter{\acfeature_{\ell}}$,
    we have $\pair{\inter{\aindname_{\gamma}}}{\inter{\aindname_{\gamma'}}} \in
    \inter{\arole}$ and $\inter{\acfeature_{\ell}}(\inter{\aindname_{\gamma'}})$
    is defined. Consequently,
    $\inter{\acfeature_{\ell}}(\inter{\aindname_{\gamma'}}) \in \inter{\arp_j}(\anode)$
    $\inter{\acfeature_{\ell}}(\inter{\aindname_{\gamma'}})$
    is equal to $\avaluation_{\anode}(\aregister_{\aindname_{\gamma'}, \ell})$,
    since $\anode = \gamma$ and $\avaluation_{\varepsilon} \models
    \acons_{\mathsf{root}}$. 
    Hence, $\avaluation_{\anode}(\aregister_{\aindname_{\gamma'}, \ell})\in
    \inter{\arp_j}(\anode)$.
\item Now, we prove the other inclusion.
    Let $\avalue \in \inter{\arp_j}(\anode)$. There is
    $\aind$ such that $\pair{\anode}{\aind} \in \inter{\arole}$
    and $\inter{\acfeature_{\ell}}(\aind) = \avalue$.
    We distinguish two cases depending how the edge
    $\pair{\anode}{\aind}$  is included in $\inter{\arole}$.
    \begin{description}
    \item[Subcase $\pair{\anode}{\anode \cdot i} \in \inter{\arole}$,
        $\alocation_{\anode \cdot i} \neq \square$
      and $i \in dir(\arole)$]
      Since $\inter{\acfeature_{\ell}}(\anode \cdot i)$ is defined,
      by definition
      of $\inter{\acfeature_{\ell}}$, $\aactvector_{\anode \cdot i}[\ell]
      = \top$.
      By definition of $\asetter_j$,
      $\aregister_{\ell}^{i} \in \asetter_j$.
      Moreover, $\avaluation_{\anode}(\aregister_{\ell}^i)
      = \avaluation_{\anode \cdot i}(\aregister_{\ell})
      = \inter{\acfeature_{\ell}}(\anode \cdot i)$,
      by definition of the valuations and
      $\inter{\acfeature_{\ell}}$.
      Consequently, 
      $\avaluation_{\anode}(\aregister_{\ell}^i) \in 
      \set{\avaluation_{\anode}(\aregisterbis)
        \mid \aregisterbis \in \asetter_j}$.
    \item[Subcase $\rassertion{\aindname_{\gamma}}{\aindname_{\gamma'}}{\arole} \in \aabox$
      and $\inter{\aindname_{\gamma'}} = \aind$]
      Since $\inter{\acfeature_{\ell}}(\aind)$ is defined,
      by definition
      of $\inter{\acfeature_{\ell}}$, $\amap(\aindname_{\gamma'})[\ell] = \top$.
      By definition of $\asetter_j$,
      $\aregister_{\aindname_{\gamma'}, \ell} \in \asetter_j$.
      Moreover, $\avaluation_{\anode}(\aregister_{\aindname_{\gamma'}, \ell})
      = \avalue$. Consequently,
      $\avaluation_{\anode}(\aregister_{\aindname_{\gamma'}, \ell}) \in 
      \set{\avaluation_{\anode}(\aregisterbis)
        \mid \aregisterbis \in \asetter_j}$.
    \end{description}
\end{enumerate}

\item[\textbf{Case:}
  $\aconcept = \cdrestriction{\forall}{v_1: \arp_1, \dots, v_k: \arp_k}{\acons}$]  

Assume that $\aconcept \in \actype_{\anode}$. 
By the definition of general transitions for
$\aautomaton$ and more specifically
the part about the constraints,
$\acons_{\anode}$ has a conjunct $\acons_{\aconcept}$
of the form
\[
\acons_{\aconcept}
= 
\bigwedge_{(\aregisterbis_1,\dots,\aregisterbis_k) \in \asetter_1 \times \cdots \times \asetter_k}
\acons(\aregisterbis_1,\dots,\aregisterbis_k).
\]
Let \(\avaluation_{\anode} : \registers{\beta}{\degree} \to
\adomain\) be the valuation defined as in the previous
case. 
  Since $\arun$ is a run, we have
  \(\avaluation_{\anode} \models \acons_{\anode}\)
  and therefore for all 
  $(\aregisterbis_1,\dots,\aregisterbis_k) \in
  \asetter_1 \times \cdots \times \asetter_k$,
  we have 
$\avaluation_{\anode} \models
  \acons(\aregisterbis_1,\dots,\aregisterbis_k)$.
  Now, by construction of the $\asetter_j$'s
  and by the interpretation of the concrete features in
  $\ainter$,  for all $j$, we have seen that (\ref{equation-rpj-to-zj-bis})
  holds true.
  Consequently,
  for all $(\avalue_1, \ldots, \avalue_k)
  \in \inter{\arp_1}(\anode) \times \cdots \times
  \inter{\arp_k}(\anode)$,
  we have 
  $[v_1 \mapsto \avalue_1, \dots,
    v_k \mapsto \avalue_k] \models \acons$.
  By the semantics of $\ALCI(\acdomain)$, we get
  $\anode \in \inter{\aconcept}$.
  
    \end{description}

    We have shown that for all \(\anode \in \aidomain\) and for all 
    \(\aconcept \in \subconcepts{\aontology}\), if \(\aconcept \in \actype_{\anode}\), 
    then \(\anode \in \inter{\aconcept}\).
    It remains to verify that the interpretation \(\ainter\) satisfies the 
    ontology \(\aontology= (\atbox, \aabox)\), which can be performed similarly
    to the proof of Lemma~\ref{lemma-encoding-soundness}.
\end{proof}

Similarly, completeness is also shown along the lines of the proof of
Lemma~\ref{lemma-encoding-completeness}.
\begin{lem}[Completeness]
\label{lemma-encoding-completeness-alci}
\(\aontology\)  is consistent with an interpretation satisfying (UNA)
  implies \(\alang(\aautomaton) \neq \emptyset\).
\end{lem}

In the inductive construction of the accepting run,
we need to define and update the parent abstraction $\pabs$, and
when we want to choose an element in the interpretation domain 
$\aidomain$ for an $\arole$-successor, we possibly choose its parent node as the
witness if it is accessible via $\arole$.
\iftoggle{versionlong}{
  \begin{proof} The proof is an adaptation of the proof of
  Lemma~\ref{lemma-encoding-completeness}
  but we provide the details to allow the reader to perform an easy check.
  Nevertheless, there are slight changes everywhere. 
    Let \(\ainter = (\aidomain, \inter{\cdot})\) be an interpretation
    such that $\ainter \models \aontology$
    and $\ainter$ satisfies (UNA) by assumption.
    Below, we define a data tree $\adatatree: \interval{0}{\degree-1}^* \to
    \aalphabet \times \adomain^{\beta}$,
    a partial function $\amapbis: \interval{0}{\degree-1}^* \rightharpoonup \aidomain$,
    a map
    $\arun^{\dag}: \interval{0}{\degree-1}^* \to
    \locations$,
    and a map $\arun: \interval{0}{\degree-1}^* \to
    \transitions$ satisfying the following properties:
    (P1)--(P4) are exactly as in the proof of
    Lemma~\ref{lemma-encoding-completeness} and 
    \begin{description}
    \cut{
    \item[(P1)] The map $\arun$ is an accepting run on $\adatatree$ and consequently,
      $\adatatree \in \alang(\aautomaton)$ and $\alang(\aautomaton) \neq \emptyset$.
    \item[(P2)] For all  $\anode \in \interval{0}{\degree-1}^*$,
      $\amapbis(\anode)$ is defined iff
      $\anode$ is not the root $\varepsilon$
      and $\arun^{\dag}(\anode)$ is different from $\square$.
    \item[(P3)] For all $\anode \in \interval{0}{\degree-1}^*$,
      $\arun(\anode)$  is of the form
      $(\arun^{\dag}(\anode), \aletter, \acons,
      \arun^{\dag}(\anode \cdot 0), \ldots, \arun^{\dag}(\anode \cdot (\degree-1)))$.
    \item[(P4)] For all $\anode \in \interval{0}{\degree-1}^*$ with $\amapbis(\anode)$ defined
      and $j \in \interval{1}{\alpha}$,
      if $\adatatree(\anode) = \pair{\aletter}{\atuple}$,
      and $\inter{\acfeature_j}(\amapbis(\anode))$ is defined,
      then $\atuple[j] = \inter{\acfeature_j}(\amapbis(\anode))$.
    }
    \item[(P5)] For all $\anode \in \interval{0}{\degree-1}^+$
      such that $\arun^{\dag}(\anode)$ is of the
      form $(\pabs, (\actype,  \aactvector))$,
      we have $\localabstraction{\amapbis(\anode)}{\ainter} = (\actype,  \aactvector)$.
    \item[(P6)] For all $\anode, \anodebis \in \interval{0}{\degree-1}^+$
      with $\anode = \anodebis \cdot i$ for some $i \in \interval{0}{\degree-1}$
      such that $\arun^{\dag}(\anode)$ is of the
      form $(\triple{\arole^{\uparrow}}{\actype^{\uparrow}}{\aactvector^{\uparrow}},
      \labs)$, we have
      $\localabstraction{\amapbis(\anodebis)}{\ainter} =
      \pair{\actype^{\uparrow}}{\aactvector^{\uparrow}}$ 
      and 
      $i \in dir(\arole^{\uparrow})$.  
    \end{description}

    By construction of the constraint automaton $\aautomaton$,
    for all the locations $\alocation, \alocation_0, \ldots, \alocation_{\degree-1}$,
    there is at most one constraint $\acons$ such that
    $(\alocation, \aletter, \acons, \alocation_0, \ldots, \alocation_{\degree-1})$
    is a transition in $\transitions$. That is why, the map $\arun$
    is uniquely determined by the map $\arun^{\dag}$. Actually we only need 
    to verify that indeed $\arun$ is a total function in order to satisfy
    the property (P3). The satisfaction of the property (P4) can be easily realised once
    the map $\amapbis$ is defined.
    Similarly, the satisfaction of the property (P5) can also be easily realised
    once the map $\amapbis$ is defined, but one needs to check that
    indeed  $\localabstraction{\amapbis(\anode)}{\ainter}$ is a local abstraction,
    which is not difficult to verify (proof omitted herein).
    The same applies for the satisfaction of (P6).

    We define $\adatatree$, $\amapbis$ and $\arun^{\dag}$ inductively
    on the depth of the nodes.
    For the base case, $\amapbis(\varepsilon)$ is undefined and
    $\arun^{\dag}(\varepsilon) \egdef \rootlocation$. For all $j \in
    \interval{0}{\eta-1}$,
    $\amapbis(j) \egdef \inter{\aindname_j}$ and
    $\arun^{\dag}(j) \egdef \pair{\pair{j}{\amap}}{\localabstraction{\inter{\aindname_j}}{\ainter}}$
    where $\amap$ is defined as follows.
    Assuming that $\pair{\actype_j}{\aactvector_j} = \localabstraction{\inter{\aindname_j}}{\ainter}$
    then $\amap(\aindname_{j}) \egdef \aactvector_j$, for all  $j \in
    \interval{0}{\eta-1}$ (remember that the sets of symbolic links have been
    removed from the local abstractions).
      For all $j \in \interval{\eta}{\degree-1}$,
      $\amapbis(j)$ is undefined and
      $\arun^{\dag}(j) = \square$.

      Let us handle now the induction step for the nodes at depth at least two.

      \proofsubparagraph{Induction hypothesis}
      For all nodes \(\anode\) at depth at most \(D \geq 1\),
      $\arun^{\dag}(\anode)$ is defined, and 
      if $\amapbis(\anode)$ is defined and
      $\arun^{\dag}(\anode) =
      (\pabs_{\anode}, (\actype_{\anode}, 
      \aactvector_{\anode}))$,
      then
      $(\actype_{\anode},  \aactvector_{\anode})
      = \localabstraction{\amapbis(\anode)}{\ainter}
      $.
      Moreover, if $\anode = \anodebis \cdot  i$ for some $i$
      and the depth of $\anodebis$ is at least one with
      $\arun^{\dag}(\anodebis)$ of the form  
      $(\pabs_{\anodebis}, (\actype_{\anodebis}, 
      \aactvector_{\anodebis}))$
      and $\pabs_{\anode} =
      \triple{\arole^{\uparrow}_{\anode}}{\actype^{\uparrow}_{\anode}}{
        \aactvector_{\anode}^{\uparrow}}$,
      then $\pair{\actype^{\uparrow}_{\anode}}{\aactvector_{\anode}^{\uparrow}}
      = \pair{\actype_{\anodebis}}{\aactvector_{\anodebis}}$
      and $i \in dir(\arole^{\uparrow}_{\anode})$.
      Moreover, if $\anode = j$ for some $j \in \interval{0}{\eta-1}$,
      $\pabs_{\anode} = \pair{j}{\amap}$.

      \proofsubparagraph{Inductive construction}
      Let \(i \in \interval{0}{\degree-1}\) such that
      the depth of $\anode \cdot i$ is at most $D +1$,
      and $\anode \cdot i$ is the smallest element in
      $\interval{0}{\degree-1}^{+}$ with respect to
      the lexicographical ordering such that $\arun^{\dag}(\anode \cdot i)$ is not yet defined.
      If $\arun^{\dag}(\anode) = \square$, then
      $\amapbis(\anode \cdot i)$ is undefined 
      and $\arun^{\dag}(\anode \cdot i) \egdef \square$. Otherwise, we proceed
      by a case analysis.

      \begin{itemize}

      \item If \(\lambda^{-1}(i) = \exrest{\arole}{\aconcept}\)
        and $\exrest{\arole}{\aconcept} \not \in \actype_{\anode}$,
        then $\amapbis(\anode \cdot i)$ is undefined 
        and $\arun^{\dag}(\anode \cdot i) \egdef \square$.

      \item If \(\lambda^{-1}(i) = \exrest{\arole}{\aconcept}\)
        and $\exrest{\arole}{\aconcept} \in \actype_{\anode}$,
        then $\amapbis(\anode) \in \inter{(\exrest{\arole}{\aconcept})}$ by the induction hypothesis.
        There is \(\aind \in \aidomain\) such that
        \(\pair{\amapbis(\anode)}{\aind} \in \inter{\arole}\) and 
        \(\aind \in \inter{\aconcept}\).
       We pick an arbitrary such a witness $\aind$,
        and $\amapbis(\anode \cdot i) \egdef \aind$ and
        $\arun^{\dag}(\anode \cdot i) \egdef
        \pair{\triple{\arole}{\actype_{\anode}}{\aactvector_{\anode}}}{\localabstraction{\aind}{\ainter}}$. 
        
      \item  If \(\lambda^{-1}(i) =
        \pair{\aconcept^{\star}
        }{j}\)
        and ($\aconcept^{\star} \not \in \actype_{\anode}$ or $j > k$)
        with $\aconcept^{\star} = \cdrestriction{\exists}{v_1: \arolepath_1, \ldots, v_k: \arolepath_k}{\acons}$,
       then $\amapbis(\anode \cdot i)$ is undefined 
       and $\arun^{\dag}(\anode \cdot i) \egdef \square$.

     \item If \(\lambda^{-1}(i) = \pair{\aconcept^{\star}
       }{j}\), 
        $\aconcept^{\star} \in \actype_{\anode}$ and $j \leq k$
        with $\aconcept^{\star} = \cdrestriction{\exists}{v_1: \arolepath_1, \ldots, v_k: \arolepath_k}{\acons}$,
        then $\amapbis(\anode) \in \inter{(\aconcept^{\star})}$ by the induction hypothesis.
        By the semantics of $\ALCI(\acdomain)$,
        there is a tuple \((\avalue_1, \dots, \avalue_k) \in \inter{\arp_1} \times \cdots \times
        \inter{\arp_k}\) such that
        \[
          [v_1 \mapsto \avalue_1, \ldots, v_k \mapsto \avalue_k]
          \models \acons. 
          \]
          We pick an arbitrary such a tuple. 
          For all $j \in \interval{1}{k}$ such that
          $\arp_j$ is of length two of the form
          $\arole_j' \cdot \acfeature'_j$,
          there is $\aind_j \in \aidomain$ such that
          \(\pair{\amapbis(\anode)}{\aind_j} \in \inter{(\arole'_j)}\) and
          $\inter{\acfeature_j'}(\aind_j) = \avalue_j$.
          For all $i'$ such that
          \(\lambda^{-1}(i') = \pair{\cdrestriction{\exists}{v_1:
              \arolepath_1, \ldots, v_k: \arolepath_k}{\acons}}{j}\) for such $j$'s,
          $\amapbis(\anode \cdot i') \egdef \aind_j$ and 
            $\arun^{\dag}(\anode \cdot i') \egdef
            \pair{\triple{\arole}{\actype_{\anode}}{\aactvector_{\anode}}}{\localabstraction{\aind_j}{
                \ainter}}$.
          For all the other $i'$, $\amapbis(\anode \cdot i')$ is undefined 
          and $\arun^{\dag}(\anode \cdot i') \egdef \square$.
          The case $i' = i$ has been necessarily handled.
          We need to go through all the indices $i'$ (and not to consider $i$ alone)
          because the same tuple $(\avalue_1, \dots, \avalue_k)$ needs to be considered.

    \end{itemize}

      By the construction of the constraint automaton $\aautomaton$, and in particular
      the part about constraints, for
      all $\anode \in \interval{0}{\degree-1}^*$,
      there is exactly one constraint $\acons_{\anode}$
      such that 
      $(\arun^{\dag}(\anode), \aletter, \acons_{\anode},
      \arun^{\dag}(\anode \cdot 0), \ldots, \arun^{\dag}(\anode \cdot (\degree-1)))$
      is a transition of $\aautomaton$ and this is precisely
      how $\arun(\anode)$ is defined.
      This holds mainly because we took care to propagate the
      location $\square$ as soon as it appears on a node.

      \proofsubparagraph{Data tree $\adatatree$}
     It remains to define the 
    data tree \(\adatatree :
    \interval{0}{\degree-1}^* \to \aalphabet \times \adomain^\beta\)
    such that for all node $\anode$,
    \(\adatatree(\anode) \egdef (\aletter, \atuple_{\anode})\) with $\atuple_{\anode}$
    defined as follows ($\aletter$ is the unique letter of the alphabet).
    We recall that 
the distinguished registers $\aregister_1, \ldots, \aregister_{\beta}$
referring to values at the current node shall be represented by
$
\aregister_1, \ldots, \aregister_{\alpha},
\aregister_1^{\uparrow}, \ldots, \aregister_{\alpha}^{\uparrow}, 
\aregister_{\aindname_0, 1}, \ldots,\aregister_{\aindname_0,\alpha},
\ldots,
\aregister_{\aindname_{\eta-1}, 1}, \ldots,\aregister_{\aindname_{\eta-1},\alpha}
$.
    \begin{itemize}
    \item $\adatatree(\varepsilon) \egdef \pair{\aletter}{\atuple}$
      for some arbitrary tuple $\atuple$. 
    \item For all $j \in \interval{1}{\alpha}$,
      if $\inter{\acfeature_j}(\amapbis(\anode))$ is undefined,
      then $\atuple_{\anode}[j]$ is arbitrary. Otherwise, 
      $\atuple_{\anode}[j] \egdef \inter{\acfeature_j}(\amapbis(\anode))$.
    \item If $\anodebis$ is the parent of $\anode$ and $\anodebis \neq \varepsilon$, then
      for all $j \in \interval{1}{\alpha}$,
      if $\inter{\acfeature_j}(\amapbis(\anodebis))$ is undefined or  $\anodebis = \varepsilon$,
      then $\atuple_{\anode}[\alpha+ j]$ is arbitrary.
      Otherwise, $\atuple_{\anode}[\alpha + j] \egdef
      \inter{\acfeature_j}(\amapbis(\anodebis))$.
    \item For all $j \in \interval{1}{\alpha}$ and $i \in \interval{0}{\eta-1}$,
      if $\inter{\acfeature_j}(\inter{\aindname_i})$ is undefined,
      then $\atuple_{\anode}[\alpha \cdot (i+2) + j]$ is arbitrary.
      Otherwise, $\atuple_{\anode}[\alpha \cdot (i+2) + j] \egdef
      \inter{\acfeature_j}(\inter{\aindname_i})$.
      These values are helpful only when $\anode \in \interval{0}{\eta-1}$. 
    \end{itemize}

    \proofsubparagraph{The map $\arun$: an accepting run on $\adatatree$} It remains to show that $\arun$ is an
    accepting run on the data tree $\adatatree$. We have to show that
    indeed (a) $\arun$ is a map from $\interval{0}{\degree-1}^*$ to $\transitions$,
    (b) $\arun^{\dag}(\varepsilon) \in \locations_{in}$,
    (c) $\arun$ is accepting and the condition (ii) for runs holds.
    The satisfaction of (b) is immediate from the base case above
    and (c) is straightforward because $F = \locations$.

    In order to check the property (a), we go through all the values $\arun(\anode)$ and check
    that it satisfies the conditions for defining transitions.
    This checking does not require the data values from $\adatatree$, unlike the verification of the
    condition (ii). Furthermore, if the conditions
    on $\arun^{\dag}(\anode), \arun^{\dag}(\anode \cdot 0), \ldots,\arun^{\dag}(\anode \cdot (\degree-1))$
    are satisfied, then the definition of $\acons_{\anode}$ is uniquely determined.
   
    \begin{itemize}
    \item If $\arun^{\dag}(\anode) = \square$, then
      we have seen that
      $\arun^{\dag}(\anode \cdot 0) = \cdots =\arun^{\dag}(\anode \cdot (\degree-1)) = \square$
      by definition of $\arun^{\dag}$,
      which allows us to use the padding transition.

    \item We have $\arun(\varepsilon)
      = (\rootlocation, \aletter, \acons_{\mathsf{root}},
      \pair{\pair{0}{\amap}}{\localabstraction{\inter{\aindname_0}}{\ainter}}, \ldots,
      \pair{\pair{\eta-1}{\amap}}{\localabstraction{\inter{\aindname_{\eta-1}}}{\ainter}},
      \square, \ldots, \square)$, which guarantees that all the requirements
      for the root transitions are satisfied. The details are omitted.
      By way of example, suppose that
      $\rassertion{\aindname_{\gamma}}{\aindname_{\gamma'}}{\arolename}
      \in \aabox$ and $\varest{\arolename^{-}}{\aconcept'} \in \actype_{\gamma'}$.
      We have $\pair{\inter{\aindname_{\gamma'}}}{\inter{\aindname_{\gamma}}} \in \inter{(\arolename^{-})}$
      and $\inter{\aindname_{\gamma'}} \in \inter{(\varest{\arolename^{-}}{\aconcept'})}$
      ($\ainter$ satisfies $\aontology$).
      Consequently,   $\inter{\aindname_{\gamma}} \in \inter{\aconcept'}$
      by the semantics of $\ALCI(\acdomain)$ 
      and we can conclude $\aconcept' \in \actype_{\gamma}$ by definition of $\actype_{\gamma}$.

    \item If $\arun^{\dag}(\anode)$ is
      of the form
      $\pair{\pabs}{\pair{\actype_{\anode}}{\aactvector_{\anode}}}$,
      then let us check that
      the conditions
    on $\arun^{\dag}(\anode), \arun^{\dag}(\anode \cdot 0), \ldots,\arun^{\dag}(\anode \cdot (\degree-1))$
    are satisfied.
    \begin{description}
    \item[Requirements from existential restrictions]
      Let $\exrest{\arole}{\aconcept} \in \actype_{\anode}$ with
      $i = \lambda(\exrest{\arole}{\aconcept})$.
      We have seen that either $\arun^{\dag}(\anode \cdot i) =
      \pair{\pabs_{\anode \cdot i}}{\pair{\actype_{\anode \cdot i}}{
          \aactvector_{\anode \cdot i}}}$ and $\aconcept \in
      \actype_{\anode \cdot i}$
      or  $\inverse{\arole} = \arole^{\uparrow}_{\anode}$ and
      $\aconcept \in \actype^{\uparrow}_{\anode}$ with $\arun^{\dag}(\anode) =
      \pair{\triple{\arole^{\uparrow}_{\anode}}{\actype_{\anode}^{\uparrow}}{\aactvector_{\anode}^{\uparrow}
      }}{\labs_{\anode}}$. 
      This is sufficient to satisfy these requirements.

    \item[Requirements from value restrictions] Similar to the previous checking.
    \item[Requirements from back-propagation]
     Let $i \in \interval{0}{\degree-1}$, 
     \(\alocation_{\anode \cdot i} = (\pabs_{\anode \cdot i},  (\actype_{\anode \cdot i},
     \aactvector_{\anode \cdot i}))\),
     $\varest{\arole}{\aconcept} \in \actype_{\anode \cdot i}$ and
     $i \in dir(\inverse{\arole})$.
     By definition of $\inter{\arole}$, $\pair{\amapbis(\anode)}{\amapbis(\anode \cdot i)}
     \in \inter{(\inverse{\arole})}$.
     By construction, $\amapbis(\anode \cdot i) \in
     \inter{(\varest{\arole}{\aconcept})}$ and therefore
     $\anode \in \inter{\aconcept}$.
     By construction, $\aconcept \in \actype_{\anode}$. 
    
  \item[Requirements from inactive children] For each $i \in \interval{0}{\degree-1}$,
    the requirements are satisfied and this can be read from
    the cases
    (\(\lambda^{-1}(i) = \exrest{\arolename}{\aconcept}\)
    and $\exrest{\arolename}{\aconcept} \not \in \actype_{\anode}$)
    or (\(\lambda^{-1}(i) =
        \pair{\cdrestriction{\exists}{v_1: \arolepath_1, \ldots, v_k: \arolepath_k}{\acons}}{j}\)
        and ($\lambda^{-1}(i) \not \in \actype_{\anode}$ or $j > k$)).
      \item[Requirements for parent information propagation]
        Let $i \in \interval{0}{\degree-1}$ such that \(\alocation_i\) is a contextual abstraction
        $(\triple{\arole^{\uparrow}_{\anode \cdot i}}{\actype^{\uparrow}_{\anode \cdot i}}{
          \aactvector^{\uparrow}_{\anode \cdot i}},
        (\actype_{\anode \cdot i},  \aactvector_{\anode \cdot i}))\).
         By definition of $\inter{\arole}$, $\pair{\amapbis(\anode)}{\amapbis(\anode \cdot i)}
         \in \inter{\arole}$. By construction, $i \in dir(\arole_{\anode \cdot i}^{\uparrow})$.
         By using the basic properties of the construction,
         we can also show that
         $\pair{\actype_{\anode}}{\aactvector_{\anode}}
    = \pair{\actype^{\uparrow}_{\anode \cdot i}}{\aactvector^{\uparrow}_{\anode \cdot i}}$.
    \end{description}
    
    \end{itemize}

   It remains to check the condition (ii), which is recalled below. 
   \begin{romanenumerate}
   \item[(ii)] For all nodes $\anode \in \interval{0}{\degree-1}^*$
     with $\adatatree(\anode) = \pair{\aletter}{\atuple_{\anode}}$,
     we define the valuation 
    \(\avaluation_{\anode} : \registers{\beta}{\degree} \to
  \adomain\) such that
  \(
  \avaluation_{\anode}(\aregister_j) = \atuple_{\anode}[j]
  \), 
  $\avaluation_{\anode}(\aregister_j^{\uparrow}) = \atuple_{\anode}[\alpha+j]$,
  $\avaluation_{\anode}(\aregister_j^i) = \atuple_{\anode \cdot i}[j]$
  and
  $\avaluation_{\anode}(\aregister_j^{\uparrow \ i}) = \atuple_{\anode}[j]$
  (equal to $\atuple_{\anode}[j]$)
  for all \(1 \le j \le \alpha,\ 0 \le i < \degree\).
  Furthermore,
  for all $j \in \interval{0}{\eta-1}$ and $\ell \in \interval{1}{\alpha}$,
  $\avaluation_{\anode}(\aregister_{\aindname_{j},\ell}) =
  \atuple_{\anode}[\alpha \times (j+2) +\ell]$
  and for all $i < \degree$, $\avaluation_{\anode}(\aregister_{\aindname_{j},\ell}^i) =
  \atuple_{\anode}[\alpha \times (j+2) +\ell]$ too (these last values are never
  used). 
  Then
  \(\avaluation_{\anode} \models \acons_{\anode}\), where $\acons_{\anode}$
  is the constraint of the transition $\arun(\anode)$.
   \end{romanenumerate}
   If $\arun^{\dag}(\anode) = \square$, then $\acons_{\anode} = \top$ and
   the property obviously holds.
   If $\arun^{\dag}(\anode) = \rootlocation$ (i.e. $\anode = \varepsilon$),
   then $\avaluation_{\anode} \models \acons_{\anode}$ for the same reason
   as in the proof of Lemma~\ref{lemma-encoding-completeness}.
   
   If $\arun^{\dag}(\anode)$ is of the
   form $\pair{\pabs_{\anode}}{\pair{\actype_{\anode}}{\aactvector_{\anode}}}$,
   with 
   $\arun(\anode) = 
   (\arun^{\dag}(\anode), \aletter, \acons_{\anode},
   \arun^{\dag}(\anode \cdot 0), \ldots, \arun^{\dag}(\anode \cdot (\degree-1)))$, then
   $\acons_{\anode}$ is equal to
   \[
   \acons_\atransition =
\left(\bigwedge_{\aconcept \in \mathrm{CD}_\exists(\actype_{\anode}) \cup \mathrm{CD}_\forall(\actype_{\anode})}
\acons_\aconcept\right)
\wedge
\left(\bigwedge_{\substack{i \in \interval{0}{\degree-1} \\ \alocation_{\anode \cdot i} \neq \square}}
\ \ 
\bigwedge_{j \in \interval{1}{\alpha}, \aactvector_{\anode}[j]=\top}
\aregister_j^{\uparrow \ i} = \aregister_j \right).
\]

Let us show that $\avaluation_{\anode} \models \acons_\atransition$.
\begin{itemize}
\item Let $i \in \interval{0}{\degree-1}$ such that
  $\arun^{\dag}(\anode \cdot i) \neq \square$,
  $j \in \interval{1}{\alpha}$, and  $\aactvector_{\anode}[j] = \top$.
  By definition of $\avaluation_{\anode}$,
  $\avaluation_{\anode}(\aregister_{j}) =
  \atuple_{\anode}[j]$.
  By definition of $\avaluation_{\anode}$,
  $\avaluation_{\anode}(\aregister_{j}^{\uparrow \ i}) =
  \atuple_{\anode}[j]$ too.
 
\item Let $\aconcept \in \mathrm{CD}_\exists(\actype_{\anode})$ with
  $\aconcept = \cdrestriction{\exists}{v_1: \arp_1, \dots, v_k: \arp_k}{\acons}$.
  Let us show that
  \[
  \avaluation_{\anode} \models
  \bigvee_{(\aregisterbis_1,\dots,\aregisterbis_k) \in \asetter_1 \times \cdots \times \asetter_k}
  \acons(\aregisterbis_1,\dots,\aregisterbis_k) .
  \]
  We have seen earlier that we have picked
  a tuple \((\avalue_1, \dots, \avalue_k) \in \inter{\arp_1} \times \cdots \times
        \inter{\arp_k}\) such that
        \[
          [v_1 \mapsto \avalue_1, \ldots, v_k \mapsto \avalue_k]
          \models \acons,
          \]
  from which $\arun^{\dag}(\anode \cdot i)$ is defined for all $i \in \interval{0}{\degree-1}$. 
 For all $j \in \interval{1}{k}$ such that
          $\arp_j$ is of length two of the form
          $\arole_j' \cdot \acfeature'_j$,
          there is $\aind_j \in \aidomain$ such that
          \(\pair{\amapbis(\anode)}{\aind_j} \in \inter{(\arole'_j)}\) and
          $\inter{\acfeature_j'}(\aind_j) = \avalue_j$.
          For all $i'$ such that
          \(\lambda^{-1}(i') = \pair{\cdrestriction{\exists}{v_1:
              \arolepath_1, \ldots, v_k: \arolepath_k}{\acons}}{j}\) for such $j$'s,
          $\amapbis(\anode \cdot i') \egdef \aind_j$ and 
          $\arun^{\dag}(\anode \cdot i) \egdef \pair{\pabs_i}{\localabstraction{\aind_j}{\ainter}}$.
          For all the other $i'$, $\amapbis(\anode \cdot i')$ is undefined 
          and $\arun^{\dag}(\anode \cdot i') \egdef \square$.
          Let us define a tuple $(\aregisterbis_1, \ldots, \aregisterbis_k)
          \in  \asetter_1 \times \cdots \times \asetter_k$
          such that $(\avaluation_{\anode}(\aregisterbis_1), \ldots,
          \avaluation_{\anode}(\aregisterbis_k)) = (\avalue_1, \dots, \avalue_k)$,
          which allows us to conclude that
          $\avaluation_{\anode} \models
  \bigvee_{(\aregisterbis_1,\dots,\aregisterbis_k) \in \asetter_1 \times \cdots \times \asetter_k}
  \acons(\aregisterbis_1,\dots,\aregisterbis_k)$.

  Let $\gamma \in \interval{1}{k}$.
  If $\arp_{\gamma} = \acfeature_{\ell}$ for some $\ell$, then
  let us take $\aregisterbis_{\gamma} = \aregister_{\ell}$
  and $\avaluation_{\anode}(\aregister_{\ell}) = \atuple_{\anode}[\ell]$
  by definition of $\avaluation_{\anode}$.
  Moreover, $\avalue_{\gamma} = \inter{\acfeature_{\ell}}(\amapbis(\anode))$
  by the semantics of $\ALCI(\acdomain)$.
    By definition of $\atuple_{\anode}[\ell]$, we have
    $\atuple_{\anode}[\ell] = \inter{\acfeature_{\ell}}(\amapbis(\anode))$.
  Consequently, $\avaluation_{\anode}(\aregisterbis_{\gamma}) = \avalue_{\gamma}$. 
  Finally, since $\inter{\acfeature_{\ell}}(\amapbis(\anode))$ is defined,
  $\aactvector_{\anode}[\ell] = \top$ and
  therefore $\aregisterbis_{\gamma} \in \asetter_{\gamma}$.

  If $\arp_{\gamma} = \arole_{\ell'} \cdot \acfeature_{\ell}$
  for some $\ell, \ell'$, then we perform a case analysis to define
  $\aregisterbis_{\gamma}$. Let $\anodebis$ be the parent of $\anode$. 
\begin{description}
  \item[Subcase $\anodebis \neq \varepsilon$]
  If $\pair{\amapbis(\anode)}{\aind_{\gamma}} \in
  \inter{(\arole_{\ell'}')}$ and $\aind_{\gamma} = \amapbis(\anodebis)$,
  then $\aregisterbis_{\gamma} = \aregister_{\ell}^{\uparrow}$.
  Otherwise ($\aind_{\gamma} \neq \amapbis(\anodebis)$),
  $\aregisterbis_{\gamma} = \aregister_{\ell}^i$ with
  $\lambda^{-1}(i) = \pair{\aecdconcept}{\gamma}$. 
  Moreover, $\avalue_{\gamma} = \inter{\acfeature_{\ell}}(\aind_{\gamma})$ by definition of
  $\avalue_{\gamma}$ and
  $\amapbis(\anode \cdot i) = \aind_{\gamma}$ by definition of $\amapbis$.
  We have $\avaluation_{\anode}(\aregister_{\ell}^i) = \atuple_{\anode \cdot i}[\ell]$
  by definition of $\avaluation_{\anode}$
  and $\atuple_{\anode \cdot i}[\ell] = \inter{\acfeature_{\ell}}(\amapbis(\anode \cdot i))$
  by definition of $\atuple_{\anode \cdot i}$. Consequently,
  $\avalue_{\gamma} = \avaluation_{\anode}(\aregisterbis_{\gamma})$.
  Since $\amapbis(\anode \cdot i)$ is defined,
  $\arun^{\dag}(\anode \cdot i) \neq \square$,
  $i \in dir(\arolename_{\ell'})$ and $\aactvector_{\anode \cdot i}[\ell] = \top$,
  we get $\aregisterbis_{\gamma} \in \asetter_{\gamma}$. 
  
\item[Subcase $\anodebis = \varepsilon$] Suppose $\anode = \inter{\aindname_{G}}$
  for some $G \in \interval{0}{\eta-1}$.
  If $\aind_{\gamma} \in \set{\inter{\aindname_0}, \ldots,\inter{\aindname_{\eta-1}}}$,
  then let us take  $\aregisterbis_{\gamma} = \aregister_{\aindname_{H},\ell}$
  for some $H$ such that $\aind_{\gamma} = \inter{\aindname_{H}}$. 
  Moreover, $\avalue_{\gamma} = \inter{\acfeature_{\ell}}(\inter{\aindname_H})$
  by the semantics of $\ALCI(\acdomain)$.
  By definition of $\avaluation_{\anode}$,
  $\avaluation_{\anode}(\aregister_{\aindname_{H},\ell}) = \atuple_{\anode}[\alpha \times (H+2) + \ell]$.
  By definition of $\atuple_{\anode}$,
  $\atuple_{\anode}[\alpha \times (H+2) + \ell] =
  \inter{\acfeature_{\ell}}(\inter{\aindname_H})$.
  Consequently, $\avaluation_{\anode}(\aregisterbis_{\gamma}) = \avalue_{\gamma}$. 
  Since  \(\pair{\amapbis(\anode)}{\aind_{\gamma}} \in \inter{(\arole_{\ell'})}\),
  $\inter{\acfeature_{\ell}}(\aind_{\gamma})$ is defined
  and $\amap(\aindname_{H})[\ell]=\top$,
  we get
  $\rassertion{\aindname_{G}}{\aindname_{H}}{\arole_{\ell'}} \in \aabox$
  and therefore $\aregisterbis_{\gamma} \in \asetter_{\gamma}$.
  Otherwise ($\aind_{\gamma} \not
  \in \set{\inter{\aindname_0}, \ldots,\inter{\aindname_{\eta-1}}}$), the treatment is exactly as
  in the previous subcase. 
 
  \end{description}

This completes the case 
  $\aconcept = \cdrestriction{\exists}{v_1: \arp_1, \dots, v_k: \arp_k}{\acons}$
  and the proof of 
  \[
  \avaluation_{\anode} \models
  \bigvee_{(\aregisterbis_1,\dots,\aregisterbis_k) \in \asetter_1 \times \cdots \times \asetter_k}
  \acons(\aregisterbis_1,\dots,\aregisterbis_k).
  \]

\item Let $\aconcept \in \mathrm{CD}_\forall(\actype_{\anode})$ with
  $\aconcept = \cdrestriction{\forall}{v_1: \arp_1, \dots, v_k: \arp_k}{\acons}$.
  In order to show that
  \[
  \avaluation_{\anode} \models
  \bigwedge_{(\aregisterbis_1,\dots,\aregisterbis_k) \in \asetter_1 \times \cdots \times \asetter_k}
  \acons(\aregisterbis_1,\dots,\aregisterbis_k),
  \]
  we proceed similarly to the
  previous case by performing a case analysis on the elements in
  the $\asetter_{\gamma}$'s. The details are omitted herein. \qedhere
\end{itemize}  
\end{proof}

}{
The tedious proof is in Appendix~\ref{appendix-proof-lemma-encoding-completeness-alci}.
  As far as  complexity is concerned, the main differences
with the construction of $\aautomaton$ for $\ALCO(\acdomain)$ include
$\beta$  still bilinear in $\alpha$ and $\eta$,
and the presence of parent abstractions.
All of this causes no harm and we can establish 
Theorem~\ref{theorem-ALCI}.
}
\cut{
As for the part of completeness proof, in the inductive construction part,
we need to define and update the parent information $\parentinfo$, and
when we want to choose an element in the domain 
$\aidomain$ for an $\arolename^{-1}$-successor, we choose its parent node as the
witness if it is accessible via $\arolename^{-1}$ and set the child node from
$\dir{\arolename^{-1}}$ to $\square$, otherwise it follows the same construction
as stated in the previous completeness proof
(witness via a child node or symbolic links).
For concrete domain restrictions involving inverse feature paths, the register 
$\aregister_j^{\uparrow}$ at node $\anode$ is set to 
$\inter{\arp_j}(\aind_\anodebis)$ if the parent node $\anodebis$
is accessible via $\arolename^{-1}$, ensuring the constraint formula is 
satisfied.
All other valuations and verifications proceed exactly as in the basic case.
}

\paragraph*{Short complexity analysis.}
As far as the computational complexity is concerned, the main differences
with the construction of $\aautomaton$ for $\ALCO(\acdomain)$ include
a  value for $\beta$ that is still bilinear in $\alpha$ and $\eta$,
and the presence of parent abstractions in locations.
This latter ingredient  amounts
to multiply the cardinal of the set of locations for
$\ALCO(\acdomain)$ by a factor exponential in $\size{\aontology}$,
which preserves a final set of locations only exponential in $\size{\aontology}$.
As global abstractions have been removed, the number of locations might be further
divided by an exponential factor. 
This allows us to establish Theorem~\ref{theorem-ALCI}
that is one of the main results.

\begin{thm} \label{theorem-ALCI} 
The consistency problem for $\ALCI(\acdomain)$ is in $k$-\exptime 
if $\acdomain$ satisfies the conditions (C1), (C2), (C3.k) and (C4), for any $k \geq 1$.
\end{thm}

As shown herein, our developments for $\ALCO(\acdomain)$
can be naturally adapted to $\ALCI(\acdomain)$, witnessing the robustness
of our approach. It is open whether there is an adaptation
for the extension $\ALCOI(\acdomain)$ with inverse role names
{\em and} nominals. Indeed, in $\ALCOI(\Rat, <)$, the ontology
\[
\aontology
=
\pair{
  \set{
  \top \sqsubseteq \cdrestriction{\exists}{
    v_1: \acfeature_1, v_2: \arolename \cdot \acfeature_1}{v_1 < v_2},
  \top \sqsubseteq \exrest{\arolename^{-}}{\anominal}
  }
}{
  \set{
    \cassertion{\aindname}{\cdrestriction{\forall}{
    v_1: \acfeature_1, v_2: \arolename \cdot \acfeature_1}{\acons}}
    }
  },
\]
can be made consistent and for any interpretation $\ainter$ satisfying it,
$\inter{\aindname}$ has an {\em infinite amount} of $\arolename$-successors, which cannot be
captured currently with the finite sets $\asetter_j$.
It is not so much the combination of inverse roles and nominals that is problematic,
see e.g. the handling of the $\mu$-calculus with converse and nominals in~\cite{Sattler&Vardi01},
but the additional presence of CD-restrictions.

\subsection{The Case of Integers}
\label{section-integers}

We have seen that the concrete domain $\triple{\Zed}{<}{=,(=_{z})_{z \in \Zed}}$
(below called $\Zed$ for simplicity) does not satisfy
the condition~(C1) and therefore $\Zed$ is not in the scope of Theorem~\ref{theorem-tgca}.
However, the consistency problem for $\ALCO(\Zed)$ can be reduced to the non-emptiness
problem for TGCA over 
$\Zed$ since it satisfies the condition~(C4), i.e.
it contains the equality relation, see Section~\ref{section-encoding-consistency}.

To establish the \exptime-membership for the consistency problem for $\ALCO(\Zed)$, it remains
to characterise the computational complexity of the nonemptiness problem for
TGCA over $\Zed$ without taking advantage of the results from
Section~\ref{section-tgca}. 
Fortunately, a {\em subclass} of TGCA is investigated in~\cite[Lemma~4.15]{Demri&Quaas25}
for which the nonemptiness problem is shown in \exptime. Below, we explain how a polynomial-time
reduction can be easily designed from  the general class of TGCA to that subclass of TGCA over $\Zed$.

A TGCA \(\aautomaton = (\locations, \aalphabet, \degree, \beta, \locations_{in}, \transitions, F)\)
is \defstyle{without sibling constraints} if all its transitions are
made of $\Zed$-constraints of the form $\acons_0 \wedge \cdots \wedge \acons_{\degree-1}$
where each $\Zed$-constraint $\acons_k$ is built over
the only registers $\aregister_1, \ldots, \aregister_{\beta},
\aregister_1^k, \ldots, \aregister_{\beta}^k$. Obviously, no constraints are expressed between
the siblings. By~\cite[Lemma~4.15]{Demri&Quaas25}, the nonemptiness problem
for TGCA without sibling constraints (over $\Zed$) can be solved in time
\[
  R_1\big(\card{\locations} \times \card{\delta}
  \times \maxconstraintsize{\aautomaton} \times
  \card{\aalphabet} \times R_2(\beta) \big)^{\mathcal{O}(R_2(\beta) \times R_3(\degree))},
  \]
  for some polynomials $R_1$, $R_2$ and $R_3$.

  Let us define a simple polynomial-time reduction from
  the general class of TGCA to the class of TGCA without sibling constraints.
  The basic idea consists in duplicating the registers of the children nodes
  and to express the constraints at the level of the parent nodes.
  The number of registers and the maximal size of the constraints increase polynomially (only),
  which is fine to establish the forthcoming \exptime-membership result. 

  Let \(\aautomaton = (\locations, \aalphabet, \degree, \beta, \locations_{in}, \transitions, F)\)
  be a TGCA over $\Zed$.
  Let us build a TGCA without sibling constraints
  \(\aautomaton' = (\locations, \aalphabet, \degree, \beta', \locations_{in}, \transitions', F)\)
  such that
  $\alang(\aautomaton) \neq \emptyset$ iff
$\alang(\aautomaton') \neq \emptyset$.
  \begin{itemize}
  \item $\beta' \egdef \beta + \degree \times \beta$.  Indeed,
    we add $\degree \times \beta$ registers whose values anticipate the register values
    of the children nodes.
    The registers are written
    \[
    \aregister_1, \ldots, \aregister_{\beta},
    \aregister_1^{child(0)}, \ldots, \aregister_{\beta}^{child(0)},
    \ldots, \aregister_1^{child(\degree-1)}, \ldots, \aregister_{\beta}^{child(\degree-1)} .
    \]
    The registers for the child $i$ are written
    \[
    \aregister_1^i, \ldots, \aregister_{\beta}^i,
    \aregister_1^{child(0),i}, \ldots, \aregister_{\beta}^{child(0),i},
    \ldots, \aregister_1^{child(\degree-1),i}, \ldots, \aregister_{\beta}^{child(\degree-1),i} .
    \]
  \item For each \((\alocation, \aletter, \acons, \alocation_0, \dots, \alocation_{\degree-1}) 
    \in \transitions\), we consider in $\transitions'$ the
    transition \((\alocation, \aletter, \acons', \alocation_0, \dots, \alocation_{\degree-1})$
    where $\acons'$ is defined as the conjunction $\acons_0 \wedge \cdots \wedge \acons_{\degree-1}$
    and each $\acons_k$ is made of the $\Zed$-constraints below.
    \begin{itemize}
    \item One can express that
      the values for  $\aregister_1^{child(k)}, \ldots, \aregister_{\beta}^{child(k)}$ anticipate
      the values of the $k$th child:
      \[
      \bigwedge_{\alpha \in \interval{1}{\beta}} \aregister_{\alpha}^{child(k)} = \aregister_{\alpha}^{k}.
      \]
    \item Similarly, expressing $\acons$ at the level of the parent node is possible by considering
      the constraint  obtained from $\acons$ by substituting each register $\aregister_{j}^i$
      ($1 \leq j \leq \beta$, $0 \leq i < \degree$) by $\aregister_{j}^{child(i)}$. 
    \end{itemize}
    The constraint $\acons_0 \wedge \cdots \wedge \acons_{\degree-1}$
    has the right shape to belong to a TGCA without sibling constraints.
  \end{itemize}
Observe that the maximal constraint size increases only by $\beta \times \degree$. 
It is then not difficult to show that $\alang(\aautomaton) \neq \emptyset$ iff
$\alang(\aautomaton') \neq \emptyset$
because essentially we have preserved the constraints while anticipating the register values for the
children.  The data trees in $\alang(\aautomaton)$ are exactly those obtained
from the ones in $\alang(\aautomaton')$ in which the extra  $\degree \times \beta$
data values per node are removed. 
Consequently, the nonemptiness problem for TGCA over $\Zed$ can be solved in time
\[
  R_1\big(\card{\locations} \times \card{\delta}
  \times (\maxconstraintsize{\aautomaton} + \beta \times \degree) \times
  \card{\aalphabet} \times R_2(\beta \times (\degree+1)) \big)^{\mathcal{O}(R_2(\beta \times (\degree+1))
    \times R_3(\degree))}.
  \]

By taking advantage of the developments in Section~\ref{section-encoding-consistency}, we can conclude. 
  
\begin{thm}\label{theorem-integers}
The consistency problem for $\ALCO(\Zed)$ is \exptime-complete. 
\end{thm}

We can also conclude that the consistency problem for $\ALCO(\Nat)$ is \exptime-complete
with the concrete domain $\triple{\Nat}{<}{=,(=_{n})_{n \in \Nat}}$.
Furthermore, one may wonder how Theorem~\ref{theorem-integers}  differs from~\cite[Theorem~28]{Labai&Ortiz&Simkus20}.
Theorem~\ref{theorem-integers}  is definitely a new result as~\cite{Labai&Ortiz&Simkus20}
handles neither assertion boxes  nor nominal concepts.
By contrast,  the description logics
in~\cite{Labai&Ortiz&Simkus20} natively admit
functional role names, which we can also add. Actually, the main difference
rests on the type of CD-restrictions. Herein, we handle tree-like CD-constraints
between an element and direct successor elements. By contrast, the CD-restrictions
for the description logic involved in~\cite[Theorem~28]{Labai&Ortiz&Simkus20} are about constraints along a {\em single} finite path. 
Besides, the extensions from the previous subsections can also be added
to   $\ALCO(\Zed)$ and $\ALCO(\Nat)$ and still preserving the
\exptime-membership of the consistency problem; typically adding functional role names.

\section{Conclusion}
\label{section-conclusion}

We have considered a class of concrete domains $\acdomain$ for which
the consistency problem for $\ALCO(\acdomain)$ can be reduced to
the nonemptiness problem for constraint automata parameterised
by $\acdomain$, leading to the \exptime-membership of the
consistency problem (Theorem~\ref{theorem-main}). This follows an automata-based approach,
whose versatility allowed us to add ingredients such as
inverse roles (while giving up nominals) and functional role names (Section~\ref{section-extensions}).
Furthermore, we carefully examined the conditions on the concrete domain $\acdomain$
to get the \exptime upper bound. In particular, we have shown
that the nonemptiness problem for constraint automata
is in \exptime whenever $\acdomain$ satisfies
the conditions (C1), (C2) and (C3.1) (Theorem~\ref{theorem-tgca}).
Below, we present a schematic presentation of the reductions,
the red conditions are those that allow the reductions,
the blue ones are related to the computational complexity. 

\begin{center}
\scalebox{1}{
\begin{tikzpicture}[node distance=1.5cm,scale=1,draw=black,rounded corners]

\node (CONS) at (0,0)  {Consistency ($\ALCO({\acdomain})$)};

 \node (NETGCA) at (6.1,0)  {\neproblem{TGCA($\acdomain$)}};

 \node (NEBTA) at (11.2,0)  {\neproblem{BTA}};

  \path[->,line width=1.8pt]
  (CONS) edge node[above,text=red] {(C4)} node[below,text=blue] {(C2)(C3.k)} (NETGCA);

\path[->,line width=1.8pt]
  (NETGCA) edge node[above,text=red] {(C1)} node[below,text=blue] {(C2)(C3.k)} (NEBTA);

\end{tikzpicture}
}
\end{center}

On the downside, the relationships between
the completion property, homomorphism $\omega$-compactness
and the amalgamation property are still to be characterised. 
The decidability/complexity status of
the logics $\ALCOI(\acdomain)$ is left open (also evoked in~\cite[Section~7]{Baaderetal25}  for some
extension). 

\cut{
The work can be continued in many ways. For instance,
is it possible to get similar complexity results for a class
of concrete domains that do not satisfy the condition (C1)
with an automata-based approach? Typically, such a class
may include $\triple{\Nat}{<}{=}$, $\triple{\Zed}{<}{=}$
or $\triple{\set{0,1}^*}{\prefix}{=}$. What about
general results for concrete domains built over  the string domain
$\set{0,1}^*$? 
}

\bibliographystyle{alphaurl}
\bibliography{bb-lmcs.bib}

@string{FI = "Fundamenta Informaticae"}

@string{IC = "Information \& Computation"}

@string{JAR = "Journal of Automated Reasoning"}

@string{JCSS = "Journal of Computer and System Sciences"}

@string{LNCS = "LNCS"}

@string{LMCS = "Logical Methods in Computer Science"}

@string{LNAI = "LNAI"}

@string{SIAM = "SIAM Journal of Computing"}

@inproceedings{CarapelleTurhan16,
  author    = {C. Carapelle and
               A.{-}Y. Turhan},
  title     = {Description {L}ogics {R}easoning w.r.t. {G}eneral {TB}oxes Is {D}ecidable for
               {C}oncrete {D}omains with the {EHD}-Property},
  booktitle = {ECAI'16},
  volume    = {285},
  pages     = {1440--1448},
  publisher = {{IOS} Press},
  year      = {2016},
}

@InProceedings{Labai&Ortiz&Simkus20,
  author = 	"N. Labai and  M. Ortiz and M. Simkus",
  title = 	"An {E}xpTime {U}pper {B}ound for $\mathcal{ALC}$ with Integers",
  booktitle = 	"KR'20",
  year = 	"2020",
  OPTeditor = 	"",
  pages = 	"425--436",
  OPTorganization = 	"",
  publisher = 	"Morgan Kaufmann",
  OPTaddress = 	"",
  OPTmonth = 	"",
}

@InProceedings{Segoufin&Torunczyk11,
  author = 	 {L. Segoufin and S. Toru{\'n}czyk},
  title = 	 {Automata based verification over linearly ordered data domains},
  booktitle = {STACS'11},
  pages = 	 {81--92},
  year = 	 {2011},
}

@InProceedings{Lutz03,
  author =       {C. Lutz},
  title =        {Description Logics with Concrete Domains---A Survey},
  OPTeditors =      {Philippe Balbiani and Nobu-Yuki Suzuki and Frank Wolter and Michael Zakharyaschev},
  booktitle =    {Advances in Modal Logics Volume 4},
  publisher =    {King's College Publications},
  year =         {2003},
  pages = "265--296",
}

@PhdThesis{Lutz02,
  author = 	"C. Lutz",
  title = 	{The Complexity of Description Logics with Concrete Domains},
  school = 	"RWTH, Aachen",
  year = 	"2002",
  OPTaddress = 	"",
  OPTmonth = 	"",
  OPTnote = 	""
}

@article{Lutz&Milicic07,
  author    = {C. Lutz and
               M. Milici{\'c}},
  title     = {A {T}ableau {A}lgorithm for {D}escription {L}ogics with {C}oncrete {D}omains and
               {G}eneral {T}Boxes},
  journal   = JAR,
  volume    = {38},
  number    = {1-3},
  pages     = {227--259},
  year      = {2007},
}

@InProceedings{Balbiani&Condotta02,
  author = 	"Ph. Balbiani and J.F. Condotta",
  title = 	"Computational Complexity of propositional linear temporal logics based on 
                  qualitative spatial or temporal reasoning",
  booktitle = 	"FroCoS'02",
  year = 	"2002",
  OPTeditor = 	"A. Armando",
  pages = 	"162--173",
  volume = "2309",
  series = LNAI,
  publisher = 	"Springer",
}

@InProceedings{Baader&Hanschke91,
  author = 	"F. Baader and P. Hanschke",
  title = 	"A scheme for integrating concrete domains into concept languages",
  booktitle = 	"IJCAI'91",
  year = 	"1991",
  pages = 	"452--457",
}

@Article{Vardi&Wolper86,
  author = 	"M.Y. Vardi and P. Wolper",
  title = 	"Automata-theoretic techniques for modal logics of programs",
  journal = 	JCSS,
  year = 	"1986",
  volume = 	"32",
  OPTnumber = 	"",
  pages = 	"183--221",
  OPTmonth = 	"",
  OPTnote = 	""
}

@book{Baaderetal17,
  author    = {F. Baader and
               I. Horrocks and
               C. Lutz and
               U. Sattler},
  title     = {An Introduction to Description Logic},
  publisher = {Cambridge University Press},
  year      = {2017},
}

@InProceedings{Buchi62,
  author = 	"J.R. B{\"u}chi",
  title = 	"On a decision method in restricted second-order arithmetic",
  booktitle = 	"International Congress on Logic, Method and Philosophical Science'60",
  year = 	"1962",
  OPTeditor = 	"",
  pages = 	"1--11",
  OPTorganization = 	"",
  OPTpublisher = 	"",
  OPTaddress = 	"",
  OPTmonth = 	"",
  OPTnote = 	""
}

@Article{Dechter92,
  author =      "R. Dechter",
  title =       "From local to global consistency",
  journal =     "Artificial Intelligence",
  year =        "1992",
  OPTvolume =   "",
  OPTnumber =   "",
  pages =    "87--107",
  OPTmonth =    "",
  OPTnote =        ""
}

@Article{Allen83,
  author = 	 {J. Allen},
  title = 	 {Maintaining knowledge about temporal intervals},
  journal = 	 {Communications of the ACM},
  year = 	 {1983},
  volume = 	 {26},
  number = 	 {11},
  pages = 	 {832--843},
  OPTmonth = 	 {},
  OPTnote = 	 {},
  OPTannote = 	 {}
}

@InProceedings{Wolter&Zakharyaschev00,
  author = 	 {F. Wolter and M. Zakharyaschev},
  title = 	 {Spatio-temporal representation and reasoning based on {RCC}-8},
  booktitle = {KR'00},
  pages = 	 {3--14},
  year = 	 {2000},
  OPTeditor = 	 {A. Cohn and F. Giunchiglia and B. Selman},
  OPTpublisher = {},
}

@PhdThesis{Labai21,
  author = 	 {N. Labai},
  title = 	 {Automata-based reasoning for decidable logics with data values},
  school = 	 {TU Wien},
  year = 	 {2021},
  month = 	 {May},
}

@inproceedings{Peterler&Quaas22,
  author    = {D. Peteler and
               K. Quaas},
  title     = {Deciding {E}mptiness for {C}onstraint {A}utomata on {S}trings with the {P}refix
               and {S}uffix {O}rder},
  booktitle = {MFCS'22},
  series    = {LIPIcs},
  volume    = {241},
  pages     = {76:1--76:15},
  publisher = {Schloss Dagstuhl - Leibniz-Zentrum f{\"{u}}r Informatik},
  year      = {2022},
}

@article{Baader&Rydval22,
  author    = {F. Baader and
               J. Rydval},
  title     = {Using Model Theory to Find Decidable and Tractable Description Logics
               with Concrete Domains},
  journal   = JAR,
  volume    = {66},
  number    = {3},
  pages     = {357--407},
  year      = {2022},
}

@inproceedings{Baader09,
  author    = {F. Baader},
  title     = {{D}escription {L}ogics},
  booktitle = {Reasoning Web. Semantic Technologies for Information Systems, 5th
               International Summer School 2009, Tutorial Lectures},
  series    = LNCS,
  volume    = {5689},
  pages     = {1--39},
  publisher = {Springer},
  year      = {2009},
}

@Misc{Kartzow&Weidner15,
  author = 	 {A. Kartzow and
               Th. Weidner},
  title = 	 {Model Checking Constraint {LTL} over Trees},
  howpublished = {CoRR, abs/1504.06105},
  year = 	 {2015},
}

@InProceedings{Lutz01b,
  author = 	"C. Lutz",
  title = 	"{NEXPTIME}-Complete Description Logics with Concrete Domains",
  booktitle = 	"IJCAR'01",
  year = 	"2001",
  pages = 	"46--60",
  OPTorganization = 	"",
  series = LNCS,
  volume = "2083",
  publisher = 	"Springer",
}

@PhdThesis{Rydval22,
  author = 	 {J. Rydval},
  title = 	 {Using Model Theory to Find Decidable and Tractable Description Logics with Concrete Domains},
  school = 	 {Dresden University},
  year = 	 {2022},
}

@Article{Baaderetal03bis,
  author = 	 {F. Baader and J. Hladik and C. Lutz and F. Wolter},
  title = 	 {From {T}ableaux to {A}utomata for {D}escription {L}ogics},
  journal = 	 FI,
  year = 	 {2003},
  volume = 	 {57},
  number = {2--4},
  pages = 	 {247--279},
  OPTmonth = 	 {},
  OPTnote = 	 {},
  OPTannote = 	 {}
}

@inproceedings{Borgwardt&DeBortoli&Koopmann24,
  author       = {S. Borgwardt and
                  F. {De Bortoli} and
                  P. Koopmann},
  title        = {The Precise Complexity of Reasoning in $\mathcal{ALC}$ with $\omega$-Admissible Concrete Domains},
  booktitle    = {DL'24},
  series       = {{CEUR} Workshop Proceedings},
  volume       = {3739},
  publisher    = {CEUR-WS.org},
  year         = {2024}
}

@inproceedings{Demri&Quaas23bis,
  author       = {S. Demri and K. Quaas},
  title        = {First Steps Towards Taming Description Logics with Strings},
  booktitle    = {JELIA'23},
  series       = LNCS,
  volume       = {14281},
  pages        = {322--337},
  publisher    = {Springer},
  year         = {2023}
}

@Article{Demri&Sattler02,
  author = 	"S. Demri and U. Sattler",
  title = 	"Automata-theoretic decision procedures for information logics",
  journal = 	FI,
  year = 	"2002",
  volume = 	"53",
  number = 	"1",
  pages = 	"1--22",
}

@Article{Demri&DSouza07,
  author = 	 {S. Demri and D. D'Souza},
  title = 	 "An automata-theoretic approach to constraint {LTL}",
  journal = 	 IC,
  year = 	 {2007},
  volume = "205",
  number = "3", 
  pages = "380--415", 
  OPTnote = 	 {},
}

@inproceedings{Chatterjee&Henzinger12,
  author       = {K. Chatterjee and
                  M. Henzinger},
  OPTeditor       = {Yuval Rabani},
  title        = {An $\mathcal{O}(n^2)$ time algorithm for alternating
                  {B}{\"{u}}chi games},
  booktitle    = {SODA'12},
  pages        = {1386--1399},
  publisher    = {{SIAM}},
  year         = {2012},
}

@InProceedings{Pratt79,
  author = 	"V. Pratt",
  title = 	"Models of program logics",
  booktitle = 	"FoCS'79",
  year = 	"1979",
  pages = 	"115--122",
}

@Book{Blackburn&deRijke&Venema01,
  author = 	"P. Blackburn and M. de Rijke  and Y. Venema",
  title = 	"Modal Logic",
  publisher = 	"Cambridge University Press",
  year = 	"2001",
}

@inproceedings{Baaderetal25,
  author       = {F. Baader and
                  S. Borgwardt and
                  F. {De Bortoli} and
                  P. Koopmann},
  OPTeditor       = {Clark W. Barrett and
                  Uwe Waldmann},
  title        = {Concrete Domains Meet Expressive Cardinality Restrictions in Description
                  Logics},
  booktitle    = {CADE'25},
  series       = LNCS,
  volume       = {15943},
  pages        = {676--695},
  publisher    = {Springer},
  year         = {2025},
}

@Book{Demri&Goranko&Lange16,
  author = 	"S. Demri and V. Goranko and M. Lange",
  title = 	"Temporal Logics in Computer Science",
  publisher = 	"Cambridge University Press",
  year = 	"2016",
}

@Article{Demri&Quaas25,
  author = 	 {S. Demri and K. Quaas},
  title = 	 {Constraint Automata on Infinite Data Trees: From {CTL(Z)/CTL*(Z)} To
               Decision Procedures},
  journal = 	 LMCS,
  year = 	 {2025},
  volume = 	 {21},
  number = 	 {2},
  pages = 	 {16:1--16:79},
}

@article{Bhaskar&Praveen24,
  author       = {A. Bhaskar and
               M. Praveen},
  title        = {Realizability Problem for Constraint {LTL}},
  journal      = IC,
  volume       = {296},
  pages        = {105126},
  year         = {2024},
}

@Book{PrattHartmann23,
  author = 	 {I. Pratt-Hartmann},
  title = 	 {Fragments of First-Order Logic},
  publisher = 	 {Oxford University Press},
  year = 	 {2023},
}

@incollection{Baader&Lutz06,
  author       = {F. Baader and C. Lutz},
  editor       = {P. Blackburn and J. van Benthem and F. Wolter},
  title        = {Description Logic},
  booktitle    = {Handbook of Modal Logic},
  pages        = {757--819},
  publisher    = {Elsevier},
  year         = {2006},
}

@incollection{Vardi06,
  author       = {M.Y. Vardi},
  editor       = {P. Blackburn and J. van Benthem and F. Wolter},
  title        = {Automata-theoretic Techniques for Temporal Reasoning},
  booktitle    = {Handbook of Modal Logic},
  pages        = {971--989},
  publisher    = {Elsevier},
  year         = {2006},
}

@incollection{Marx06,
  author       = {M. Marx},
  editor       = {P. Blackburn and J. van Benthem and F. Wolter},
  title        = {Complexity of Modal Logic},
  booktitle    = {Handbook of Modal Logic},
  pages        = {139--179},
  publisher    = {Elsevier},
  year         = {2006},
}

@incollection{Baader&Horrocks&Sattler08,
  author       = {F. Baader and
                  I. Horrocks and
                  U. Sattler},
  editor       = {F. {van Harmelen} and
                  V. Lifschitz and
                  B.W. Porter},
  title        = {Description Logics},
  booktitle    = {Handbook of Knowledge Representation},
  series       = {Foundations of Artificial Intelligence},
  volume       = {3},
  pages        = {135--179},
  publisher    = {Elsevier},
  year         = {2008},
}

@Article{Tobies00,
  author = 	"S. Tobies",
  title = 	"The Complexity of Reasoning with cardinality restrictions
  and nominals in expressive description logics",
  journal = 	"Journal of Artificial Intelligence Research",
  year = 	"2000",
  volume = 	"12",
  pages = 	"199--217",
}

@InProceedings{Sattler&Vardi01,
  author = 	"U. Sattler and M.Y. Vardi",
  title = 	{The hybrid mu-calculus},
  booktitle = 	"IJCAR'01",
  OPTeditor =	 "A. Leitsch and R. Gor{\'e} and T. Nipkow",
  year = 	2001,
  pages = 	"76--91",
  series =      LNAI,
  volume =      "2083",
  publisher = 	"Springer"
}

@inproceedings{Dauvier&Filiot&Reynier26,
  author       = {N. Dauvier and
                  E. Filiot and
                  P.{-}A. Reynier},
  OPTeditor       = {Stefano Guerrini and
                  Barbara K{\"{o}}nig},
  title        = {Register-Bounded Synthesis from Constraint {LTL}},
  booktitle    = {CSL'26},
  series       = {LIPIcs},
  volume       = {363},
  pages        = {8:1--8:18},
  publisher    = {Schloss Dagstuhl - Leibniz-Zentrum f{\"{u}}r Informatik},
  year         = {2026},
}

@inproceedings{Demri&Gu26,
  author       = {S. Demri and
                  T. Gu},
  OPTeditor       = {Stefano Guerrini and
                  Barbara K{\"{o}}nig},
  title        = {Robustness of Constraint Automata for Description Logics with Concrete
                  Domains},
  booktitle    = {CSL'26},
  series       = {LIPIcs},
  volume       = {363},
  pages        = {42:1--42:18},
  publisher    = {Schloss Dagstuhl - Leibniz-Zentrum f{\"{u}}r Informatik},
  year         = {2026},
}

\end{document}